\documentclass[11pt]{article}
\usepackage[utf8]{inputenc}
\usepackage{graphicx}
\usepackage{amsmath}
\usepackage{tabularx}
\usepackage{amssymb}
\usepackage{parskip}
\usepackage{textcomp}
\usepackage{bbm} 
\usepackage[dvipsnames]{xcolor}
\usepackage{hyperref}
\usepackage{cancel}
\usepackage{soul}
\usepackage{ragged2e}
\usepackage{bm}
\usepackage{romannum}
\usepackage{mathtools}
\usepackage{float}
\usepackage{tikz}
\usetikzlibrary{positioning}
\usetikzlibrary{decorations.text}
\usetikzlibrary{decorations.pathmorphing}
\usetikzlibrary{decorations.pathreplacing}
\usetikzlibrary{patterns}
\tikzset{snake it/.style={decorate, decoration={snake, segment length=1.5mm, amplitude=1pt}}}

\graphicspath{ {images/} }

\usepackage{geometry}
\AtBeginDocument{\pagenumbering{arabic}}

\title{\vspace{2mm}\LARGE Timelike bounce hypersurfaces in charged null dust collapse\vspace{-0mm}}
\author{David Bick\\ \\ \footnotesize
\textit{Department of Pure Mathematics and Mathematical Statistics,}\\ \footnotesize \textit{
University of Cambridge, Wilberforce Road, Cambridge CB3 0WB, United Kingdom}\vspace{5mm}}
\date{\today}

\begin{document}
\maketitle
\renewcommand{\labelenumi}{(\roman{enumi})}

\begin{center}
\begin{minipage}{0.85\textwidth}
\begin{center}
\footnotesize
\vspace{4mm}\textbf{Abstract}\vspace*{2mm}\\
\justifying
\hspace*{4mm}We establish results on the dynamics of interacting charged null fluids in general relativity, specifically in the context of the bouncing continuation proposed in \hyperlink{Ori91}{[Ori91]}. In this model---the setting for a number of prominent case studies on black hole formation---charged massless particles may instantaneously change direction (bounce) after losing all their 4-momentum due to electrostatic repulsion. We initiate the study of \textit{timelike} bounce hypersurfaces in spherical symmetry: scenarios in which an incoming beam of charged null dust changes direction along a timelike surface $\mathcal{B}$, which is the (free) boundary of an interacting 2-dust region. We identify a novel decoupling of the equations of motion in this region.\\
\hspace*{4mm}First, it is shown that every timelike curve segment $\gamma$ in the spherically symmetric quotient of Minkowski or Reissner-Nordstr\"om spacetimes arises as the bounce hypersurface $\mathcal{B}$ of a charged null dust beam incident from past null infinity $\mathcal{I}^-$. We construct a spacetime $(\mathcal{M},g_{\mu\nu})$ describing the full trajectory of the beam, which includes gluing to Reissner-Nordstr\"om and Vaidya regions. Across $\mathcal{B}$ the metric has regularity $g_{\mu\nu}\in C^{2,1}$ and satisfies Einstein's equation classically, while $C^\infty$ gluing may be achieved across all other interfaces. We also obtain examples of timelike bounce hypersurfaces terminating in a null point.\\
\hspace*{4mm}Since these constructions are teleological, we secondly consider a given charged incoming beam from past null infinity. We formulate and solve a free boundary problem which represents the formation of a timelike bounce hypersurface. The result is conditional, applying only in the exterior region of a Reissner-Nordstr\"om spacetime, and subject to a technical regularity condition. 
\end{center}
\end{minipage}
\end{center}
\vspace{1cm}
\Large\textbf{Contents}\small\vspace*{4mm}\\
\hyperlink{sec:1}{\textbf{1\hspace{3mm}Introduction}}\hfill \textbf{\pageref{1}} \\
\hyperlink{sec:1.1}{\hspace{5mm}\text{1.1\hspace{3mm}Null dust models}}\dotfill \pageref{1.1}\\
\hyperlink{sec:1.2}{\hspace{5mm}\text{1.2\hspace{3mm}The charged Vaidya metric}}\dotfill \pageref{1.2}\\
\hyperlink{sec:1.3}{\hspace{5mm}\text{1.3\hspace{3mm}Ori's bouncing continuation}}\dotfill \pageref{1.3}\\
\hyperlink{sec:1.4}{\hspace{5mm}\text{1.4\hspace{3mm}Prescribed timelike bounce hypersurfaces}}\dotfill \pageref{1.4} \\
\hyperlink{sec:1.5}{\hspace{5mm}\text{1.5\hspace{3mm}The formation problem for timelike bounces}}\dotfill \pageref{1.5}\vspace*{2mm} \\
\hyperlink{sec:2}{\textbf{2\hspace{3mm}Preliminaries}}\hfill \textbf{\pageref{2}}\\
\hyperlink{sec:2.1}{\hspace{5mm}\text{2.1\hspace{3mm}Spherical symmetry and double null gauge}}\dotfill \pageref{2.1}\\
\hyperlink{sec:2.2}{\hspace{5mm}\text{2.2\hspace{3mm}Electromagnetic fields and the Lorentz force}}\dotfill \pageref{2.2}\\
\hyperlink{sec:2.3}{\hspace{5mm}\text{2.3\hspace{3mm}Equations of motion in the interacting region}}\dotfill \pageref{2.3} \\
\hyperlink{sec:2.4}{\hspace{5mm}\text{2.4\hspace{3mm}The $(r,\kappa,Q)$ system}}\dotfill \pageref{2.4} \\
\hyperlink{sec:2.5}{\hspace{5mm}\text{2.5\hspace{3mm}The $(r,\varpi,\phi,Q)$ system}}\dotfill \pageref{2.5}\vspace*{2mm} \\
\hyperlink{sec:3}{\textbf{3\hspace{3mm}Scattering results}}\hfill \textbf{\pageref{3}}\\
\hyperlink{sec:3.1}{\hspace{5mm}\text{3.1\hspace{3mm}Examples with prescribed timelike bounce hypersurface}}\dotfill \pageref{3.1}\\
\hyperlink{sec:3.2}{\hspace{5mm}\text{3.2\hspace{3mm}Examples of termination at a null point}}\dotfill \pageref{3.2}\vspace*{2mm}\\
\hyperlink{sec:4}{\textbf{4\hspace{3mm}Formation of a timelike bounce hypersurface}}\hfill \textbf{\pageref{4}}\\
\hyperlink{sec:4.1}{\hspace{5mm}\text{4.1\hspace{3mm}Setup of the free boundary problem}}\dotfill \pageref{4.1}\\
\hyperlink{sec:4.2}{\hspace{5mm}\text{4.2\hspace{3mm}Description of the iteration scheme}}\dotfill \pageref{4.2}\\
\hyperlink{sec:4.3}{\hspace{5mm}\text{4.3\hspace{3mm}Existence of the iteration scheme}}\dotfill \pageref{4.3}\\
\hyperlink{sec:4.4}{\hspace{5mm}\text{4.4\hspace{3mm}Contraction estimates}}\dotfill \pageref{4.4}\\
\hyperlink{sec:4.5}{\hspace{5mm}\text{4.5\hspace{3mm}Proof of Theorem 1.3}}\dotfill \pageref{4.5}\vspace*{2mm} \\ \\ \\

\Large\hypertarget{sec:1}{\textbf{1\hspace{4mm}Introduction}}\label{1}\normalsize\\ \\
This article examines \textit{null dust} models in general relativity: Einstein-matter models consisting of self-gravitating, pressureless and massless perfect fluids. Null dust models have provided theorists with a fundamental tool for probing the behaviour of solutions to the Einstein equation 
\begin{equation}
\text{Ric}[g]_{\mu\nu}-\tfrac{1}{2}R[g]g_{\mu\nu}=2 T_{\mu\nu} \tag{\hypertarget{eqn:1.1}{1.1}}
\end{equation}
clarifying issues of principle in the study of gravitational collapse, singularity formation, black hole interiors, strong cosmic censorship, and the role played by energy conditions, to name only a few. We provide answers to some basic outstanding questions about the behaviour and well-posedness of this fundamental system, specifically in spherically symmetric scenarios involving overlapping ingoing and outgoing  \textit{charged} flows: see already \hyperlink{thm:1.1}{Theorems~1.1} and \hyperlink{thm:1.3}{1.3}.\\ \\ \\
\large\hypertarget{sec:1.1}{\textbf{1.1\hspace{4mm}Null dust models}}\label{1.1}\normalsize\\ \\
Null dust models provide a rich family of examples pertinent to gravitational collapse. More generally, self-gravitating null dusts may be considered a proxy for high-frequency behaviour of the \textit{vacuum} Einstein equations, providing a powerful motivation for their study. This has been made precise in recent studies (see [\hyperlink{LR20}{LR20}; \hyperlink{Tou25}{Tou25}] and references therein) in connection to the Burnett conjecture \hyperlink{Bur89}{[Bur89]}. \\ \\
When studying a single null dust, the matter equations take the form 
\begin{equation*}
k^\nu\nabla_\nu k^\mu=0\tag{\hypertarget{eqn:1.2}{1.2}}
\end{equation*}
\begin{equation*}
\nabla_\mu\big(\rho k^\mu\big)=0\tag{\hypertarget{eqn:1.3}{1.3}}
\end{equation*}
for a null 4-velocity vector field $k$ and non-negative energy density $\rho$. These are coupled to \hyperlink{eqn:1.1}{(1.1)} via the relation
\begin{equation*}
T_{\mu\nu}=\rho k_\mu k_\nu\tag{\hypertarget{eqn:1.4}{1.4}}
\end{equation*}
In spherical symmetry, an explicit solution to \hyperlink{eqn:1.1}{(1.1)}-\hyperlink{eqn:1.4}{(1.4)} is available, introduced by \textsc{Vaidya} \hyperlink{Vai51}{[Vai51]}. Covering the quotient manifold $\mathcal{Q}=\mathcal{M}/SO(3)$ by ingoing Eddington-Finkelstein-type coordinates $(v,r)$ (valued in $v\in\mathbb{R}$, $r\in(0,\infty)$), we have
\begin{equation*}
g=-\bigg(1-\frac{2m(v)}{r}\bigg)dv^2+2dvdr+r^2g_{S^2}\tag{\hypertarget{eqn:1.5}{1.5}}
\end{equation*}
\begin{equation*}
k=-\partial_r\qquad\rho=\frac{m'(v)}{r^2}\tag{\hypertarget{eqn:1.6}{1.6}}
\end{equation*}
Note that, throughout the article, we will perform our analysis on $\mathcal{Q}$, and speak interchangeably of curves and hypersurfaces, because curves in $\mathcal{Q}$ correspond to hypersurfaces `upstairs' in $\mathcal{M}$. This \textit{Vaidya metric} \hyperlink{eqn:1.5}{(1.5)} (sometimes \textit{radiating Schwarzschild metric}), in which $m(v)$ is an arbitrarily prescribed nondecreasing function, provides an infinite-dimensional family of explicit solutions to \hyperlink{eqn:1.1}{(1.1)}. Physically, \hyperlink{eqn:1.5}{(1.5)} corresponds to a stream of radiation injected radially from infinity. Being even simpler to set up and study than Tolman-Bondi (i$.$e$.$ massive) dust, the metric \hyperlink{eqn:1.5}{(1.5)} and its outgoing counterpart became the subject of a vast number of studies on gravitational collapse and singularity formation. 
\\ \\
Though outside the class of explicit solutions, the case of two overlapping null dusts remains tractable, and allows for the modelling of energy transfer and backreaction. The seminal paper of \textsc{Poisson-Israel} \hyperlink{PI90}{[PI90]} presented a study of ingoing and outgoing null dust flows in a charged black hole interior, discovering the phenomenon of \textit{mass inflation} at the Cauchy horizon, now essential in discussion of strong cosmic censorship. Their study added nuance to an earlier result of \textsc{Hiscock} \hyperlink{His81}{[His81]} which examined a single ingoing null flow on a charged background (see below). We commend to the reader the discussion in \hyperlink{Daf14}{[Daf14]}. As another prominent example of interacting null dusts, we mention also the AdS instability mechanism introduced by \textsc{Moschidis} \hyperlink{Mos17}{[Mos17]}, predicated on a catastrophic energy transfer between carefully arranged null dust beams.
\\ \\ \\
\large\hypertarget{sec:1.2}{\textbf{1.2\hspace{4mm}The charged Vaidya metric}}\label{1.2}\normalsize\\ \\
The Vaidya metric \hyperlink{eqn:1.5}{(1.5)} admits a straightforward generalization to Einstein-Maxwell theory. The null dust particles, now possessing a fundamental charge $\mathfrak{e}\in\mathbb{R}_{>0}$, source a Maxwell 2-form $F$ according to
\begin{equation*}
\nabla^\alpha F_{\mu\alpha}=\mathfrak{e}\rho k_\mu\tag{\hypertarget{eqn:1.7}{1.7}}
\end{equation*}
which in turn provides its usual electromagnetic energy-momentum contribution:
\begin{equation*}
T_{\mu\nu}=\rho k_\mu k_\nu+{F_\mu}^\alpha F_{\nu\alpha}-\tfrac{1}{4}g_{\mu\nu}F_{\alpha\beta}F^{\alpha\beta}\tag{\hypertarget{eqn:1.8}{1.8}}
\end{equation*}
The \textit{charged Vaidya metric} (sometimes \textit{Vaidya-Bonnor metric})
\begin{equation*}
g=-\bigg(1-\frac{2\varpi(v)}{r}+\frac{Q(v)^2}{r^2}\bigg)dv^2+2dvdr+r^2g_{S^2}\tag{\hypertarget{eqn:1.9}{1.9}}
\end{equation*}
solves equations \hyperlink{eqn:1.1}{(1.1)}-\hyperlink{eqn:1.3}{(1.3)} and \hyperlink{eqn:1.7}{(1.7)}-(\hyperlink{eqn:1.8}{1.8)}, together with
\begin{equation*}
F=-\frac{Q}{r^2}dv\wedge dr\tag{\hypertarget{eqn:1.10}{1.10}}
\end{equation*}
\begin{equation*}
k=-\partial_r,\qquad \rho=\frac{1}{r^2}\bigg(\varpi'(v)-\frac{Q(v)Q'(v)}{r}\bigg)\tag{\hypertarget{eqn:1.11}{1.11}}
\end{equation*}
The line element \hyperlink{eqn:1.9}{(1.9)} and its outgoing counterpart were first obtained through an algebraic classification of possible Einstein tensors in spherical symmetry \hyperlink{PS68}{[PS68]}, before receiving their physical interpretation by \textsc{Vaidya-Bonnor} \hyperlink{BV70}{[BV70]}. While the metric thus arises as a solution to a self-consistent system of Einstein and matter equations, the matter source is exotic, consisting of `charged photons'\footnote[1]{In the words of \hyperlink{SI80}{[SI80]}, the line element \hyperlink{eqn:1.9}{(1.9)} describes ``the radial infall of a coherent stream of charged ‘photons’. The use of lightlike particles automatically avoids shell-crossing singularities and obviates the need for radial stresses to prevent their occurrences. This eliminates all our problems except how to charge a photon!''}. However, to make one remark in its defence: in view of results on the reverse Burnett conjecture mentioned above, it is reasonable to expect that \textit{charged} null dusts may arise as suitable high-frequency limits of solutions to more fundamental models of charged matter.
\\ \\
As before, $\varpi(v)$ is nondecreasing, but $Q(v)$ may be either nondecreasing or nonincreasing, according to whether the radiation is like- or oppositely-charged to the background spacetime\footnote[2]{The special case in which $Q$ is a non-zero constant is sometimes called the \textit{Reissner-Nordstr\"om-Vaidya metric}. This is the case examined in \hyperlink{His81}{[His81]}.}. We consider only the former possibility, since the bouncing phenomenon to be examined only occurs in the like-charged case. These functions, known hereafter as \textit{(ingoing) Vaidya seed data}, are otherwise freely prescribed, making \hyperlink{eqn:1.9}{(1.9)} an especially convenient tool for studying the third law of black hole thermodynamics. Consideration of the charged Vaidya metric by \textsc{Sullivan-Israel} \hyperlink{SI80}{[SI80]} led the latter author to formulate the third law in a way which explicitly imposes the weak energy condition. In \hyperlink{SI80}{[SI80]}, the authors produced counterexamples to the third law using the charged Vaidya metric, but observed that the weak energy condition is violated near the apparent horizon in all such examples. Indeed, immediately from \hyperlink{eqn:1.11}{(1.11)}, the energy density $\rho$ becomes negative in the region 
\begin{equation*}
r<r_b(v):=\frac{Q(v)Q'(v)}{\varpi'(v)}\tag{\hypertarget{eqn:1.12}{1.12}}
\end{equation*}
Ironically, these examples were interpreted as an argument \textit{for} the third law, if suitably qualified by imposing an energy condition. In fact, on the contrary, recent examples [\hyperlink{KU22}{KU22}; \hyperlink{KU24}{KU24}] have shown that this qualification is not enough to salvage the third law: see \hyperlink{sec:1.3}{Section~1.3}. In any case, the puzzling fact remained that violations of the weak energy condition seemed to arise readily in the dynamics of charged null dust, even starting from initial data with $\rho>0$. This would make charged null dust pathological, whether or not its equations of motion are self-consistent.
\\ \\ \\
\large\hypertarget{sec:1.3}{\textbf{1.3\hspace{4mm}Ori's bouncing continuation}}\label{1.3}\normalsize\\ \\
These concerns were resolved by \textsc{Ori}, who showed that the problems with \hyperlink{eqn:1.9}{(1.9)} lay in its interpretation in the contemporary literature [\hyperlink{SI80}{SI80}; \hyperlink{LZ91}{LZ91}]. We briefly review the author's arguments, but commend to the reader the full discussion in \hyperlink{Ori91}{[Ori91]} and in $[$\hyperlink{KU24}{KU24}, Section 4$]$. Ori's insight was that the same metric \hyperlink{eqn:1.9}{(1.9)} also arises in the solution to another system of matter equations. Namely, when the geodesic equation \hyperlink{eqn:1.2}{(1.2)} is replaced by a generalization of the Lorentz force law for massless particles\footnote[3]{Integral curves of $k$ satisfying \hyperlink{eqn:1.13}{(1.13)} are called \textit{electromagnetic geodesics}.}
\begin{equation*}
k^\nu\nabla_\nu k^\mu=\mathfrak{e}{F^\mu}_{\nu}k^\nu\tag{\hypertarget{eqn:1.13}{1.13}}
\end{equation*}
then a solution is obtained by taking the same metric \hyperlink{eqn:1.9}{(1.9)}, together with
\begin{equation*}
k=\frac{\mathfrak{e}}{Q'(v)}\bigg(\varpi'(v)-\frac{Q(v)Q'(v)}{r}\bigg)(-\partial_r),\qquad \rho=\frac{Q'(v)^2}{\mathfrak{e}^2r^2}\bigg(\varpi'(v)-\frac{Q(v)Q'(v)}{r}\bigg)^{-1}\tag{\hypertarget{eqn:1.14}{1.14}}
\end{equation*}
In contrast to the predictions of \hyperlink{eqn:1.11}{(1.11)}, the particles' 4-momentum $k$ decays to zero as the fluid shells contract. Thus electrostatic charge causes the particles to `decelerate', even while their worldlines remain null. In fact, $k$ vanishes at a positive value of $r$, the same $r=r_b(v)$ where the energy density $\rho$ vanished in \hyperlink{eqn:1.11}{(1.11)}. After incorporating a Lorentz force term, $\rho$ now \textit{diverges} instead! Note also that the \textit{number current} $N:=\rho k$ tends to a non-zero finite limit, which we will continue to see in the sequel.\\ \\
In addition to this new singular behaviour, one also confronts an ambiguity in solving \hyperlink{eqn:1.13}{(1.13)}. Let $\gamma$ be a radial null electromagnetic geodesic on a fixed charged spherically symmetric background, with tangent vector $k\to0$ at $p\in\mathcal{M}$. Then the subsequent solution beyond $p$ is not uniquely determined, admitting both ingoing and outgoing continuations consistent with \hyperlink{eqn:1.13}{(1.13)}. Since the outgoing case is selected by perturbations of $\gamma$ with small angular momentum, Ori concluded that this \textit{bouncing continuation} is correct for $\gamma$ itself. Remarkably, this idea was, at the same time, proposed independently by \textsc{Dray} \hyperlink{Dra90}{[Dra90]} in the context of charged \textit{thin shells} of null dust\footnote[4]{In that author's words, ``Rather than allowing the energy density of the shell to become negative, it seems more reasonable to expect that the shell should bounce at the radius where the energy density becomes zero.''}.\\ \\
Returning to the charged Vaidya metric, which is predicated on a \textit{congruence} of ingoing null fluid trajectories, how then shall the picture be modified to incorporate a bouncing continuation into each worldline? The locations of each bounce glue together into a \textit{bounce hypersurface} $\mathcal{B}$. The Vaidya metric, which contains no outgoing flow, must then be discarded in the causal future of $\mathcal{B}$. However, a sensitive issue is at  play, seeing as the \textit{location} of $\mathcal{B}$ itself is a prediction of the solution \hyperlink{eqn:1.9}{(1.9)},\hyperlink{eqn:1.14}{(1.14)}. \\ \\
In the Vaidya metric \hyperlink{eqn:1.9}{(1.9)}, the 4-velocity vanishes at $\{r=r_b(v)\}$ (see \hyperlink{eqn:1.12}{(1.12)}). In the sequel, we will refer to this hypersurface as a \textit{curve}, since it projects to one in the spherically symmetric\vspace{1mm}
\begin{minipage}{0.59\textwidth}
quotient. When this curve is spacelike with respect to (\hyperlink{eqn:1.9}{1.9)}---that is, when
\begin{equation*}
r'_b(v)>\tfrac{1}{2}\Big(1-\frac{2\varpi(v)}{r_b(v)}+\frac{2Q^2(v)}{r^2_b(v)}\Big)
\end{equation*}
then excising its causal future leaves the region $\{r>r_b(v)\}$ untouched, and $\mathcal{B}$ coincides with $\{r=r_b(v)\}$. An outgoing Vaidya metric may be surgically attached at this boundary, and the bouncing continuation is then realized for the full congruence of fluid trajectories, as represented in the opposite Figure. Despite the blowup of $\rho$, the new metric is $C^2$ regular across $\mathcal{B}$---see [\hyperlink{Ori91}{Ori91}, Section~7]. \end{minipage}
\hspace{0.00\textwidth}
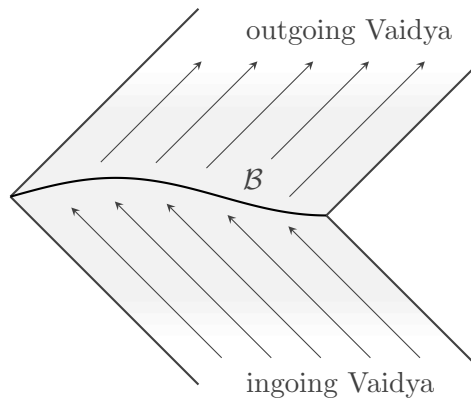
\begin{minipage}{0.41\textwidth}
\vspace{0mm}
\begin{figure}[H]
\begin{center}
\begin{tikzpicture}[scale=1]
\fill [fill=gray, draw=none,opacity=0.1]
(0,0) -- (1.2,1.2) -- (5.65,1.2) -- (4.2,-0.25) -- (5.35,-1.4) -- (1.4,-1.4) -- cycle;
\fill [fill=gray, draw=none,opacity=0.075]
(1.2,1.2) -- (5.65,1.2) -- (5.8,1.35) -- (1.35,1.35) -- cycle;
\fill [fill=gray, draw=none,opacity=0.05]
(1.35,1.35) -- (5.8,1.35) -- (5.95,1.5) -- (1.5,1.5) -- cycle;
\fill [fill=gray, draw=none,opacity=0.025]
(1.5,1.5) -- (5.95,1.5) -- (6.1,1.65) -- (1.65,1.65) -- cycle;
\fill [fill=gray, draw=none,opacity=0.075]
(5.35,-1.4) -- (1.4,-1.4) -- (1.55,-1.55) -- (5.5,-1.55) -- cycle;
\fill [fill=gray, draw=none,opacity=0.05]
(5.65,-1.7) -- (1.7,-1.7) -- (1.55,-1.55) -- (5.5,-1.55) -- cycle;
\fill [fill=gray, draw=none,opacity=0.025]
(5.65,-1.7) -- (1.7,-1.7) -- (1.85,-1.85) -- (5.85,-1.85) -- cycle;
\draw [darkgray, -stealth] (2.809, -2.144) -- (0.809, -0.144);
\draw [darkgray, -stealth] (3.468, -2.144) -- (1.388, -0.064);
\draw [darkgray, -stealth] (4.127, -2.144) -- (2.077, -0.094);
\draw [darkgray, -stealth] (4.786, -2.144) -- (2.886, -0.244);
\draw [darkgray, -stealth] (5.445, -2.144) -- (3.695, -0.394);
\draw [darkgray, -stealth] (1.202, 0.46) -- (2.542, 1.8);
\draw [darkgray, -stealth] (1.934, 0.45) -- (3.284, 1.8);
\draw [darkgray, -stealth] (2.606, 0.38) -- (4.026, 1.8);
\draw [darkgray, -stealth] (3.468, 0.5) -- (4.768, 1.8);
\draw [darkgray, -stealth] (3.71, 0) -- (5.51, 1.8);
\draw [darkgray, thick] (0,0) -- (2.5, 2.5);
\draw [darkgray, thick] (0,0) -- (2.5, -2.5);
\draw [darkgray, thick] (4.2,-0.25) -- (6.2, 1.75);
\draw [darkgray, thick] (4.2,-0.25) -- (6.2, -2.25);
\draw [thick, domain=0:3, samples=150] plot ({1.4*\x},{0.25*sin(\x*90)});
\node[darkgray] at (3, 2.2) [anchor = west] {outgoing Vaidya};
\node[darkgray] at (3, -2.5) [anchor = west] {ingoing Vaidya};
\node[darkgray] at (2.96, 0.24) [anchor = west]{$\mathcal{B}$};
\end{tikzpicture}
\end{center}
\end{figure}
\end{minipage}
\vspace*{3mm} \\
\color{black}Following \hyperlink{Ori91}{[Ori91]}, the bouncing continuation has been revisited by several authors, extending the construction, with essentially the same features, to the more general Husain null fluids \hyperlink{CB17}{[CB17]}; and to higher dimensions, AdS settings, and modified theories of gravity \hyperlink{CGV16}{[CGV16]}. We also mention especially the recent \textsc{Kehle-Unger} study \hyperlink{KU24}{[KU24]}, which applies Ori's bouncing continuation to the study of extremal black hole formation---a key application of charged null dust, as mentioned in \hyperlink{sec:1.2}{Section~1.2}. Various extremal collapse scenarios proposed in the literature involve the injection of charged matter [\hyperlink{DI67}{DI67}; \hyperlink{FH79}{FH79}; \hyperlink{Pro83}{Pr\'o83}], specifically charged thin shells. These examples were dismissed as serious counterexamples to the third law, due to their low regularity. Using Ori's bouncing continuation, the authors of \hyperlink{KU24}{[KU24]} follow (and improve upon) these collapse scenarios, prescribing a (totally geodesic\footnote[5]{This convenient choice makes the ingoing and outgoing regions exact time reflections of each other.}) spacelike $\mathcal{B}$ and attaching ingoing and outgoing Vaidya spacetimes as described above, with an interior Minkowski background. A family of spacetimes is obtained which interpolates between dispersion and black hole formation, with extremality arising on the threshold between the two [\hyperlink{KU24}{KU24}, Theorem~4 and Figure~7]. Dubbing this phenomenon \textit{extremal critical collapse}, the authors then desingularize their examples, reproducing them in the Einstein-Maxwell-Vlasov system of which charged null dust is the hydrodynamic limit [\hyperlink{KU24}{KU24}, Theorem~1]. The resulting spacetimes have $C^\infty$ regularity, so are robust against the criticisms raised against the previous singular examples.
\\ \\
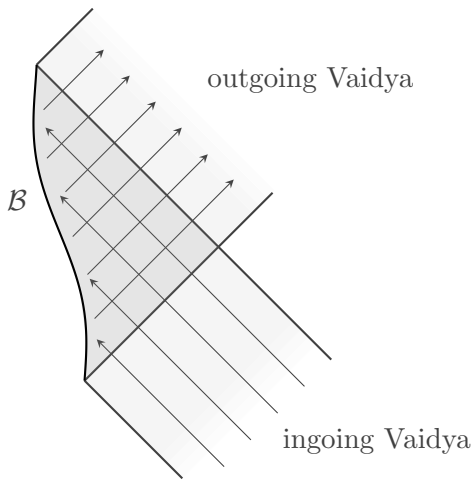
\begin{minipage}{0.45\textwidth}
\vspace{-6mm}
\begin{figure}[H]
\begin{center}
\begin{tikzpicture}[scale=1]
\draw [darkgray, -stealth] (3.856, -2.144) -- (2.162, -0.45);
\draw [darkgray, -stealth] (4.212, -1.788) -- (2.068, 0.356);
\draw [darkgray, -stealth] (4.568, -1.432) -- (1.674, 1.462);
\draw [darkgray, -stealth] (4.924, -1.076) -- (1.48, 2.368);
\draw [darkgray, -stealth] (1.456,2.604) -- (2.256,3.404);
\draw [darkgray, -stealth] (1.501,1.959) -- (2.601,3.059);
\draw [darkgray, -stealth] (1.747,1.513) -- (2.947,2.713);
\draw [darkgray, -stealth] (1.843,0.917) -- (3.293,2.367);
\draw [darkgray, -stealth] (2.039,0.412) -- (3.639,2.012);
\draw [darkgray, -stealth] (2.134,-0.174) -- (3.984,1.676);
\fill [fill=gray, draw=none,opacity=0.08]
(1.36,3.2) -- (1.81,3.65) -- (4.23,1.23) -- (3.78,0.78) -- cycle;
\fill [fill=gray, draw=none,opacity=0.06]
(1.91,3.75) -- (1.81,3.65) -- (4.23,1.23) -- (4.33,1.33) -- cycle;
\fill [fill=gray, draw=none,opacity=0.04]
(1.91,3.75) -- (2.01,3.85) -- (4.43,1.43) -- (4.33,1.33) -- cycle;
\fill [fill=gray, draw=none,opacity=0.02]
(2.11,3.95) -- (2.01,3.85) -- (4.43,1.43) -- (4.53,1.53) -- cycle;
\fill [fill=gray, draw=none,opacity=0.08]
(3.78,0.78) -- (4.78,-0.22) -- (3,-2) -- (2,-1) -- cycle;
\fill [fill=gray, draw=none,opacity=0.06]
(4.78,-0.22) -- (4.88,-0.32) -- (3.1,-2.1) -- (3,-2) -- cycle;
\fill [fill=gray, draw=none,opacity=0.04]
(4.98,-0.42) -- (4.88,-0.32) -- (3.1,-2.1) -- (3.2,-2.2) -- cycle;
\fill [fill=gray, draw=none,opacity=0.02]
(4.98,-0.42) -- (5.08,-0.52) -- (3.3,-2.3) -- (3.2,-2.2) -- cycle;
\draw [darkgray, thick] (2, -1) -- (3.3, -2.3);
\draw [darkgray, thick] (1.36, 3.2) -- (5.28, -0.72);
\draw [darkgray, thick] (1.36,3.2) -- (2.11, 3.95);
\draw [darkgray, thick] (2,-1) -- (4.5,1.5);
\fill [fill=gray, draw=gray,opacity=0.20]
(2,-1) -- ({2-0.2*(0.5)+0.15*sin(0.5*110)},{-1+1.2*0.5}) -- ({2-0.2*(1)+0.15*sin(1*110)},{-1+1.2*1}) -- ({2-0.2*(1.5)+0.15*sin(1.5*110)},{-1+1.2*1.5}) -- ({2-0.2*(2)+0.15*sin(2*110)},{-1+1.2*2}) -- ({2-0.2*(2.5)+0.15*sin(2.5*110)},{-1+1.2*2.5}) -- ({2-0.2*(3)+0.15*sin(3*110)},{-1+1.2*3}) -- (1.36,3.2) -- (3.78,0.78) -- cycle;
\draw [thick, domain=0:3.5, samples=150] plot ({2-0.2*(\x)+0.15*sin(\x*110)},{-1+1.2*\x});
\node[darkgray] at (1.37, 1.4) [anchor = east] {$\mathcal{B}$};
\node[align=left, darkgray] at (3.5, 3) [anchor = west] {outgoing Vaidya};
\node[align=left, darkgray] at (4.5, -1.8) [anchor = west] {ingoing Vaidya};
\end{tikzpicture}
\end{center}
\end{figure}
\end{minipage}
\hspace{0.01\textwidth}
\begin{minipage}{0.52\textwidth}
In summary, the bouncing continuation is well-understood when $\{r=r_b(v)\}$ is spacelike. However, $\{r=r_b(v)\}$ may have a timelike (or even mixed) causal character, arising from physically reasonable Vaidya seed data\footnotemark[6]. In this case, assuming that a self-consistent continuation can be realized, the bounce hypersurface $\mathcal{B}$ becomes the boundary of an interacting region, in which outgoing trajectories, having bounced already in the past, overlap with ingoing ones, yet to bounce in the future. We emphasize that this $\mathcal{B}$ no longer coincides in general with the curve $\{r=r_b(v)\}$, with the predictions of the Vaidya metric discarded in the putative interacting region (dark shaded region in the opposite Figure).
\end{minipage}\footnotetext[6]{The authors of \hyperlink{CGV16}{[CGV16]} point out that timelike $\{r=r_b(v)\}$ do not arise when the ratio $Q(v)/\varpi(v)$ is a constant $\nu$ lying in the ``physically reasonable'' range $\nu\in[0,1)$, a designation presumably expressing commitment to the third law. In any case, the ansatz $Q/\varpi=\text{const}.$ is not physically contentful, only algebraically convenient. Moreover, it undesirably entails $Q'/\varpi'\in[0,1)$, which is not necessary for subextremality, and creates a restriction to high-frequency, low-density beams through \hyperlink{eqn:1.14}{(1.14)}. Vaidya seed data $\varpi(v),Q(v)$ need not become (super)extremal to undergo a timelike bounce: see \hyperlink{ex:4.1}{Example~4.1}.}
\vspace{4mm} \\
Ori leaves the resolution of the timelike case outstanding, observing that a simple gluing procedure will not be sufficient for the construction\footnote[7]{``A
more general solution, which includes two simultaneous null flows, is required in this
case, but as yet the analytic form of such a solution is unknown.''}. \textbf{It is the aim of this article to address this outstanding case.} Though we do not identify an explicit metric generalizing \hyperlink{eqn:1.9}{(1.9)} for the interacting region (and are not aware of any reason to believe in the existence of one), we identify a remarkable decoupling of the equations of motion, which we hope may be regarded as the next best outcome. We then use this decoupling to resolve the timelike case, performing several new constructions. Saving the full details for \hyperlink{sec:2.3}{Section~2.3}, we now sketch the details of this decoupling.
\\ \\ 
In the absence of any known exact solution, one may nevertheless attempt to study the PDE system that governs the dynamics of two charged, self-gravitating null dusts. In terms of ingoing and outgoing dust variables $\rho_\text{in},\rho_\text{out}$ and $k_\text{in}, k_\text{out}$, the energy-momentum tensor and Maxwell equation now have two hydrodynamic contributions:
\begin{equation*}
T_{\mu\nu}=\rho_\text{in}(k_\text{in})_\mu(k_\text{in})_\nu+\rho_\text{out}(k_\text{out})_\mu(k_\text{out})_\nu+{F_\mu}^\alpha F_{\nu\alpha}-\tfrac{1}{4}g_{\mu\nu}F_{\alpha\beta}F^{\alpha\beta}\tag{\hypertarget{eqn:1.15}{1.15}}
\end{equation*}
\begin{equation*}
\nabla^\alpha F_{\mu\alpha}=\mathfrak{e}\rho_\text{in} (k_\text{in})_\mu+\mathfrak{e}\rho_\text{out} (k_\text{out})_\mu\tag{\hypertarget{eqn:1.16}{1.16}}
\end{equation*}
and equations \hyperlink{eqn:1.3}{(1.3)},\hyperlink{eqn:1.13}{(1.13)} are imposed separately for the ingoing and outgoing dust species.
A well-adapted coordinate choice for $\mathcal{Q}=\mathcal{M}/SO(3)$ is \textit{double null coordinates} $(u,v)$, under which the metric assumes the form 
$$g=-\Omega(u,v)^2dudv+r(u,v)^2g_{S^2} $$
We will find that the charge, now a function $Q=Q(u,v)$ of both coordinates, satisfies the equation
\begin{equation*}
\partial_u\partial_vQ=0\tag{\hypertarget{eqn:1.17}{1.17}}
\end{equation*}
To the author's knowledge, equation \hyperlink{eqn:1.17}{(1.17)} has not been identified before for this system. The interacting dusts in \hyperlink{PI90}{[PI90]}, for instance, were not charge-carrying, despite propagating on a charged background. We remark that two special cases of this equation, $\partial_uQ=0$ and $\partial_vQ=0$, characterize respectively the ingoing and outgoing charged Vaidya spacetimes, as will be made precise in \hyperlink{lem:2.2}{Lemma~2.2}. \\ \\
Further to this, the equations to be solved for $r,\Omega^2$ are
\begin{equation*}\partial_u\partial_vr=-\frac{\Omega^2}{4r}-\frac{\partial_ur\partial_vr}{r}+rT_{uv}\tag{\hypertarget{eqn:1.18}{1.18}}
\end{equation*}
\begin{equation*}
\partial_u\partial_v\log\Omega^2=\frac{\Omega^2}{2r^2}+\frac{2\partial_ur\partial_vr}{r^2}-2T_{uv}-\tfrac{1}{2}\Omega^2g^{AB}T_{AB}\tag{\hypertarget{eqn:1.19}{1.19}}
\end{equation*}
where $A,B$ denote arbitrary angular components (with summation assumed). Meanwhile, because both dusts are null, the only energy-momentum components that involve any hydrodynamic terms are $T_{uu},T_{vv}$. So the $T_{uv}$, $T_{AB}$ components in \hyperlink{eqn:1.18}{(1.18)}-\hyperlink{eqn:1.19}{(1.19)} can be computed from $Q,r,\Omega^2$ only. It follows that, at least for the `bulk' equations of motion, the electromagnetic, gravitational, and hydrodynamic parts of the system decouple. One may first solve equation \hyperlink{eqn:1.17}{(1.17)} to determine $Q$, then solve \hyperlink{eqn:1.18}{(1.18)}-\hyperlink{eqn:1.19}{(1.19)} for $r,\Omega^2$, and then finally recover the dust variables from the $Q,r,\Omega^2$ already obtained. \\ \\
What conditions are to be imposed at $\mathcal{B}$ itself? For a timelike bounce, the other side of $\mathcal{B}$ is electrovacuum. The rigidities of Birkhoff's theorem imply that this is locally isometric to a Reissner-Nordstr\"om spacetime, and so in particular $Q$ is constant. For a reasonable gluing across $\mathcal{B}$, we must impose that 
\begin{equation*}
Q\big|_\mathcal{B}=Q_0\tag{\hypertarget{eqn:1.20}{1.20}}
\end{equation*}
As for the metric, the variables $r,\Omega^2$ are to satisfy also the Raychaudhuri equations
\begin{align*}\partial_u\bigg(\frac{\partial_ur}{\Omega^2}\bigg)  &=-\tfrac{1}{4}r\Omega^2T^{vv}\equiv-\tfrac{1}{4}r\Omega^2N^v_\text{out}k^v_\text{out}\tag{\hypertarget{eqn:1.21}{1.21}}\\
\partial_v\bigg(\frac{\partial_vr}{\Omega^2}\bigg)&=- \tfrac{1}{4}r \Omega^2 T^{uu}\equiv-\tfrac{1}{4}r\Omega^2N^u_\text{in}k^u_\text{in}\tag{\hypertarget{eqn:1.22}{1.22}}
\end{align*}
The content of the identities `$\equiv$' above are that the `$uu$' and `$vv$' components of $T_{\mu\nu}$ have no electromagnetic part, as may be seen from examining \hyperlink{eqn:1.15}{(1.15)}. At the bounce $\mathcal{B}$, the 4-velocities $k_\text{in},k_\text{out}$ vanish. Meanwhile, conservation of the number currents mean that the terms $N^u_\text{in},N^v_\text{out}$ remain bounded. It follows that the right-hand sides of \hyperlink{eqn:1.21}{(1.21)}-\hyperlink{eqn:1.22}{(1.22)} vanish at $\mathcal{B}$. The boundary conditions for the metric are thus simply
\begin{equation*}
\partial_u\bigg(\frac{\partial_ur}{\Omega^2}\bigg)\bigg|_\mathcal{B}=0\qquad\partial_v\bigg(\frac{\partial_vr}{\Omega^2}\bigg)\bigg|_\mathcal{B}=0\tag{\hypertarget{eqn:1.23}{1.23}}
\end{equation*}
which again does not involve hydrodynamic terms.
\newpage
\large\hypertarget{sec:1.4}{\textbf{1.4\hspace{4mm}Prescribed timelike bounce hypersurfaces}}\label{1.4}\normalsize \\ \\
While the above reductions make the situation far more tractable than might be expected, there still remains the free-boundary aspect of the problem, namely that the location of the timelike bounce hypersurface is \textit{a priori} unknown. Since we are at first concerned simply to produce \textit{examples} of timelike hypersurfaces, we can begin by taking a scattering approach. That is, if we fix the location of the bounce $\gamma$ in advance, we can attempt to reconstruct an interacting region whose boundary is precisely $\gamma$. This is a similar strategy to the author's previous work on timelike dust \textit{caustics} \hyperlink{Bic25}{[Bic25]}. We state our results in a coordinate-free manner. In the sequel, we invoke the Reissner-Nordstr\"om spacetime $\mathcal{M}_{RN[\varpi,Q]}$ with parameters $\varpi,Q\geq0$. This does not refer to the maximal analytic extension, but only to a subset pertinent to the physical problem of injecting matter from past null infinity. See \hyperlink{sec:2.2}{Section~2.2} for discussion. The spherical symmetry of $\mathcal{M}_{RN[\varpi,Q]}$ and $\gamma$ enable us to refer to the latter as a \textit{curve}, even though it is of course a hypersurface `upstairs' in $\mathcal{M}_{RN[\varpi,Q]}$.\\ \\
\hypertarget{thm:1.1}{\textbf{Theorem 1.1}} (Prescribed timelike bounce hypersurface)\textbf{.} \textit{Let $\mathcal{M}_{\text{RN}[\varpi_0,Q_0]}$ be the Reissner-Nordstr\"om spacetime with parameters $\varpi_0,Q_0\geq0$. Let $(\gamma(\tau))_{\tau\in[a,b]}$ be a smooth radial timelike curve segment in $\mathcal{M}_{\text{RN}[\varpi_0,Q_0]}$ with $r(\gamma(\tau))$ bounded below ($\tau$ is not necessarily proper time). Let $\sigma(\tau)$ be a smooth function on $[a,b]$, positive on $(a,b)$, and bounded above by a constant $\varepsilon>0$ depending on $\gamma,\varpi_0,Q_0$. Then there exists a spherically symmetric spacetime $(\mathcal{M},g_{\mu\nu})$ describing the full trajectory of a charged null dust beam in Ori's bouncing model, whose bounce hypersurface occurs precisely at $\gamma$. More specifically, $(\mathcal{M},g_{\mu\nu})$ consists of five regions:
\begin{itemize}
\item[(\Romannum{1})] is an interacting 2-dust region bounded by a pair of null hypersurfaces and a timelike hypersurface identified, through spherical symmetry, with $\gamma$. The (one-sided) derivative of $Q$ along the spacelike unit normal to $\gamma$ (pointing into region (I)), at $\gamma(\tau)$, is $\sigma(\tau)$. 
\end{itemize}\vspace{-2mm}
\begin{minipage}{0.51\textwidth}
\begin{itemize}
\item[(I\hspace{-1.2pt}I)] is isometric to an ingoing Vaidya spacetime.\vspace{1mm}
\item[(I\hspace{-1.2pt}I\hspace{-1.2pt}I)] is isometric to (a portion of) $\mathcal{M}_{\text{RN}[\varpi_0,Q_0]}$.\vspace{1mm}
\item[(I\hspace{-1.2pt}V)] is isometric to an outgoing Vaidya spacetime.\vspace{1mm}
\item[(\Romannum{5})] is isometric to (a portion of) another $\mathcal{M}_{\text{RN}[\varpi_1,Q_1]}$ with $\varpi_1>\varpi_0$, $Q_1>Q_0$.
\end{itemize}
\vspace{5mm}$\rho_\text{in},k_\text{in}$ are supported on regions $(I)$ and $(I\hspace{-2.5pt}I)$, and $\rho_\text{out},k_\text{out}$ are supported on regions $(I)$ and $(I\hspace{-1.2pt}V)$. $(\mathcal{M},g_{\mu\nu})$ is smooth away from the boundaries of (I)-(V). Across $\gamma$ the gluing is $C^{2,1}$, but the dust energy densities diverge as $\gamma$ is approached from within (I). Across all other boundaries, the gluing is $C^{k+1}$, where $k\in\mathbb{N}_0\cup\{\infty\}$ is the order of the zeroes (if any) of $\sigma$ at the endpoints $\tau=a,b$. If $\sigma$ does not vanish at the endpoints, the gluing is merely $C^1$, and we say that the beam has a \ul{hard edge}.\end{minipage}
\hspace{0.01\textwidth}
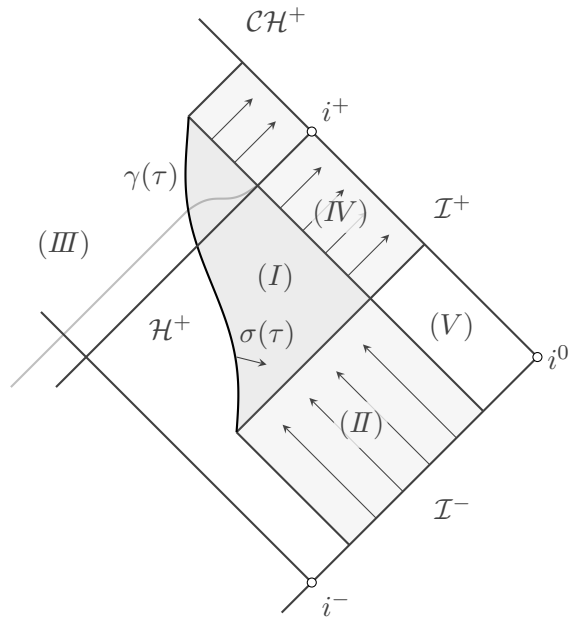
\begin{minipage}{0.48\textwidth}
\vspace{-6mm}
\begin{figure}[H]
\begin{center}
\begin{tikzpicture}[scale=1]
\draw [darkgray, -stealth] (3.856, -2.144) -- (2.612, -0.9);
\draw [darkgray, -stealth] (4.212, -1.788) -- (2.968, -0.544);
\draw [darkgray, -stealth] (4.568, -1.432) -- (3.324, -0.188);
\draw [darkgray, -stealth] (4.924, -1.076) -- (3.68, 0.168);
\draw [darkgray, -stealth] (1.973,2.587) -- (2.523,3.137);
\draw [darkgray, -stealth] (1.666,2.893) -- (2.216,3.443);
\draw [darkgray, -stealth] (2.28+0.3,2.28-0.3) -- (2.28+0.3+0.55,2.28-0.3+0.55);
\draw [darkgray, -stealth] (2.28+0.6,2.28-0.6) -- (2.28+0.6+0.55,2.28-0.6+0.55);
\draw [darkgray, -stealth] (2.28+0.9,2.28-0.9) -- (2.28+0.9+0.55,2.28-0.9+0.55);
\draw [darkgray, -stealth] (2.28+1.2,2.28-1.2) -- (2.28+1.2+0.55,2.28-1.2+0.55);
\draw [fill=white,white,opacity=0.8] (3.3,-1.2) rectangle (4.0,-0.6);
\node[darkgray] at (4.1, -0.9) [anchor = east] {$(I\hspace{-2.5pt}I)$};
\draw [fill=white,white,opacity=0.8] (3.01,1.66) rectangle (3.75,2.13);
\node[darkgray] at (0.2, 1.5) [anchor = east] {$(I\hspace{-2.5pt}I\hspace{-2.5pt}I)$};
\node[darkgray] at (3.9, 1.9) [anchor = east] {$(I\hspace{-1.5pt}V)$};
\fill [fill=gray, draw=gray,opacity=0.07]
(2.08,3.92) -- (4.5,1.5) -- (3.78,0.78) -- (1.36,3.2) -- cycle;
\fill [fill=gray, draw=gray,opacity=0.07]
(3.78,0.78) -- (5.28,-0.72) -- (3.5,-2.5) -- (2,-1) -- cycle;
\draw [gray, thick, opacity=0.5] (-1, -0.4) -- (1.35, 1.95);
\draw [gray, thick, domain=0:0.95, samples=150, opacity=0.5] plot ({1.35+\x},{1.95+0.35*\x+0.05*sin(\x*400)});
\draw [darkgray, thick] (2, -1) -- (3.5, -2.5);
\draw [darkgray, thick] (1.36, 3.2) -- (5.28, -0.72);
\draw [darkgray, thick] (2,-1) -- (4.5,1.5);
\fill [fill=gray, draw=gray,opacity=0.15]
(2,-1) -- ({2-0.2*(0.5)+0.15*sin(0.5*110)},{-1+1.2*0.5}) -- ({2-0.2*(1)+0.15*sin(1*110)},{-1+1.2*1}) -- ({2-0.2*(1.5)+0.15*sin(1.5*110)},{-1+1.2*1.5}) -- ({2-0.2*(2)+0.15*sin(2*110)},{-1+1.2*2}) -- ({2-0.2*(2.5)+0.15*sin(2.5*110)},{-1+1.2*2.5}) -- ({2-0.2*(3)+0.15*sin(3*110)},{-1+1.2*3}) -- (1.36,3.2) -- (3.78,0.78) -- cycle;
\draw [thick, domain=0:3.5, samples=150] plot ({2-0.2*(\x)+0.15*sin(\x*110)},{-1+1.2*\x});
\draw [darkgray, thick] (-0.4, -0.4) -- (3, 3);
\draw [darkgray, -stealth] (2, 0) -- (2.4, -0.1);
\draw [darkgray, thick] (1.5,4.5) -- (6,0);
\draw [darkgray, thick] (6,0) -- (2.6, -3.4);
\draw [darkgray, thick] (3,-3) -- (-0.6,0.6);
\draw [darkgray, thick] (1.36,3.2) -- (2.08,3.92);
\node[darkgray] at (3, 3) [anchor = south west] {$i^+$};
\node[darkgray] at (6, 0) [anchor = west]{$i^0$};
\node[darkgray] at (3, -3) [anchor = north west] {$i^-$};
\filldraw[color=black, fill=white](6,0) circle (0.06);
\filldraw[color=black, fill=white](3,3) circle (0.06);
\filldraw[color=black, fill=white](3,-3) circle (0.06);
\node[darkgray] at (0.7, 0.4) [anchor = west] {$\mathcal{H}^+$};
\node[darkgray] at (2.9, 1.05) [anchor = east] {$(I)$};
\node[darkgray] at (5.3, 0.4) [anchor = east] {$(V)$};
\node[darkgray] at (2.92, 0.3) [anchor = east] {$\sigma(\tau)$};
\node[darkgray] at (1.37, 2.4) [anchor = east] {$\gamma(\tau)$};
\node[align=left, darkgray] at (4.5, 2) [anchor = west] {$\mathcal{I}^+$};
\node[align=left, darkgray] at (4.5, -2) [anchor = west] {$\mathcal{I}^-$};
\node[align=left, darkgray] at (2, 4.5) [anchor = west] {$\mathcal{CH}^+$};
\end{tikzpicture}
\end{center}
\end{figure}
\end{minipage}}\vspace*{-1.8mm}\\
\\ \\
The above Figure shows a particular choice of $\gamma$ which straddles the interior and exterior regions of a subextremal $\mathcal{M}_{RN[\varpi_0,Q_0]}$. The grey curve is the apparent horizon, which, if the $\varepsilon>0$ in the statement is reduced further, can be guaranteed to be spacelike in the interacting region.\\ \\
As predicted in \hyperlink{Ori91}{[Ori91]}, the geometric regularity across the bounce is fairly high: the metric is $C^2$ regular, satisfying the Einstein equation in a classical sense. Even though a blowup of fluid energy density $\rho$ necessarily accompanies the bounce, the vanishing of 4-velocities $k$ makes the energy-momentum tensor continuous across $\mathcal{B}$. Meanwhile, the regularity at the edges of the beam is as good as one prescribes, and can be $C^\infty$.\\ \\ 
Each example obtained by this theorem realises a bounce hypersurface which remains timelike throughout the beam's full trajectory. However, \textit{a priori} the causal character of a bounce hypersurface could be mixed, with, say, spacelike and timelike components. We supplement the above result by providing examples of a timelike bounce terminating at a null point. This is not available straightforwardly as an application of \hyperlink{thm:1.1}{Theorem~1.1}, for instance by truncating the putative bounce curve $\gamma$ near the null point, because the smallness assumption on $\sigma(t)$ may not stabilize at a positive value. Nevertheless, by reformulating the system of variables, eliminating the lapse $\Omega^2$ in particular (see already \hyperlink{sec:2.4}{Section~2.4}), we obtain the following.
\\ \\
\hypertarget{thm:1.2}{\textbf{Theorem 1.2}} (Timelike bounce hypersurface terminating at null point)\textbf{.} \textit{Let $\mathcal{M}_{\text{RN}[\varpi_0,Q_0]}$ be the Reissner-Nordstr\"om spacetime with parameters $\varpi_0,Q_0\geq0$. Let $(\gamma(\tau))_{\tau\in[a,b]}$ be a smooth radial curve segment in $\mathcal{M}_{\text{RN}[\varpi_0,Q_0]}$, with $r(\gamma(\tau))$ bounded below, for which $\gamma'(\tau)$ is timelike when $\tau\in[a,b)$, and outgoing null at $\tau=b$. Then there exists a spherically symmetric spacetime $(\mathcal{M},g_{\mu\nu})$ describing the trajectory of a charged null dust beam in Ori's bouncing model, including the formation of a timelike bounce hypersurface, up to its termination at a null point. Moreover, $(\mathcal{M},g_{\mu\nu})$ consists of three regions:\vspace*{2mm}\\
\begin{minipage}{0.51\textwidth}
\begin{itemize}
\item[(\Romannum{1})]is an interacting 2-dust region bounded by a pair of null hypersurfaces and a causal curve identified with $\gamma$, at whose limit point the outgoing number current $N_\text{out}$ diverges, indicated by the shading in the opposite Figure. \vspace{1mm}
\item[$(I\hspace{-2.5pt}I)$]is isometric to an ingoing Vaidya spacetime, throughout which $dQ\neq0$.\vspace{1mm}
\item[$(I\hspace{-2.5pt}I\hspace{-2.5pt}I)$]is isometric to (a portion of) $\mathcal{M}_{\text{RN}[\varpi_0,Q_0]}$.\vspace{3mm}
\end{itemize}
The regularity and gluing properties are the same as in \hyperlink{thm:1.1}{Theorem~1.1}.\end{minipage}
\hspace{0.01\textwidth}
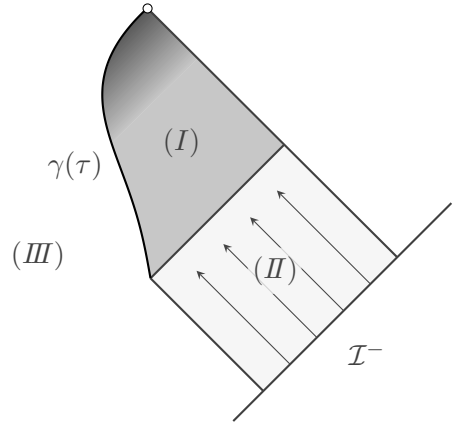
\begin{minipage}{0.48\textwidth}
\vspace{-4mm}
\begin{figure}[H]
\begin{center}
\begin{tikzpicture}[scale=1]
\draw [darkgray, -stealth] (3.856, -2.144) -- (2.612, -0.9);
\draw [darkgray, -stealth] (4.212, -1.788) -- (2.968, -0.544);
\draw [darkgray, -stealth] (4.568, -1.432) -- (3.324, -0.188);
\draw [darkgray, -stealth] (4.924, -1.076) -- (3.68, 0.168);
\draw [fill=white,white,opacity=0.8] (3.3,-1.2) rectangle (4.0,-0.6);
\node[darkgray] at (4.1, -0.9) [anchor = east] {$(I\hspace{-2.5pt}I)$};
\node[darkgray] at (1, -0.7) [anchor = east] {$(I\hspace{-2.5pt}I\hspace{-2.5pt}I)$};
\fill [fill=gray, draw=gray,opacity=0.07]
(3.78,0.78) -- (5.28,-0.72) -- (3.5,-2.5) -- (2,-1) -- cycle;
\shade [left color=black, right color = white,shading angle=45,opacity=0.7] ({2-0.5*(1.5)+0.15*sin(1.5*110)+0.057*(1.5)^3},{-1+1.2*1.5}) -- ({2-0.5*(1.75)+0.15*sin(1.75*110)+0.057*(1.75)^3},{-1+1.2*1.75}) -- ({2-0.5*(2)+0.15*sin(2*110)+0.057*(2)^3},{-1+1.2*2}) -- ({2-0.5*(2.25)+0.15*sin(2.25*110)+0.057*(2.25)^3},{-1+1.2*2.25}) -- ({2-0.5*(2.5)+0.15*sin(2.5*110)+0.057*(2.5)^3},{-1+1.2*2.5}) -- ({2-0.5*(2.75)+0.15*sin(2.75*110)+0.057*(2.75)^3},{-1+1.2*2.75}) -- ({2-0.5*(3)+0.15*sin(3*110)+0.057*(3)^3},{-1+1.2*3}) -- (2.63,1.93) -- cycle;
\fill [fill=gray, draw=none,opacity=0.44] ({2-0.5*(0)+0.15*sin(0*110)+0.057*(0)^3},{-1+1.2*0}) -- ({2-0.5*(0.5)+0.15*sin(0.5*110)+0.057*(0.5)^3},{-1+1.2*0.5}) -- ({2-0.5*(1)+0.15*sin(1*110)+0.057*(1)^3},{-1+1.2*1}) -- ({2-0.5*(1.5)+0.15*sin(1.5*110)+0.057*(1.5)^3},{-1+1.2*1.5}) -- (2.63,1.93) -- (3.78,0.78) -- cycle;
\draw [darkgray, thick] (2, -1) -- (3.5, -2.5);
\draw [darkgray, thick] (1.961, 2.59) -- (5.28, -0.72);
\draw [darkgray, thick] (2,-1) -- (3.78,0.78);
\draw [thick, domain=0:3, samples=150] plot ({2-0.5*(\x)+0.15*sin(\x*110)+0.057*(\x)^3},{-1+1.2*\x});
\filldraw[color=black, fill=white](1.961,2.59) circle (0.06);
\draw [darkgray, thick] (6,0) -- (3.1, -2.9);
\node[darkgray] at (2.8, 0.8) [anchor = east] {$(I)$};
\node[darkgray] at (1.5, 0.5) [anchor = east] {$\gamma(\tau)$};
\node[align=left, darkgray] at (4.5, -2) [anchor = west] {$\mathcal{I}^-$};
\end{tikzpicture}
\end{center}
\end{figure}
\end{minipage}}\vspace*{4mm}\\
The significance of $dQ\neq0$ is that it precludes trivial examples where $(\mathcal{M},g_{\mu\nu})$ is isometric to $\mathcal{M}_{RN[\varpi_0,Q_0]}$ in a neighbourhood of the null point. These may easily be constructed by extending $\gamma$ in \hyperlink{thm:1.1}{Theorem~1.1} up to a null point. We also emphasize that `termination' only refers to termination \textit{as a timelike bounce}. In fact, we expect that a much broader family of examples is possible, where $\gamma$ may have spacelike and timelike components, and that the divergence of $N_\text{out}$ near any null point of $\gamma$ would be strong enough to prevent $C^2$ regularity of $g_{\mu\nu}$, but mild enough for Einstein's equation to be satisfied in a weak sense.\\ \\
The proofs of both theorems are relatively straightforward. The decoupled equations are solved `sideways' from the prescribed bounce curve $\gamma$. It is easy to obtain a \textit{local} solution (near $\gamma$) and the only concern is whether the solution extends to the full putative interacting region. To do this, an elementary Cauchy stability argument suffices: the Reissner-Nordstr\"om solution provides a solution to the equations of motion on the full triangular characteristic domain, around which we perturb. \\ \\ \\
\large\hypertarget{sec:1.5}{\textbf{1.5\hspace{4mm}The formation problem for timelike bounces}}\label{1.5}\normalsize\\ \\
Having produced in \hyperlink{thm:1.1}{Theorem~1.1} a large family of examples of timelike bounce hypersurfaces, we turn to the \textit{forwards} problem. That is, given Vaidya seed data $\varpi(v),Q(v)$---which encode initial data at past null infinity for an incoming charged null beam---does there always exist a unique development of these data in terms of Ori's bouncing model? This is well-understood when the curve $\{r=r_b(v)\}$ is spacelike with respect to the charged Vaidya metric \hyperlink{eqn:1.9}{(1.9)}. If $\{r=r_b(v)\}$ is timelike, then, even with an affirmative answer to this forwards problem, we would not expect $\mathcal{B}$ to coincide with $\{r=r_b(v)\}$ (see again discussion in \hyperlink{sec:1.3}{Section~1.3}). Instead, we may consider $\{r=r_b(v)\}$ to be a naive \textit{first approximation} to the true bounce $\mathcal{B}$. For instance, while they certainly agree on the location of the formation point, \textit{a priori} the true $\mathcal{B}$ need not remain timelike. As we shall show, however, if the naive $\{r=r_b(v)\}$ is timelike, then the true $\mathcal{B}$ is guaranteed to also be timelike over some initial interval of advanced time.\\ \\
Of course, \hyperlink{thm:1.1}{Theorem 1.1} already yields, in particular, a large family of Vaidya seed data $\varpi(v),Q(v)$ whose developments lead to timelike bounces. These seed data may be read off from region $(I\hspace{-2.5pt}I)$\footnote[8]{But care must be taken over the parameter $v$---see \hyperlink{sec:2.1}{Section 2.1}.}. The forwards problem, in other words, is whether the Vaidya seed data obtained in this way \textit{exhaust} all possible $\varpi(v),Q(v)$ which ``want'' to form a timelike bounce, in the sense that their $\{r=r_b(v)\}$ is timelike with respect to \hyperlink{eqn:1.9}{(1.9)}. The following result provides a conditional solution to the forwards problem, for Vaidya seed data $\varpi(v),Q(v)$ which have a \textit{hard edge}, meaning that the derivatives $\varpi'(v),Q'(v)$ do not vanish at the edge of the beam. This implies that the 4-momentum $k_\text{in}$ and energy density $\rho_\text{in}$ jump discontinuously to zero at the edge of the beam. The result is also only capable of treating the case where the background charge $Q_0$ is positive, and where the bounce occurs in the Reissner-Nordstr\"om \textit{exterior}.\\ \\
\hypertarget{thm:1.3}{\textbf{Theorem 1.3}} (Hard-edge timelike bounce formation in Reissner-Nordstr\"om exterior)\textbf{.} \textit{Let ingoing Vaidya seed data $\varpi(v),Q(v)$ be defined on an interval containing 0, and belong to $C^2$ when restricted to $\mathbb{R}_{\geq0}$. Suppose they encode a hard-edge beam forming a timelike bounce in the exterior region. That is, suppose $\varpi,Q$ are constant on $v\leq0$:
$$\varpi(v)=\varpi_0>0\qquad Q(v)=Q_0>0 $$
and have a `hard edge':
$$\lim_{v\searrow0}\varpi'(v)>0\qquad\lim_{v\searrow0}Q'(v)>0$$
and the curve $\{r=r_b(v)\}$, in the original metric \hyperlink{eqn:1.9}{(1.9)}, begins in the exterior region, with a timelike tangent vector:
$$\lim_{v\searrow0}r_b(v)>\begin{cases}\varpi_0+\sqrt{\varpi^2_0-Q^2_0}\quad\text{if }\varpi_0\geq Q_0\\ 0\hspace{3cm}\text{otherwise}\end{cases} \qquad\qquad\lim_{v\searrow0} r'_b(v)<\tfrac{1}{2}\Big(1-\frac{2\varpi_0}{r_b(0)}+\frac{Q^2_0}{r_b(0)^2}\Big)\vspace{0mm} $$
\begin{minipage}{0.51\textwidth}
Then there exists a spherically symmetric spacetime $(\mathcal{M},g_{\mu\nu})$ describing the trajectory of a charged null dust beam in Ori's model, including the formation of a timelike bounce hypersurface and continuation up to an advanced time $v=\varepsilon>0$, and starting from the data $\varpi(v),Q(v)$ at $\mathcal{I}^-$. The decomposition of $(\mathcal{M},g_{\mu\nu})$ is the same as in \hyperlink{thm:1.2}{Theorem~1.2}. The gluing across the boundary between region $(I)$ and $(I\hspace{-2.5pt}I)$ is $C^1$ only.\end{minipage}
\hspace{0.01\textwidth}
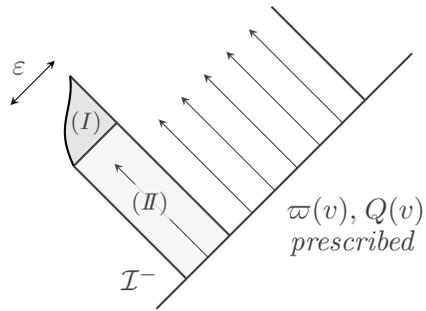
\begin{minipage}{0.48\textwidth}
\vspace{0mm}
\begin{figure}[H]
\begin{center}
\begin{tikzpicture}[scale=1]
\draw [darkgray, -stealth] (3.786, -2.214) -- (2.542, -0.97);
\draw [darkgray, -stealth] (4.38, -1.63) -- (3.13, -0.38);
\draw [darkgray, -stealth] (4.68, -1.33) -- (3.43, -0.08);
\draw [darkgray, -stealth] (4.98, -1.03) -- (3.73, 0.22);
\draw [darkgray, -stealth] (5.28, -0.73) -- (4.03, 0.52);
\draw [darkgray, -stealth] (5.58, -0.43) -- (4.33, 0.82);
\draw [darkgray, thick] (5.88, -0.13) -- (4.63, 1.12);
\draw [fill=white,white,opacity=0.8] (2.75,-1.7) rectangle (3.3,-1.25);
\fill [fill=gray, draw=gray,opacity=0.07]
(2.58,-0.42) -- (4.08,-1.93) -- (3.5,-2.5) -- (2,-1) -- cycle;
\fill [fill=gray, draw=none,opacity=0.2] ({2-0.2*(0)-0.06*sin(0*300)+0.1*(0)^2},{-1+1.2*0}) -- ({2-0.2*(0.25)-0.06*sin(0.25*300)+0.1*(0.25)^2},{-1+1.2*0.25}) -- ({2-0.2*(0.5)-0.06*sin(0.5*300)+0.1*(0.5)^2},{-1+1.2*0.5}) -- ({2-0.2*(0.75)-0.06*sin(0.75*300)+0.1*(0.75)^2},{-1+1.2*0.75}) -- ({2-0.2*(1)-0.06*sin(1*300)+0.1*(1)^2},{-1+1.2*1}) -- (2.58,-0.42) -- cycle;
\draw [<->,>=stealth] (1.75,0.4) -- (1.16,-0.19);
\draw [darkgray, thick] (2, -1) -- (3.5, -2.5);
\draw [darkgray, thick] (2,-1) -- (2.58,-0.42);
\draw [darkgray, thick] (1.95,0.2) -- (4.08,-1.93);
\draw [thick, domain=0:1, samples=150] plot ({2-0.2*(\x)-0.06*sin(\x*300)+0.1*(\x)^2},{-1+1.2*\x});
\draw [darkgray, thick] (6.5,0.5) -- (3.1, -2.9);
\node[darkgray] at (1.5, 0.08) [anchor = south east] {$\varepsilon$};
\node[align=left, darkgray] at (2.5, -2.5) [anchor = west] {$\mathcal{I}^-$};
\node[align=left, darkgray] at (4.7, -1.6) [anchor = west] {$\varpi(v),\hspace{1pt}Q(v)$};
\node[align=left, darkgray] at (4.74, -2.03) [anchor = west] {\textit{prescribed}};
\node[darkgray] at (2.51, -0.43) [anchor = east] {\footnotesize$(I)$};
\node[darkgray] at (3.4, -1.5) [anchor = east] {\footnotesize$(I\hspace{-2.5pt}I)$};
\end{tikzpicture}
\end{center}
\end{figure}
\end{minipage}}\vspace*{4mm}\\
The advanced time $\varepsilon>0$ obtained in the theorem depends on various quantities computed from $\varpi(v),Q(v)$: see \hyperlink{sec:4.1}{Section~4.1}. In particular, it depends quantitatively on how close the inequality 
$$r'_b(0)<\tfrac{1}{2}\Big(1-\frac{2\varpi_0}{r_b(0)}+\frac{Q^2_0}{r_b(0)^2}\Big) $$is to equality: if the curve $\{r=r_b(v)\}$ is very close to being null, then the true bounce curve $\mathcal{B}$ cannot be guaranteed to remain timelike for long. \\ \\
We expect the methods used to establish \hyperlink{thm:1.3}{Theorem~1.3} to require significant re-working to address the case of $Q_0=0$. On closer inspection of \hyperlink{eqn:1.14}{(1.14)}, one sees that a beam injected onto a Minkowski background, and bouncing at a positive radius, will have a divergent energy density $\rho$ already at the edge of the beam, even far away from the bounce. Thus the behaviour of the dynamical variables is qualitatively different at the formation point. However, the bouncing continuation is still tenable in this case: \hyperlink{thm:1.1}{Theorem~1.1} has no such restriction, and the examples constructed in \hyperlink{KU24}{[KU24]} are all injections onto a Minkowski background. 
\\ \\
The proof is considerably more difficult than for \hyperlink{thm:1.1}{Theorems 1.1}-\hyperlink{thm:1.2}{1.2}, since the boundary conditions \hyperlink{eqn:1.23}{(1.23)} imposed on $r,\Omega^2$ at the bounce are not of a standard type. In finding a successful iterative scheme with which to construct the solution, we follow Christodoulou's strategy in his series of papers [\hyperlink{Chr95}{Chr95}; \hyperlink{Chr96a}{Chr96a}; \hyperlink{Chr96b}{Chr96b}; \hyperlink{CL14}{CL14}] on the \textit{two-phase model}. In these works, Christodoulou studies a free boundary problem with some similar features, corresponding to the point of transition between two phases of a relativistic fluid. In his problem, the system is reformulated into a set of 6 first-order propagation equations, including a transport equation for Hawking mass. Following Christodoulou, we again eliminate the lapse function $\Omega^2$ and study a new collection of variables $(r,\varpi,\phi,Q)$ (see already \hyperlink{sec:2.5}{Section~2.5}). In particular, the difficult boundary conditions \hyperlink{eqn:1.23}{(1.23)} are reinterpreted as boundary data for the Hawking mass $\varpi$, which is propagated separately with its own transport equation. The geometric variable $\phi$ obeys a wave equation of a sensitive character, and it is here that the hard-edge caveat plays a role, guaranteeing that the characteristic data $\phi_+$ is bounded below in its second derivative: $\inf_{v}\phi''_+(v)>0$.
\\ \\
Lastly, we comment on the non-generic case of \textit{null} bounce hypersurfaces. If, over some interval of advanced time $v$, the seed data $\varpi(v),Q(v)$ obey the non-trivial condition 
\begin{equation*}
r'_b(v)=\tfrac{1}{2}\Big(1-\frac{2\varpi(v)}{r_b(v)}+\frac{2Q^2(v)}{r^2_b(v)}\Big)\tag{1.24}
\end{equation*}
then the dust spheres bounce \textit{simultaneously}---in the sense of retarded time $u$. The outgoing continuation would consist of a thin shell of matter, on which positive mass and charge would be concentrated. Constructing this continuation in detail would require solving a gluing problem using the techniques laid out in [\hyperlink{Isr65}{Isr65}; \hyperlink{BI91}{BI91}; \hyperlink{CD87}{CD87}; \hyperlink{Kha11}{Kha11}]. We do not discuss this exceptional possibility further, but expect the construction to be straightforward. \\ \\
\textbf{Acknowledgements. }The author would like to express his gratitude to his advisor Mihalis Dafermos for his support, encouragement and guiding insight; and to Christoph Kehle for many clarifying discussions. This work was funded by EPSRC.
\\ \\ \\
\Large\hypertarget{sec:2}{\textbf{2\hspace{4mm}Preliminaries}}\label{2}\normalsize\\ \\
\large\hypertarget{sec:2.1}{\textbf{2.1\hspace{4mm}Spherical symmetry and double null gauge}}\label{2.1}\normalsize\\ \\
We choose the metric signature $(-++\hspace{2pt}+)$, and choose units such that the speed of light is $c=1$ and Newton's constant is $G=1/4\pi$. Thus the Einstein equation reads
\begin{equation}
\text{Ric}[g]_{\mu\nu}-\tfrac{1}{2}R[g]g_{\mu\nu}=2 T_{\mu\nu} \tag{\hypertarget{eqn:2.1}{2.1}}
\end{equation}
We now state with more precision the notion of spherical symmetry that we will need. Since we are solely concerned with behaviour away from the centre of symmetry, we are able to omit discussion of the centre entirely.\\ \\
\textbf{Definition 2.1} (Spherical symmetry)\textbf{.} \textit{We say that a connected, time-oriented (3+1)-dimensional Lorentzian manifold $(\mathcal{M},g_{\mu\nu})$ is \ul{spherically symmetric} if it admits an $SO(3)$-action by isometries, the orbits of which are spacelike 2-spheres, and moreover if $\mathcal{M}$ splits diffeomorphically as $\mathcal{M}=\mathcal{Q}\times S^2$, where $(\mathcal{Q},g_\mathcal{Q})$ is a connected, time-oriented (1+1)-dimensional Lorentzian manifold, with the metric taking the form 
$$g=g_\mathcal{Q}+r^2g_{S^2} $$
Here, $r:\mathcal{Q}\to\mathbb{R}_{>0}$ is the \ul{area radius} of the 2-spheres. We will speak interchangeably of \ul{radial} curves in $\mathcal{M}$, which are constant in the $S^2$ factor, and their projections to $\mathcal{Q}$.}\\ \\
We are interested in spherically symmetric spacetimes for which the quotient manifold $\mathcal{Q}$ admits a global \textit{double null} coordinate system $(u,v)$ with respect to which $g_\mathcal{Q}$ takes the form
$$g_\mathcal{Q}=-\Omega^2(u,v)dudv $$
and such that $\partial_u$ and $\partial_v$ are future-directed. This can always be achieved \textit{locally} on any (1+1)-dimensional Lorentzian manifold $\mathcal{Q}$. The level curves of $u$ and $v$ are null hypersurfaces which are uniquely determined in the (1+1)-dimensional setting. However, the \textit{labelling} of the null hypersurfaces is arbitrary: given null coordinates $(u,v)$, we obtain  a new coordinate system by setting
$$\tilde u =f(u)\qquad \tilde v=g(v)$$
where $f,g:\mathbb{R}\to\mathbb{R}$ are smooth functions with strictly positive derivative. In the new coordinate system, we have
$$g=-\tilde{\Omega}^2(\tilde u,\tilde v)d\tilde ud\tilde v+\tilde{r}^2(\tilde u,\tilde v)g_{S^2}$$
where
$$\tilde{\Omega}^2(\tilde u,\tilde v)=(f'g')^{-1}\Omega^2(f^{-1}(\tilde u),g^{-1}(\tilde v)),\qquad \tilde{r}(\tilde u,\tilde v)=r(f^{-1}(\tilde u),g^{-1}(\tilde v)) $$
Note that, in contrast, the Vaidya metric comes with a \textit{canonical} null coordinate $v$: the only reparametrizations preserving the form \hyperlink{eqn:1.9}{(1.9)} are overall translations of the $v$ coordinate, since it is insisted that $g_{vr}\equiv1$.\\ \\
We exploit the considerable flexibility in the choice of the null coordinates to study our problem on an exactly triangular domain. In \hyperlink{thm:1.1}{Theorems~1.1}-\hyperlink{thm:1.3}{1.3}, our focus is on region $(I)$. By a suitable\vspace*{2pt}\\
\begin{minipage}{0.75\textwidth} choice of null coordinates $(u,v)$, region $(I)$ corresponds precisely to 
$$\triangleright:=\{a\leq u\leq v\leq b\} $$
This does not exhaust the gauge freedom, and in fact the remaining gauge choice is precisely the subgroup of transformations acting as 
$$\tilde u=f(u)\qquad \tilde v=f(v)$$
where $f$ is as above, under which the domain maps to
$$\triangleright:=\{f(a)\leq\tilde u\leq\tilde v\leq f(b)\} $$\vspace{0mm}
\end{minipage}
\hfill
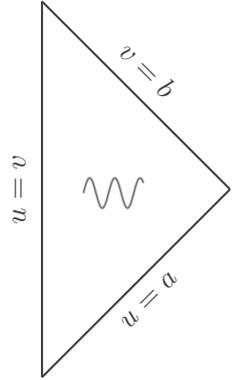
\begin{minipage}{0.22\textwidth}
\vspace{-8mm}
\begin{figure}[H]
\begin{center}
\begin{tikzpicture}
\draw [darkgray, thick] (0.0, 0) -- (0.0, 5.0);
\draw [darkgray, thick] (0, 0) -- (2.5,2.5);
\draw [darkgray, thick] (2.5,2.5) -- (0, 5.0);
\node[darkgray,rotate=90] at (-0.3, 1.9) [anchor = west] {$u=v$};
\node[darkgray, align=left, rotate=45] at (1.0,0.6) [anchor = west] {$u=a$};
\node[align=left, darkgray, rotate=-45] at (1.0,4.5) [anchor = west] {$v=b$};
\draw [gray, thick, domain=0.0:1, samples=150] plot ({(0.55+0.8*\x)},{2.45+0.2*sin(15*\x r)});
\end{tikzpicture}
\end{center}
\end{figure}
\end{minipage} \\
\textbf{Gauge choices.}
\begin{itemize}
\item[$\circ$] In \hyperlink{thm:1.1}{Theorem~1.1}, data are posed on $\mathcal{B}$. In view of this, we use the \textit{lapse-normalized} gauge choice, under which $\Omega^2|_{u=v}=1$. This leaves only a 1-dimensional gauge subgroup of vertical translations left over. We will consistently use a subscript `0' in reference to quantities evaluated on the $\{u=v\}$ part of the boundary. Hence for example we would write $\Omega^2_0(t)\equiv1$. 
\item[$\circ$] In \hyperlink{thm:1.2}{Theorem~1.2}, transforming region $(I)$ to an exactly triangular domain is more complicated because the null endpoint of $\gamma$ leads to a geometric irregularity. Nevertheless, we shall do this---see already \hyperlink{sec:2.4}{Section~2.4}.
\item[$\circ$] In \hyperlink{thm:1.3}{Theorem~1.3}, data are posed at $\{u=a\}$, and we may set $a=0$. The natural gauge choice in this case is to preserve the null Vaidya coordinate $v$ which parametrizes our characteristic data. We refer to this as the \textit{Vaidya-normalized} gauge choice. We will consistently use a subscript `$+$' in reference to quantities evaluated on the $\{u=0\}$ part of the boundary. Hence for example, since we intend to glue across $u=0$ to a given ingoing Vaidya region, we have $Q_+(v)=Q(v)$, where $Q(v)$ is as given in the theorem statement.
\end{itemize}
In this article, we will encounter situations of finite regularity, as is already seen in the statements of \hyperlink{thm:1.1}{Theorems~1.1}-\hyperlink{thm:1.3}{1.3}. The double null coordinate charts we will construct are such that the Reissner-Nordstr\"om regions present in all our examples are smooth, and the differential structure of $\mathcal{M}$ is taken to be the one generated by such charts. We may thus speak unambiguously about the exact regularity of gluing.\\ \\
Recall the \textit{Hawking mass} function $m:\mathcal{Q}\to\mathbb{R}$ which is defined by 
\begin{equation*}1-\frac{2m}{r}=g(\nabla r,\nabla r)\vspace{-2mm}\tag{\hypertarget{eqn:2.2}{2.2}}\end{equation*}
In the $(u,v)$ coordinates, we have the expression
\begin{equation*}
m=\frac{r}{2}\bigg(1+\frac{4\partial_ur\partial_vr}{\Omega^2}\bigg)\tag{\hypertarget{eqn:2.3}{2.3}}
\end{equation*}\newpage
\large\hypertarget{sec:2.2}{\textbf{2.2\hspace{4mm}Electromagnetic fields and the Lorentz force}}\label{2.2}\normalsize\\ \\
We consider spherically symmetric spacetimes which admit a spherically symmetric
electromagnetic field with no magnetic charge. This is described by a closed 2-form $F$ given by the expression
$$F=-\frac{\Omega^2Q}{2r^2}du\wedge dv $$
Here $Q:\mathcal{Q}\to\mathbb{R}$ is the \textit{electric charge}, which has an invariant meaning as the total charge content in the symmetry sphere labelled by coordinates $(u,v)$. The electromagnetic energy-momentum tensor, seen already in equations \hyperlink{eqn:1.8}{(1.8)} and \hyperlink{eqn:1.15}{(1.15)}, takes the usual form
$$T^{EM}_{\mu\nu}={F_\mu}^\alpha F_{\nu\alpha}-\tfrac{1}{4}g_{\mu\nu}F_{\alpha\beta}F^{\alpha\beta}$$
Written in double null coordinates $(u,v)$, and in terms of $Q$ above, we have
$$T^{EM}=\frac{\Omega^2Q^2}{2r^4}dudv+\frac{Q^2}{2r^2}g_{S^2} $$
In the special case of \textit{electrovacuum}, $Q$ is constant, and by Birkhoff's theorem spacetime is locally isometric to a member $\mathcal{M}_{RN[\varpi,Q]}$ of the Reissner-Nordstr\"om family. The Reissner-Nordstr\"om spacetimes mentioned in \hyperlink{thm:1.1}{Theorems~1.1}-\hyperlink{thm:1.2}{1.2} are taken to refer to the following restricted regions of interest.
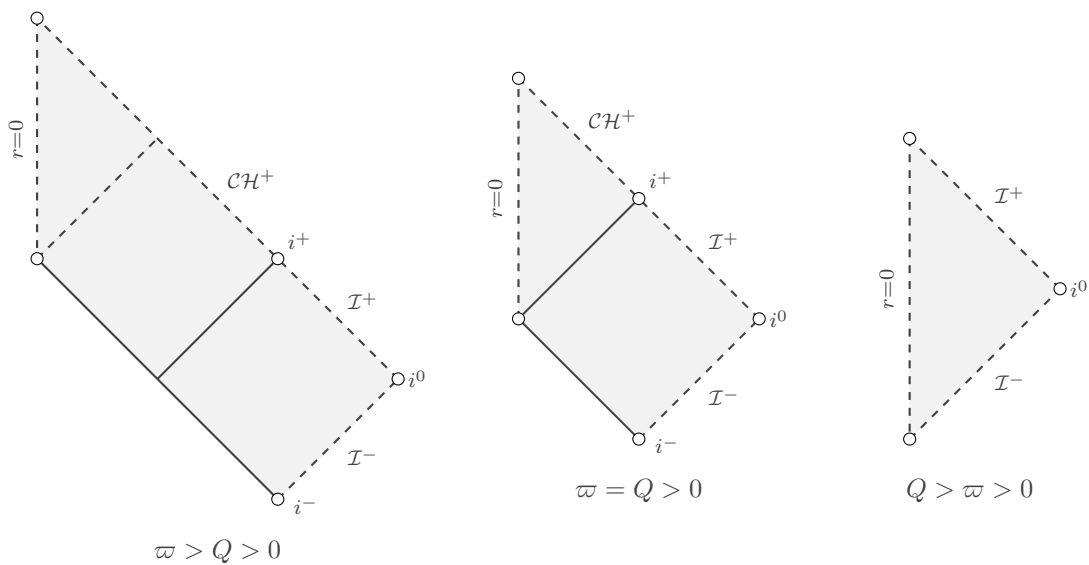
\vspace{2mm}
\begin{figure}[H]
\begin{center}
\begin{tikzpicture}[scale=0.8]
\fill [fill=gray, draw=gray,opacity=0.1]
(-8,2) -- (-6,4) -- (-2,0) -- (-4,-2) -- cycle;
\fill [fill=gray, draw=gray,opacity=0.1]
(-8,2) -- (-8,6) -- (-6,4) -- cycle;
\draw [darkgray, thick] (-6, 0) -- (-4,-2);
\draw [darkgray, thick] (-6, 0) -- (-4,2);
\draw [darkgray, thick, dashed] (-4,-2) -- (-2,0);
\draw [darkgray, thick, dashed] (-8,2) -- (-6,4);
\draw [darkgray, thick, dashed] (-2,0) -- (-8,6);
\draw [darkgray, thick] (-6,0) -- (-8,2);
\draw [darkgray, thick, dashed] (-8,6) -- (-8,2);
\node[darkgray] at (-3, -1) [anchor = north west] {$_{\mathcal{I}^-}$};
\node[darkgray] at (-3, 1) [anchor = south west] {$_{\mathcal{I}^+}$};
\node[darkgray, rotate=90] at (-8.35, 4.5) [anchor = east] {$_{r=0}$};
\node[darkgray] at (-5, 3) [anchor = south west] {$_{\mathcal{CH}^+}$};
\node[darkgray] at (-5, -2.5) [anchor = north] {\small$\varpi>Q>0$};
\node[darkgray] at (-4, 2) [anchor = south west] {$_{i^+}$};
\node[darkgray] at (-2, 0) [anchor = west] {$_{i^0}$};
\node[darkgray] at (-3.9, -1.8) [anchor = north west] {$_{i^-}$};
\filldraw[color=black, fill=white](-4,-2) circle (0.1);
\filldraw[color=black, fill=white](-2,0) circle (0.1);
\filldraw[color=black, fill=white](-4,2) circle (0.1);
\filldraw[color=black, fill=white](-8,6) circle (0.1);
\filldraw[color=black, fill=white](-8,2) circle (0.1);
\draw [darkgray, thick, dashed] (0,1) -- (0,5);
\draw [darkgray, thick] (0, 1) -- (2,-1);
\draw [darkgray, thick] (0, 1) -- (2,3);
\draw [darkgray, thick, dashed] (2,-1) -- (4,1);
\draw [darkgray, thick, dashed] (4,1) -- (0,5);
\fill [fill=gray, draw=gray,opacity=0.1]
(0,1) -- (0,5) -- (4,1) -- (2,-1) -- cycle;
\node[darkgray] at (3, 0) [anchor = north west] {$_{\mathcal{I}^-}$};
\node[darkgray] at (3, 2) [anchor = south west] {$_{\mathcal{I}^+}$};
\node[darkgray] at (1, 4) [anchor = south west] {$_{\mathcal{CH}^+}$};
\node[darkgray] at (2, 3) [anchor = south west] {$_{i^+}$};
\node[darkgray] at (4,1) [anchor = west] {$_{i^0}$};
\node[darkgray] at (2.1, -0.8) [anchor = north west] {$_{i^-}$};
\node[darkgray,rotate=90] at (-0.1, 2.5) [anchor = south west] {$_{r=0}$};
\node[darkgray] at (2, -1.5) [anchor = north] {\small$\varpi=Q>0$};
\filldraw[color=black, fill=white](2,-1) circle (0.1);
\filldraw[color=black, fill=white](4,1) circle (0.1);
\filldraw[color=black, fill=white](2,3) circle (0.1);
\filldraw[color=black, fill=white](0,1) circle (0.1);
\filldraw[color=black, fill=white](0,5) circle (0.1);
\fill [fill=gray, draw=gray,opacity=0.1]
(6.5,4) -- (6.5,-1) -- (9,1.5) -- cycle;
\draw[darkgray, thick, dashed] (6.5,4) -- (6.5,-1);
\draw [darkgray, thick, dashed] (6.5,-1) -- (9,1.5);
\draw [darkgray, thick, dashed] (9,1.5) -- (6.5,4);
\node[darkgray] at (7.75, 0.25) [anchor = north west] {$_{\mathcal{I}^-}$};
\node[darkgray] at (7.75, 2.75) [anchor = south west] {$_{\mathcal{I}^+}$};
\node[darkgray] at (9, 1.5) [anchor = west] {$_{i^0}$};
\node[darkgray] at (7.5, -1.5) [anchor = north] {\small$Q>\varpi>0$};
\node[darkgray,rotate=90] at (6.4, 1) [anchor = south west] {$_{r=0}$};
\filldraw[color=black, fill=white](6.5,-1) circle (0.1);
\filldraw[color=black, fill=white](9,1.5) circle (0.1);
\filldraw[color=black, fill=white](6.5,4) circle (0.1);
\filldraw[color=white, fill=white](10,0) circle (0.04);
\end{tikzpicture}
\end{center}
\end{figure}
\vspace{-4mm}That is, in the subextremal case $\varpi> Q>0$, we take a neighbouring pair of interior and exterior regions, together with a `deep interior' region foliated by untrapped symmetry spheres. In the extremal case $\varpi=Q>0$ (and also for the Schwarzschild subfamily $Q=0$, not pictured), we take a neighbouring pair of interior and exterior regions. 
In the superextremal case, including the subfamily of point charges $\varpi=0$, $Q\neq0$ (also not pictured), and in the Minkowski case, we may take the usual maximal extension. See [\hyperlink{HE73}{HE73}, Chapter 6] for a general exposition of the Reissner-Nordstr\"om spacetime and its subfamilies.\\ \\
These restrictions are guided by two considerations. Firstly, as earlier mentioned, we are interested in the physical problem of injecting matter from past null infinity $\mathcal{I^-}$, and so we are concerned only with Reissner-Nordstr\"om regions occurring in the causal future $J^+(\mathcal{I}^-)$\footnote[9]{Even spheres in the deep interior region, if possessing non-negative Hawking mass, arise in physical gravitational collapse: Cauchy surfaces may be chosen which enter this region and also extend to $i^0$.}.  In particular, this removes those regions whose symmetry spheres are anti-trapped\footnote[10]{The bouncing continuation is, in any case, a physical consequence of the contraction of fluid spheres, so it is reasonable to study the portion of $\mathcal{M}_{RN[\varpi,Q]}$ with $\partial_ur<0$.}. Secondly, we exclude extensions beyond the Cauchy horizon $\mathcal{CH}^+$ (if present) from our study. This ensures that the following subsets of $\mathcal{M}_{RN[\varpi,Q]}$, essential to our construction of region $(I)$, always exist:\\ \\
\begin{minipage}{0.75\textwidth}
\hypertarget{def:2.2}{\textbf{Definition 2.2}} (Characteristic triangle exterior to $\gamma$)\textbf{.} \textit{Let $\varpi,Q\geq0$ be Reissner-Nordstr\"om parameters, and let $\gamma(\tau)_{\tau\in[a,b]}$ be a smooth radial curve segment in $\mathcal{M}_{RN[\varpi,Q]}$, timelike in the range $\tau\in(a,b)$. Viewing $\gamma$ as a curve segment in $\mathcal{Q}=\mathcal{M}/SO(3)$, consider the region in $\mathcal{Q}$ bounded by $\gamma$, the outgoing future-directed null curve from $\gamma(a)$, and the ingoing past-directed null curve from $\gamma(b)$. We refer to this region (and its `upstairs' equivalent in $\mathcal{M}$) as the \ul{characteristic triangle exterior to $\gamma$}}.
\end{minipage}
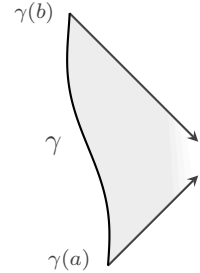
\begin{minipage}{0.25\textwidth}
\vspace{-6mm}
\begin{figure}[H]
\begin{center}
\begin{tikzpicture}[scale=0.8]
\draw [darkgray, thick, -stealth] (1.36, 3.2) -- (3.5, 1.06);
\draw [darkgray, thick, -stealth] (2,-1) -- (3.5,0.5);
\fill [fill=gray, draw=none,opacity=0.14]
(2,-1) -- ({2-0.2*(0.5)+0.15*sin(0.5*110)},{-1+1.2*0.5}) -- ({2-0.2*(1)+0.15*sin(1*110)},{-1+1.2*1}) -- ({2-0.2*(1.5)+0.15*sin(1.5*110)},{-1+1.2*1.5}) -- ({2-0.2*(2)+0.15*sin(2*110)},{-1+1.2*2}) -- ({2-0.2*(2.5)+0.15*sin(2.5*110)},{-1+1.2*2.5}) -- ({2-0.2*(3)+0.15*sin(3*110)},{-1+1.2*3}) -- (1.36,3.2) -- (3.0,1.56) -- (3.1,0.1) -- cycle;
\fill [fill=gray, draw=none,opacity=0.1]
 (3.2,0.2) -- (3.1,1.46) -- (3.0,1.56) -- (3.1,0.1) -- cycle;
\fill [fill=gray, draw=none,opacity=0.06]
 (3.2,0.2) -- (3.1,1.46) -- (3.2,1.36) -- (3.3,0.3) -- cycle;
\fill [fill=gray, draw=none,opacity=0.03]
 (3.4,0.4) -- (3.3,1.26) -- (3.2,1.36) -- (3.3,0.3) -- cycle;
\draw [thick, domain=0:3.5, samples=150] plot ({2-0.2*(\x)+0.15*sin(\x*110)},{-1+1.2*\x});
\node[darkgray] at (1.4, 1) [anchor = east] {$\gamma$};
\node[darkgray] at (1.9, -0.9) [anchor = east] {$_{\gamma(a)}$};
\node[darkgray] at (1.3, 3.2) [anchor = east] {$_{\gamma(b)}$};
\end{tikzpicture}
\end{center}
\end{figure}
\end{minipage}
\vspace{3mm} \\
The desired region $(I)$ in \hyperlink{thm:1.1}{Theorems~1.1}-\hyperlink{thm:1.2}{1.2} is to be constructed as a perturbation of the characteristic triangle exterior to $\gamma$ in $\mathcal{M}_{RN[\varpi,Q]}$. Our above restrictions of $\mathcal{M}_{RN[\varpi,Q]}$ ensure that, for any $\gamma$, these null curves always intersect, so that the region exists.\\ \\
In the Reissner-Nordstr\"om spacetimes, the charge $Q$ makes a contribution to the Hawking mass, so that $m$ does not coincide with the constant $\varpi$ except when $Q=0$, and in fact $m$ is non-constant. To adjust for this contribution, one defines the \textit{renormalized} Hawking mass
\begin{equation*}
\varpi:=m+\frac{Q^2}{2r}\equiv\frac{r}{2}\bigg(1+\frac{Q^2}{r^2}+\frac{4\partial_ur\partial_vr}{\Omega^2}\bigg)\tag{\hypertarget{eqn:2.4}{2.4}}
\end{equation*}
By construction, this quantity is constant in the Reissner-Nordstr\"om family, justifying the (abuse of) notation, but is well-defined on general charged spherically symmetric spacetimes. Analysis of $\varpi$ plays a fundamental role in the present article. \\ \\
As mentioned in \hyperlink{sec:1.3}{Section~1.3}, the Lorentz force law \hyperlink{eqn:1.13}{(1.13)} is taken as the equation of motion for the null dust 4-velocities $k$. For discussion of the implications of this equation for the fluid trajectories, see \hyperlink{Ori91}{[Ori91]} and [\hyperlink{KU24}{KU24}, Section~2.5].\\ \\ \\
\large\hypertarget{sec:2.3}{\textbf{2.3\hspace{4mm}Equations of motion in the interacting region}}\label{2.3}\normalsize\\ \\
We may now give the full system of equations in the interacting region. Following [\hyperlink{Mos17}{Mos17}; \hyperlink{Mos20}{Mos20}; \hyperlink{KU24}{KU24}], we write our equations with respect to the energy-momentum $T$ and number currents $N$ of the null dusts, instead of the original $\rho$ and $k$ variables. This is insignificant, as our decoupling will soon dispense with them also, but will be helpful later when we recover the hydrodynamic variables, as $T$ and $N$ do not experience the blowup associated to $\rho$ at $\mathcal{B}$.\\ \\
Thus, the dynamical variables, in total, consist of two positive smooth functions $r,\Omega^2$, and five nonnegative smooth functions $Q$, $N^v_\text{out}$, $N^u_\text{in}$ , $T_\text{out}^{vv}$ and $T^{uu}_\text{in}$ on $\mathcal{Q}$. This is a notational reminder of the fact that $T^{uu},T^{vv}$ only have contributions from the ingoing and outgoing dusts respectively.\\ \\
The system satisfies the wave equations
\begin{align*}\partial_u\partial_v r&= - \frac{\Omega^2}{2r^2}\bigg(\varpi-\frac{Q^2}{r}\bigg)\tag{\hypertarget{eqn:2.5}{2.5}}\\
\partial_u\partial_v \log\Omega^2&=\frac{\Omega^2}{r^3}\bigg(\varpi-\frac{3Q^2}{2r}\bigg)\tag{\hypertarget{eqn:2.6}{2.6}}
\end{align*}
the Raychaudhuri equations 
\begin{align*}\partial_u\bigg(\frac{\partial_ur}{\Omega^2}\bigg)  &=-\tfrac{1}{4}r\Omega^2T^{vv}_\text{out},\tag{\hypertarget{eqn:2.7}{2.7}}\\
\partial_v\bigg(\frac{\partial_vr}{\Omega^2}\bigg)&=- \tfrac{1}{4}r \Omega^2 T^{uu}_\text{in}\tag{\hypertarget{eqn:2.8}{2.8}}
\end{align*}
and, from \hyperlink{eqn:1.16}{(1.16)}, the Maxwell equations
\begin{align*}
\partial_u Q&=-\tfrac{1}{2}\mathfrak{e} r^2\Omega^2N^v_\text{out}\tag{\hypertarget{eqn:2.9}{2.9}}\\
\partial_v Q&= \tfrac{1}{2}\mathfrak{e}r^2\Omega^2 N^u_\text{in}\tag{\hypertarget{eqn:2.10}{2.10}}
\end{align*}
Finally, the number currents satisfy\footnote[11]{These are naturally postulated for the two dust species  separately. Conservation of the energy-momentum tensor \hyperlink{eqn:1.15}{(1.15)} alone does not entail them both, just as, in the Einstein-Vlasov system, conservation of energy-momentum alone does not entail that the Vlasov equation holds pointwise on the tangent bundle.} the conservation laws 
\begin{align*}
\partial_u(r^2\Omega^2N^u_\text{in})=0\tag{\hypertarget{eqn:2.11}{2.11}}\\  
\partial_v(r^2\Omega^2N^v_\text{out})=0\tag{\hypertarget{eqn:2.12}{2.12}}
\end{align*}
From \hyperlink{eqn:2.5}{(2.5)}-\hyperlink{eqn:2.8}{(2.8)}, one easily derives the mass evolution equations
\begin{align*}
    \partial_u\varpi &= -\tfrac{1}{2}r^2\Omega^2T_\text{out}^{vv}\partial_vr-\tfrac 12\mathfrak e r\Omega^2QN_\text{out}^v\tag{\hypertarget{eqn:2.13}{2.13}}\\
    \partial_v\varpi&= -\tfrac 12 r^2\Omega^2T_\text{in}^{uu}\partial_ur +\tfrac 12\mathfrak e r\Omega^2QN_\text{in}^u\tag{\hypertarget{eqn:2.14}{2.14}}
\end{align*}
Alternative forms are written below in \hyperlink{eqn:2.22}{(2.22)}-\hyperlink{eqn:2.23}{(2.23)}.\\ \\
Given a solution to the above system, and if $N^v_\text{out},N^u_\text{in},T^{vv}_\text{out},T^{uu}_\text{in}$ are non-zero, then we may recover the original hydrodynamic variables. Indeed, if we set
\begin{equation*}\rho_\text{out} := \frac{(N_\text{out}^v)^2}{T_\text{out}^{vv}}, \qquad \rho_\text{in} := \frac{(N_\text{in}^u)^2}{T_\text{in}^{uu}},\qquad k_\text{out}^v:= \frac{T_\text{out}^{vv}}{N_\text{out}^v},\qquad k_\text{in}^u:=\frac{T_\text{in}^{uu}}{N_\text{in}^u}\tag{\hypertarget{eqn:2.15}{2.15}}\end{equation*}
then
\begin{align*}
k_\text{out}^v \partial_v k_\text{out}^v +\partial_v\log\Omega^2(k^v_\text{out})^2 &= \mathfrak e\frac{Q}{r^2}k^v_\text{out}\tag{\hypertarget{eqn:2.16}{2.16}}\\
k_\text{in}^u \partial_u k_\text{in}^u +\partial_u\log\Omega^2(k^u_\text{in})^2 &= -\mathfrak e\frac{Q}{r^2}k^u_\text{in}\tag{\hypertarget{eqn:2.17}{2.17}}
\end{align*}
We stress that the denominators must be non-zero to recover the original variables. Since this \textit{a priori} may not hold, one may view the system (\hyperlink{eqn:2.5}{2.5)}-\hyperlink{eqn:2.12}{(2.12)} as a set of \textit{formal} equations: see [\hyperlink{KU24}{KU24}, Section~4.6].
\\ \\
\textbf{Observation 2.1} \textit{We note that equations \hyperlink{eqn:2.9}{(2.9)}-\hyperlink{eqn:2.12}{(2.12)} entail that
\begin{equation*}
    \partial_u \partial_v Q =0\tag{\hypertarget{eqn:2.18}{2.18}}
\end{equation*}
This means that the variables $(r,\Omega^2,Q)$ form a closed system, obeying \hyperlink{eqn:2.5}{(2.5)}, \hyperlink{eqn:2.6}{(2.6)} and \hyperlink{eqn:2.18}{(2.18)}.}\\ \\
Recall the boundary conditions \hyperlink{eqn:1.20}{(1.20)} to be imposed on $Q$ at the bounce $\mathcal{B}$. One may expect an analogous boundary condition for $\varpi$, requiring continuity across $\mathcal{B}$. This in fact follows from the conditions \hyperlink{eqn:1.23}{(1.23)}, and can serve as a replacement for either, as the following lemma makes precise. \\ \\
\hypertarget{lem:2.1}{\textbf{Lemma 2.1}} (Exchanging geometric boundary conditions)\textbf{.} \textit{Suppose $\mathcal{B}$ is a timelike curve in double null coordinates $(u,v)$ along which $Q=Q_0$, and that $\partial_ur,\partial_vr\neq0$ holds on $\mathcal{B}$ except at finitely many points. Suppose $r,\Omega^2$ satisfy equations \hyperlink{eqn:2.5}{(2.5)}-\hyperlink{eqn:2.6}{(2.6)} in a neighbourhood of $\mathcal{B}$. Then any two of the following conditions implies the third:
\begin{align*}
(i)\hspace{2.4cm}\varpi\big|_\mathcal{B}=&\hspace{3pt}\text{const.}\hspace{10cm}\tag{\hypertarget{eqn:2.19}{2.19}}\\[0.5em]
(ii)\hspace{1.5cm}\partial_u\bigg(\frac{\partial_ur}{\Omega^2}\bigg)\bigg|_\mathcal{B}&=0\tag{\hypertarget{eqn:2.20}{2.20}}\\[0.5em]
(iii)\hspace{1.5cm}\partial_v\bigg(\frac{\partial_vr}{\Omega^2}\bigg)\bigg|_\mathcal{B}&=0\tag{\hypertarget{eqn:2.21}{2.21}}
\end{align*}}
\hspace{-1.5mm}\textit{Proof.} All the hypotheses of the lemma, and conditions \hyperlink{eqn:2.19}{(2.19)}-\hyperlink{eqn:2.21}{(2.21)} themselves, are invariant under a change of null coordinates. Thus we can assume $\mathcal{B}\subset\{u=v\}$ without loss of generality. Using \hyperlink{eqn:2.7}{(2.7)}-\hyperlink{eqn:2.10}{(2.10)}, the mass evolution equations \hyperlink{eqn:2.13}{(2.13)}-\hyperlink{eqn:2.14}{(2.14)} may be written as 
\begin{align*}
\partial_u\varpi&=2r\partial_vr\partial_u\bigg(\frac{\partial_ur}{\Omega^2}\bigg)+\frac{Q\partial_uQ}{r}\tag{\hypertarget{eqn:2.22}{2.22}}\\
\partial_v\varpi&=2r\partial_ur\partial_v\bigg(\frac{\partial_vr}{\Omega^2}\bigg)+\frac{Q\partial_vQ}{r}\tag{\hypertarget{eqn:2.23}{2.23}}
\end{align*}
Adding these equations and evaluating at $\mathcal{B}$ gives
\begin{equation*}
\varpi'_0(t)=2r_0(\partial_ur)_0\partial_v\bigg(\frac{\partial_vr}{\Omega^2}\bigg)_0+2r_0(\partial_vr)_0\partial_u\bigg(\frac{\partial_ur}{\Omega^2}\bigg)_0+\frac{Q_0Q'_0}{r_0}
\end{equation*}
Since $Q_0$ is constant, the latter term always vanishes. It is then clear that any two of conditions (i)-(iii) implies the third, since this makes two further terms of the equation vanish, implying that the remaining term vanishes too. If it is condition (ii) or (iii) to be inferred from the other two, then we can only do this at points $t$ where $(\partial_vr)_0(t)\neq0$ or $(\partial_ur)_0(t)\neq0$ respectively. However, it is enough, by continuity, to verify any \hyperlink{eqn:2.19}{(2.19)}-\hyperlink{eqn:2.21}{(2.21)} at all but finitely many points.\hfill$\square$\\ \\
\textbf{Remark 2.1.} \textit{We note that the first term on the right-hand side of equation \hyperlink{eqn:2.23}{(2.23)} vanishes on the bounce $\mathcal{B}$, and after rearranging for $r$ we have 
$$r_0=\frac{Q_0(\partial_vQ)_0}{(\partial_v\varpi)_0} $$
Comparison with equation \hyperlink{eqn:1.12}{(1.12)} reveals that that formula still characterizes the location of the bounce $\mathcal{B}$, if the charge and mass are computed \ul{locally}.}\\ \\ 
We refer to the reduced system of equations \hyperlink{eqn:2.5}{(2.5)}-\hyperlink{eqn:2.6}{(2.6)} and \hyperlink{eqn:2.18}{(2.18)}, together with boundary conditions \hyperlink{eqn:1.20}{(1.20)} and \hyperlink{eqn:2.19}{(2.19)}-\hyperlink{eqn:2.21}{(2.21)}, as the \ul{$(r,\Omega^2,Q)$ system}. By the above lemma, we may judiciously ignore one of conditions \hyperlink{eqn:2.19}{(2.19)}-\hyperlink{eqn:2.21}{(2.21)} when constructing solutions to the $(r,\Omega^2,Q)$ system.\\ \\
Recall that our construction in \hyperlink{thm:1.1}{Theorems 1.1}-\hyperlink{thm:1.2}{1.2} is to involve a perturbation of the characteristic triangle exterior to the prescribed curve $\gamma$ in $\mathcal{M}_{RN[\varpi,Q]}$. In \hyperlink{thm:1.1}{Theorem 1.1}, we use the lapse-normalized gauge choice, under which this region is coordinatized precisely as $\triangleright=\{a\leq u\leq v\leq b\}$ and such that $\Omega^2|_{u=v}=1$ (see \hyperlink{sec:2.1}{Section 2.1}). Without loss of generality, suppose that the $\tau\in[a,b]$ parametrizing $\gamma$ coincides with $t=\tfrac{1}{2}(u+v)$, and is in particular proper time. We consider the derivative of $r$, in the unperturbed $\mathcal{M}_{RN[\varpi,Q]}$, along the unit spacelike normal to $\gamma$ (pointing right towards $i^0$). Denoting this $(\partial_xr)_0(t)$, this is a smooth function of $t$. Here $\Omega^2|_{u=v}=1$ again justifies the notation, with $\partial_x:=\partial_v-\partial_u$. We may then also obtain an expression for the transverse derivative of $\Omega^2$, by taking a suitable combination of equations \hyperlink{eqn:2.5}{(2.5)}-\hyperlink{eqn:2.8}{(2.8)}:
\begin{equation*}
(\partial_x\log\Omega^2)_0(t)=\frac{(\partial_xr)'_0(t)}{r'_0(t)}\tag{\hypertarget{eqn:2.24}{2.24}}
\end{equation*}
If $\gamma$ lies in the exterior region only, then these transverse derivatives may be written in terms of $r_0(t)$, without remembering extra `transverse information'. By computing \hyperlink{eqn:2.2}{(2.2)} in the orthonormal frame given by $\gamma'(t)$ and the unit spacelike normal to $\gamma$, we have in this case
\begin{equation*}
(\partial_xr)_0(t)=\sqrt{1-\frac{2\varpi_0}{r_0(t)}+\frac{Q^2_0}{r_0(t)^2}+r'_0(t)^2}>0\tag{\hypertarget{eqn:2.25}{2.25}}
\end{equation*}
and, for the transverse derivative of $\Omega^2$,
\begin{equation*}
(\partial_x\log\Omega^2)_0(t)=\frac{1}{(\partial_xr)_0(t)}\bigg(r''_0(t)+\frac{\varpi_0}{r_0(t)^2}-\frac{Q^2_0}{r_0(t)^3}\bigg)\tag{\hypertarget{eqn:2.26}{2.26}}
\end{equation*}
This also shows that $(\partial_x\log\Omega^2)_0(t)$ is smooth, even when $r'_0(t)=0$, which was not immediately clear from \hyperlink{eqn:2.24}{(2.24)} (note also that $r'_0(t)=0$ is only possible in the exterior). Expressions \hyperlink{eqn:2.25}{(2.25)}-\hyperlink{eqn:2.26}{(2.26)} are not available in the interior, because the sign of $(\partial_xr)_0(t)$ cannot be read off from $r'_0(t)$ alone. Note however that the square-rooted expression in \hyperlink{eqn:2.25}{(2.25)} itself is greater than or equal to 0, with equality when $\gamma'$ is the unit normal to level curves of $r$.\\ \\
In the proof of \hyperlink{thm:1.1}{Theorem~1.1}, we view the metric components of Reissner-Nordstr\"om, in this chart, as the unique solution of equations \hyperlink{eqn:2.5}{(2.5)}-\hyperlink{eqn:2.6}{(2.6)} on $\triangleright$, with the above data posed at $\{u=v\}$. The perturbation is obtained by taking the \textit{same} data for $r,\Omega^2$, but introducing non-trivial data for $Q$, and solving the system \hyperlink{eqn:2.5}{(2.5)}, \hyperlink{eqn:2.6}{(2.6)} and \hyperlink{eqn:2.18}{(2.18)} jointly on $\triangleright$.\\ \\
The following useful identities may be derived from the wave equations \hyperlink{eqn:2.5}{(2.5)}-\hyperlink{eqn:2.6}{(2.6)}:
\begin{equation*}\partial_v\Big(r\Omega^2\partial_u\bigg(\frac{\partial_ur}{\Omega^2}\bigg)\Big)\equiv \partial_v\big(r\partial^2_ur-r\partial_ur\partial_u\log\Omega^2\big)=\frac{\Omega^2}{2r^2}Q\partial_uQ\tag{\hypertarget{eqn:2.27}{2.27}}
\end{equation*}
\begin{equation*}
\partial_u\Big(r\Omega^2\partial_v\bigg(\frac{\partial_vr}{\Omega^2}\bigg)\Big)\equiv \partial_u\big(r\partial^2_vr-r\partial_vr\partial_v\log\Omega^2\big)=\frac{\Omega^2}{2r^2}Q\partial_vQ\tag{\hypertarget{eqn:2.28}{2.28}}
\end{equation*}
These will be of fundamental use in the proofs of \hyperlink{thm:1.1}{Theorems~1.1}-\hyperlink{thm:1.3}{1.3}, in showing that the energy-momentum and number current components obtained in our examples are indeed positive and finite, allowing meaningful hydrodynamic functions to be recovered as above.\\ \\
We also record a useful lemma, according to which the Vaidya metric arises precisely as the special case of the $(r,\Omega^2,Q)$ system for which either $\partial_uQ=0$ or $\partial_vQ=0$.\\ \\
\hypertarget{lem:2.2}{\textbf{Lemma 2.2 }}(Characterization of Vaidya spacetimes as a special case of the $(r,\Omega^2,Q)$ system)\textbf{.}\vspace*{1mm}
\hspace*{0.02\textwidth}\begin{minipage}{0.73\textwidth}
\textit{\ul{Ingoing case:} Let $\Sigma\subset\mathbb{R}^2_{u,v}$ be a bounded open subset. Let $S\subset\partial\Sigma$ have the property that $\Sigma$ is the union of $u=\text{const}.$ line segments which have past endpoint on $S$. Suppose $r,\Omega^2,Q\in C^2(\Sigma)$ solve equations \hyperlink{eqn:2.5}{(2.5)}-\hyperlink{eqn:2.6}{(2.6)} and \hyperlink{eqn:2.18}{(2.18)} on $\Sigma$ with $\partial_ur<0$. Suppose moreover that $r,\Omega^2,Q$ extend as $C^2$ functions to $S$ with $$\partial_uQ\big|_{S}=0;\qquad \partial_u\bigg(\frac{\partial_ur}{\Omega^2}\bigg)\bigg|_S=0 $$
Then $(\Sigma\times S^2,-\Omega^2dudv+r^2g_{S^2})$ is isometric to an ingoing Vaidya spacetime, with the isometry provided by a reparametrization of the null coordinate $v$.}\vspace*{1mm} \\
\textit{\ul{Outgoing case:} Let $\Sigma\subset\mathbb{R}^2_{u,v}$ be a bounded open subset. Let $S\subset\partial\Sigma$ have the property that $\Sigma$ is the union of $v=\text{const}.$ line segments which have future endpoint on $S$. Suppose $r,\Omega^2,Q\in C^2(\Sigma)$ solve equations \hyperlink{eqn:2.5}{(2.5)}-\hyperlink{eqn:2.6}{(2.6)} and \hyperlink{eqn:2.18}{(2.18)} on $\Sigma$ with $\partial_vr>0$. Suppose moreover that $r,\Omega^2,Q$ extend as $C^2$ functions to $S$ with $$\partial_vQ\big|_{S}=0;\qquad \partial_v\bigg(\frac{\partial_vr}{\Omega^2}\bigg)\bigg|_S=0 $$
Then $(\Sigma\times S^2,-\Omega^2dudv+r^2g_{S^2})$ is isometric to an outgoing Vaidya spacetime, with the isometry provided by a reparametrization of the null coordinate $u$.}
\end{minipage}
\begin{minipage}{0.25\textwidth}
\vspace{-2mm}
\begin{figure}[H]
\begin{center}
\begin{tikzpicture}
\draw [darkgray, thick] (0.22, 4.78) -- (2.83,2.17);
\draw [darkgray, thick] (0,3) -- (2.28, 0.73);
\draw [darkgray, thick] (0.6,4.1) -- (2.3,2.4);
\draw [darkgray, thick] (0.6,2.7) -- (2.1,1.2);
\draw [darkgray, thick, -stealth] (1.3,2) -- (1.4,1.9);
\draw [darkgray, thick, -stealth] (1.4,3.3) -- (1.5,3.2);
\draw [darkgray, thick] (0.26, 5.46) -- (2.86,8.06);
\draw [darkgray, thick] (0.13,7.33) -- (2.43, 9.63);
\draw [darkgray, thick] (0.6,6.1) -- (2.3,7.8);
\draw [darkgray, thick] (0.6,7.5) -- (2.1,9);
\draw [darkgray, thick, -stealth] (1.3,8.2) -- (1.4,8.3);
\draw [darkgray, thick, -stealth] (1.4,6.9) -- (1.5,7);
\node[darkgray] at (1.7, 2) [anchor = west] {$\Sigma$};
\node[darkgray] at (1.7, 8.2) [anchor = west] {$\Sigma$};
\node[darkgray, align=left] at (0.1,3.8) [anchor = east] {$S$};
\node[darkgray, align=left] at (0.15,6.3) [anchor = east] {$S$};
\draw [darkgray, thick, domain=0.0:1.79, samples=150] plot ({(0.1*sin(150*\x)+0.1*(\x)^2)},{3+\x});
\draw [darkgray, thick, domain=0.26:2.13, samples=150] plot ({(0.2+0.1*sin(150*\x)-0.0*(\x)^2)},{5.2+\x});
\draw [darkgray, thick, domain=0.35:1.94, samples=150] plot ({2.8-0.13*(\x)+0.15*sin(120*(\x))},{7.7+\x});
\draw [darkgray, thick, domain=0.72:2.19, samples=150] plot ({2.2+0.13*(\x)^2)},{\x});
\end{tikzpicture}
\end{center}
\end{figure}
\end{minipage}\vspace*{3.5mm}\\
\textit{Proof. }\underline{Ingoing case:} By equation \hyperlink{eqn:2.18}{(2.18)}, and the geometry of $\Sigma$, $\partial_uQ=0$ propagates from $S$ to the whole of $\Sigma$. It then follows that the right-hand side of equation \hyperlink{eqn:2.27}{(2.27)} vanishes, that is,
$$\partial_v\Big(r\Omega^2\partial_u\bigg(\frac{\partial_ur}{\Omega^2}\bigg)\Big)=0$$
Hence $\partial_u(\partial_ur/\Omega^2)=0$ also propagates from $S$ to the whole of $\Sigma$. Finally by equation \hyperlink{eqn:2.22}{(2.22)},\vspace{-2pt}
$$\partial_u\varpi=2r\partial_vr\partial_u\bigg(\frac{\partial_ur}{\Omega^2}\bigg)+\frac{Q\partial_uQ}{r}=0$$
and so both $\varpi=\varpi(v)$ and $Q=Q(v)$ are functions of $v$ only.\\ \\
We now reparametrize the null coordinate $v$ by setting, implicitly, $v=f(\tilde v)$, for some $f$ to be chosen, and fixing some $u=\text{const.}$ line segment, say $u=u_0$. Geometrically, the lapse function transforms as 
$$\tilde{\Omega}^2(u,\tilde{v})=f'(\tilde v)\Omega^2(u,f(\tilde v)) $$
while $r,\varpi,Q$ are gauge invariants that we refer to with their original symbols. We choose $f$ so that 
$$ -\frac{\tilde{\Omega}^2(u_0,\tilde v)}{2\partial_ur(u_0,\tilde v)}\equiv1$$
which is achieved by solving the ODE\vspace{-2pt}
$$-f'(\tilde v)\frac{\Omega^2(u_0,f(\tilde v))}{2\partial_ur(u_0,f(\tilde v))}=1 $$ with an arbitrary integration constant. The reparametrization $f$ then has strictly positive derivative because $\partial_ur<0$, and so $\tilde\Omega^2$ is positive. However, now that $-\tilde{\Omega}^2/2\partial_ur=1$ holds on \textit{some} $u=u_0$, this too propagates from $u=u_0$ to the whole of $\Sigma$, using that $\partial_u(\partial_ur/\Omega^2)=0$. In the new chart $(\tilde v,r)$, one computes from the definition \hyperlink{eqn:2.4}{(2.4)} of $\varpi$ that
\begin{align*}
g&=-\bigg(1-\frac{2\varpi(\tilde v)}{r}+\frac{Q(\tilde v)^2}{r^2}\bigg)\frac{1}{(g_{\tilde v r})^2}d\tilde v^2+2g_{\tilde vr}d\tilde vdr+r^2g_{S^2}\\
&=-\bigg(1-\frac{2\varpi(\tilde v)}{r}+\frac{Q(\tilde v)^2}{r^2}\bigg)d\tilde v^2+2d\tilde vdr+r^2g_{S^2}
\end{align*}
since we have$$g_{\tilde v r}=-\frac{\tilde\Omega^2}{2\partial_ur}=1 $$
This is precisely the ingoing Vaidya metric, so we have identified the required isometry.
\\ \\
\ul{Outgoing case:} The same argument holds \textit{mutatis mutandis} as for the ingoing case, swapping the roles of the null coordinates $u,v$.\hfill$\square$\\ \\ \\
\large\hypertarget{sec:2.4}{\textbf{2.4\hspace{4mm}The $\bm{(r,\kappa,Q)}$ system}}\label{2.4}\normalsize\\ \\
We now discuss the reformulation of equations pertinent to \hyperlink{thm:1.2}{Theorem~1.2}. Recall that, in this theorem, we have a curve segment $\gamma$ in $\mathcal{M}_{RN[\varpi_0,Q_0]}$ terminating in an (outgoing) null point. To establish the theorem, we will again choose a system of double null coordinates which straighten $\gamma$, i$.$e$.$ so that $\gamma=\{u=v\}$. Let $(\tilde u,v)$ be double null coordinates covering the characteristic triangle exterior to $\gamma$ in $\mathcal{M}_{RN[\varpi_0,Q_0]}$. In this chart, $\gamma$ is represented by $\{\tilde u=f(v)\}$ for some $f\in C^\infty[a,b]$ with $f'(b)=0$. We can achieve the straightening of $\gamma$ by implicitly defining $u$ through $\tilde u=f(u)$. However, this is not a valid reparametrization of null coordinates as defined in \hyperlink{sec:2.1}{Section~2.1}, where we required $f$ to have strictly positive derivative: with this choice, the lapse function $\Omega^2$ vanishes at the null point. Such coordinates $(u,v)$ constitute an \textit{improper} chart (see [\hyperlink{Chr95}{Chr95}, Section~6])\footnote[12]{If we instead used a lapse-normalized chart $(u,v)$, we would achieve $\Omega^2\neq0$ at the cost of having $r'_0\to\infty$ at the null point: in fact, we would have $(\partial_vr)_0\to\infty$ while $(\partial_ur)_0\to0$.}. Of course, this is expected: $\{u=v\}$ is always timelike, so $\gamma=\{u=v\}$ cannot hold up to the null point without degeneracy.\\ \\
Following Christodoulou, we eliminate the lapse function $\Omega^2$ in favour of a variable that remains regular up to the null point, namely
$$\kappa:=\frac{\partial_vr}{1-\frac{2\varpi}{r}+\frac{Q^2}{r^2}}\equiv -\frac{\Omega^2}{4\partial_ur} $$
This function $\kappa$ is invariant under a reparametrization of the $u$ coordinate, which is precisely the transformation undertaken above. So, in the new (improper) chart $(u,v)$, the values of $\kappa$ remain positive and finite, as they are in the $\partial_ur<0$ region of $\mathcal{M}_{RN[\varpi_0,Q_0]}$. The new variable $\kappa$ satisfies its own wave equation which may be derived from \hyperlink{eqn:2.5}{(2.5)}-\hyperlink{eqn:2.6}{(2.6)}. Rewriting both \hyperlink{eqn:2.5}{(2.5)}-\hyperlink{eqn:2.6}{(2.6)} in terms of $\kappa$, we have
\begin{align*}
\partial_u\partial_vr&=\frac{\kappa\partial_ur}{r}-\frac{\partial_ur\partial_vr}{r}-\frac{\kappa\partial_ur Q^2}{r^3}\tag{\hypertarget{eqn:2.29}{2.29}}\\
\partial_u\partial_v\kappa&= \frac{\partial_u\kappa\partial_v\kappa}{\kappa}+\frac{2\kappa^2Q\partial_uQ}{r^3}-\frac{\kappa\partial_u\kappa}{r}+\frac{\kappa\partial_u\kappa Q^2}{r^3} \tag{\hypertarget{eqn:2.30}{2.30}}
\end{align*}
The boundary conditions \hyperlink{eqn:1.20}{(1.20)},\hyperlink{eqn:1.23}{(1.23)} are moreover equivalent to
\begin{equation*}
Q_0=Q_0\qquad(\partial_u\kappa)_0=0\qquad(\partial_vr)_0=\kappa_0\bigg(1-\frac{2\varpi_0}{r_0}+\frac{Q^2_0}{r^2_0}\bigg) \tag{\hypertarget{eqn:2.30}{2.30}}
\end{equation*}
We refer to the equations \hyperlink{eqn:2.18}{(2.18)} and \hyperlink{eqn:2.29}{(2.29)}-\hyperlink{eqn:2.30}{(2.30)}, taken together with the conditions \hyperlink{eqn:2.30}{(2.30)}, as the \ul{$(r,\kappa,Q)$ system}. The system makes sense as long as both $\kappa$ and $r$ are kept bounded away from zero. \\ \\
Where $\partial_ur<0$, we may reverse the above definition of $\kappa$ to uniquely recover $\Omega^2$ from a solution to the $(r,\kappa,Q)$ system. The resulting variables obey the $(r,\Omega^2,Q)$ system. In the proof of \hyperlink{thm:1.2}{Theorem~1.2}, we will obtain a solution to the $(r,\kappa,Q)$ system on a triangular domain $\triangleright$, representing an improper coordinatization of the region $(I)$. After transforming coordinates again so that $\partial_ur<0$ throughout the domain, we obtain the solution in a \textit{bona fide} coordinate chart.
\\ \\ \\
\large\hypertarget{sec:2.5}{\textbf{2.5\hspace{4mm}The $\bm{(r,\varpi,\phi,Q)}$ system}}\label{2.5}\normalsize\\ \\
Yet another reformulation of equations is useful for \hyperlink{thm:1.3}{Theorem~1.3}, which we first motivate by discussing in more detail the characteristic data imposed in the problem. Recall from \hyperlink{sec:2.1}{Section~2.1} the Vaidya-normalized gauge choice. We start from Vaidya seed data $\varpi(v),Q(v)$, i$.$e$.$ we start with the 
\begin{minipage}{0.62\textwidth}\vspace{0.8mm}region $(I\hspace{-2.5pt}I)$ which is exactly described by the metric \hyperlink{eqn:1.9}{(1.9)}. Because the causal future of $\mathcal{B}$ is excised, region $(I\hspace{-2.5pt}I)$ is cut off along the outgoing radial null curve emanating from the formation point $(v,r)=(0,r_b(0))$. We write this null curve as $v\mapsto (v,r_+(v))$, parametrizing with the Vaidya $v$ coordinate. (Again, all quantities evaluated at $u=0$ have the subscript `+'.) The condition that the tangent vector $\partial_v+r'_+(v)\partial_v$ is null reads
\begin{equation*}r'_+(v)=\tfrac{1}{2}\Big(1-\frac{2\varpi(v)}{r_+(v)}+\frac{Q^2(v)}{r^2_+(v)}\Big) \tag{\hypertarget{eqn:2.32}{2.32}}
\end{equation*}
\end{minipage}
\hfill
\begin{minipage}{0.35\textwidth}
\vspace{-4mm}
\begin{figure}[H]
\begin{center}
\begin{tikzpicture}
\draw [darkgray, thick] (0,0) -- (-2.0, 2);
\draw [darkgray, thick] (3.5,0.5) -- (0.5, 3.5);
\draw [darkgray, thick, -stealth] (-2,2) -- (-0.5, 3.5);
\node[darkgray, align=left] at (-1.5,2.3) [anchor = west] {$(v,r_+(v))$};
\node[darkgray, align=left] at (0.3,1.4) [anchor = west] {$(I\hspace{-2.5pt}I)$};
\fill [fill=gray, draw=none,opacity=0.1] (-1,1) -- (-2,2) -- (-1,3) -- (1,3) -- (3,1) -- cycle;
\fill [fill=gray, draw=none,opacity=0.08] (-1,1) -- (3,1) -- (3.2,0.8) -- (-0.8,0.8) -- cycle;
\fill [fill=gray, draw=none,opacity=0.08] (-1,3) -- (1,3) -- (0.9,3.1) -- (-0.9,3.1) -- cycle;
\fill [fill=gray, draw=none,opacity=0.06] (-0.8,0.8) -- (3.2,0.8) -- (3.4,0.6) -- (-0.6,0.6) -- cycle;
\fill [fill=gray, draw=none,opacity=0.06] (-0.8,3.2) -- (0.8,3.2) -- (0.9,3.1) -- (-0.9,3.1) -- cycle;
\fill [fill=gray, draw=none,opacity=0.04] (-0.4,0.4) -- (3.6,0.4) -- (3.4,0.6) -- (-0.6,0.6) -- cycle;
\fill [fill=gray, draw=none,opacity=0.04] (-0.8,3.2) -- (0.8,3.2) -- (0.7,3.3) -- (-0.7,3.3) -- cycle;
\fill [fill=gray, draw=none,opacity=0.02] (-0.4,0.4) -- (3.6,0.4) -- (3.8,0.2) -- (-0.2,0.2) -- cycle;
\fill [fill=gray, draw=none,opacity=0.02] (-0.6,3.4) -- (0.6,3.4) -- (0.7,3.3) -- (-0.7,3.3) -- cycle;
\filldraw[color=black, fill=white](-2,2) circle (0.04);
\end{tikzpicture}
\end{center}
\end{figure}
\end{minipage}\vspace{4mm}\\
Together with the initial condition $r_+(0)=r_b(0)$, and the prescribed functions $\varpi(v),Q(v)$, this uniquely determines $r_+(v)$. Let double null coordinates $(u,v)$ be chosen in a neighbourhood of the curve $\{r=r_+(v)\}$, labelling $\{r=r_+(v)\}$ as $\{u=0\}$. The above relation determining $r$ along outgoing null curves now reads
\begin{align*}
\partial_vr&\equiv\tfrac{1}{2}\Big(1-\frac{2\varpi(v)}{r}+\frac{Q^2(v)}{r^2}\Big) \\
\implies \Omega^2&\equiv-2\partial_vr
\end{align*}
using the definition \hyperlink{eqn:2.2}{(2.2)}. This identity holds in region $(I\hspace{-2.5pt}I)$ in the Vaidya normalization. Equation \hyperlink{eqn:2.5}{(2.5)} then becomes a propagation equation for $\Omega^2_+(v)$:
\begin{align*}
(\Omega^2_+)'(v)&=\Omega^2_+(v)\bigg(\frac{\varpi(v)}{r^2_+(v)}-\frac{Q^2(v)}{r^3_+(v)}\bigg)\tag{\hypertarget{eqn:2.33}{2.33}}\\
\implies \Omega^2_+(v)&=\Omega^2_+(0)\exp\bigg(\int^v_0\frac{\varpi(\tilde v)}{r^2_+(\tilde v)}-\frac{Q^2(\tilde v)}{r^3_+(\tilde v)}d\tilde v\bigg)
\end{align*}
so that $\Omega^2_+(v)$ is uniquely determined, up to an (important) missing multiplicative constant which encodes the initial tangent vector of the bounce $\mathcal{B}$.\\ \\
Turning now to the forwards problem, which is to construct region $(I)$, we uniquely define the null cooordinate $u$ to be such that $\mathcal{B}=\{u=v\}$. The above discussion yields characteristic data $r_+,\Omega^2_+,Q_+$ (modulo the missing multiplicative constant for $\Omega^2$) at $\{u=0\}$, and we continue to impose \hyperlink{eqn:1.20}{(1.20)}, \hyperlink{eqn:1.23}{(1.23)} at the bounce $\{u=v\}$. Solving for $Q$ will be immediate:
\begin{equation*}
Q(u,v)=Q_0+Q_+(v)-Q_+(u)\tag{\hypertarget{eqn:2.34}{2.34}}
\end{equation*}
However, we must find a way to propagate the other characteristic data from $u=0$. A key insight, following Christodoulou \hyperlink{Chr95}{[Chr95]}, is to give $\varpi$ the status of an independent variable. By \hyperlink{lem:2.1}{Lemma~2.1}, this takes care of one boundary condition at $\{u=v\}$. It also suggests a means of propagating $r$, namely, to make $\partial_ur$ the subject of equation \hyperlink{eqn:2.4}{(2.4)}:
$$\partial_ur=-\bigg(\frac{\Omega^2}{4\partial_vr}\bigg)\bigg(1-\frac{2\varpi}{r}+\frac{Q^2}{r^2}\bigg) $$
On the right-hand side, we have independent knowledge of $\varpi$ (through its evolution equation) and $Q$ (through the above expression), so the only other unknown is the ratio $\Omega^2/4\partial_vr$. This latter quantity also has a boundary condition at $\{u=v\}$, and---up to multiplicative constant---characteristic data at $\{u=0\}$. We therefore choose this as our other dynamical quantity, in favour of $\Omega^2$. In fact, in view of the missing multiplicative constant, we choose
$$\phi:=\log\bigg(\frac{\Omega^2}{4\partial_vr}\bigg) $$
so that $\phi_+$ has a missing \textit{additive} constant---in other words, $\phi'_+$ is fully expressible in terms of seed data. The $\log$ transformation also eliminates an undesirable term from its equation of motion.
\\ \\
We arrive at the equivalent \ul{$(r,\varpi,\phi,Q)$ system}:
\begin{align*}
\partial_u r&=-e^{\phi}\bigg(1-\frac{2\varpi}{r}+\frac{Q^2}{r^2}\bigg)\tag{\hypertarget{eqn:2.35}{2.35}}\\
\partial_v \varpi&=\tfrac{1}{2}r\partial_v\phi\bigg(1-\frac{2\varpi}{r}+\frac{Q^2}{r^2}\bigg)+\frac{Q\partial_vQ}{r}\tag{\hypertarget{eqn:2.36}{2.36}}\\
\partial_u\partial_v\phi&=e^{\phi}\bigg(-\frac{2Q\partial_vQ}{r^3}+\frac{\partial_v\phi}{r}-\frac{Q^2\partial_v\phi}{r^3}\bigg)\tag{\hypertarget{eqn:2.37}{2.37}}
\end{align*}
taken together with equation \hyperlink{eqn:2.18}{(2.18)} and with boundary conditions and characteristic data
$$\varpi\big|_{u=v}=\varpi_0\qquad\partial_v\phi\big|_{u=v}=0\qquad r\big|_{u=0}=r_+\qquad \partial_v\phi\big|_{u=0}=\phi'_+  \qquad Q\big|_{u=0}=Q_0$$
Equation \hyperlink{eqn:2.37}{(2.37)} is a nonlinear wave equation for $\phi$. Solving the equation, even with $r,Q$ fixed, is sensitive because $\partial_v\phi$ is specified \textit{at both boundaries} $\{u=0\}$ and $\{u=v\}$. If, for example, \hyperlink{eqn:2.37}{(2.37)} were instead $\partial_u\partial_v\phi=0$, this would be an overdetermined problem. We will see, however, that a unique solution to \hyperlink{eqn:2.37}{(2.37)} is available, in the case where $\phi''_+$ is bounded away from zero, which is entailed by our hard-edge condition.\\ \\
We finally demonstrate that solutions to the original $(r,\Omega^2,Q)$ system may be recovered from solutions to the $(r,\varpi,\phi,Q)$ system. Whenever $\partial_vr>0$, we may simply define
$$\Omega^2:=4e^\phi\partial_vr $$
and, using equations \hyperlink{eqn:2.35}{(2.35)}-\hyperlink{eqn:2.37}{(2.37)}, a little algebra yields
\begin{align*}
\partial_u\partial_v\log\Omega^2&=\frac{4e^\phi\partial_vr}{r^3}\bigg(\varpi-\frac{3Q^2}{2r}\bigg) \\
&=\frac{\Omega^2}{r^3}\bigg(\varpi-\frac{3Q^2}{2r}\bigg) 
\end{align*}
where, in the second line, we inserted the definition of $\Omega^2$. This is precisely equation \hyperlink{eqn:2.6}{(2.6)}, so $\Omega^2$ obeys its defining equation. The propagation equation \hyperlink{eqn:2.35}{(2.35)} for $r$ then rearranges to give equation \hyperlink{eqn:2.3}{(2.3)}, so that $\varpi$ indeed coincides with the renormalized Hawking mass. Differentiating \hyperlink{eqn:2.35}{(2.35)} in $\partial_v$ and applying \hyperlink{eqn:2.36}{(2.36)} recovers the wave equation \hyperlink{eqn:2.5}{(2.5)} for $r$. Finally, the boundary condition for $\varpi$ and $\partial_v\phi$ yield precisely conditions \hyperlink{eqn:2.19}{(2.19)} and \hyperlink{eqn:2.21}{(2.21)} postulated for the $(r,\Omega^2,Q)$ system. So we recover, as claimed, a solution to the $(r,\Omega^2,Q)$ system.
\\ \\ \\ \\
\Large\hypertarget{sec:3}{\textbf{3\hspace{4mm}Scattering results}}\label{3}\normalsize\\ \\
\large\hypertarget{sec:3.1}{\textbf{3.1\hspace{4mm}Examples with prescribed timelike bounce hypersurface}}\label{3.1}\normalsize\\ \\
We now proceed towards the proof of \hyperlink{thm:1.1}{Theorem 1.1}. Recall that a timelike curve segment $\gamma$ is prescribed in $\mathcal{M}_{RN[\varpi,Q]}$, which in \hyperlink{sec:2.3}{Section~2.3} we reparametrized (without loss of generality) by proper time. $\gamma$ is then associated with a smooth data set $r_0,(\partial_xr)_0,(\partial_x\log\Omega^2)_0$ recording the values of $r,\Omega^2$ and their derivatives at $\{u=v\}$, in a lapse-normalized double null chart. As null derivatives, we have
$$(\partial_vr)_0(t)=\tfrac{1}{2}r'_0(t)+\tfrac{1}{2}(\partial_xr)_0(t) \qquad (\partial_v\log\Omega^2)_0(t)=\tfrac{1}{2}(\partial_x\log\Omega^2)_0$$
When $\gamma$ is prescribed in the exterior region of $\mathcal{M}_{RN[\varpi,Q]}$, these transverse derivatives have explicit expressions \hyperlink{eqn:2.25}{(2.25)}-\hyperlink{eqn:2.26}{(2.26)} in terms of $r_0$. In view of \hyperlink{eqn:2.25}{(2.25)}-\hyperlink{eqn:2.26}{(2.26)}, $\Omega^2$ is naturally treated at one regularity level lower than $r$, as we will see in the following statements.\\ \\
First, we make use of the following result, allowing us to solve on a short interval and also characterizing termination. The (omitted) proof is standard and elementary: see for instance [\hyperlink{Eva10}{Eva10}, Section~12.2, Theorem~3].\newpage
\hypertarget{pro:3.1}{\textbf{Proposition 3.1 }}(Local well-posedness and blowup criterion for the $(r,\Omega^2,Q)$ system)\textbf{.} \\ \hspace*{4mm}\textit{Let $k\in\mathbb{N}_0$ be a non-negative integer.
\begin{itemize}
\item[(i)] Let data $r_0\in C^{k+2}[a,b]$, $(\log\Omega^2)_0,Q_0,(\partial_vr)_0\in C^{k+1}[a,b]$ and $(\partial_v\log\Omega^2)_0,(\partial_vQ)_0\in C^k[a,b]$ be given with $r_0>0$ on $[a,b]$. Then there exists $\delta>0$ and a unique solution $(r,\Omega^2,Q)$ on $\{0\leq x<\delta,\hspace{2pt}a\leq u,\hspace{2pt}b\leq v\}$ with $r\in C^{k+2}$, $\Omega^2,Q\in C^{k+1}$ to the problem
\begin{align*}
\partial_u\partial_vr&=-\frac{\Omega^2}{4r}-\frac{\partial_ur\partial_vr}{r}+\frac{\Omega^2Q^2}{4r^3}\tag{\hypertarget{eqn:3.1}{3.1}}\\[0.3em]
\partial_u\partial_v\log\Omega^2&=\frac{\Omega^2}{2r^2}+\frac{2\partial_ur\partial_vr}{r^2}-\frac{\Omega^2Q^2}{r^4}\tag{\hypertarget{eqn:3.2}{3.2}}\\[0.7em]
\partial_u\partial_vQ&=0\tag{\hypertarget{eqn:3.3}{3.3}}\\[0.4em]
r|_{u=v}=r_0;\hspace{7mm}&\hspace{8.2mm}\log\Omega^2|_{u=v}=(\log\Omega^2)_0;\hspace{11.7mm} Q|_{u=v}=Q_0;\\[0.2em]
\partial_vr|_{u=v}=(\partial_vr)_0;&\quad\partial_v\log\Omega^2|_{u=v}=(\partial_v\log\Omega^2)_0;\quad\partial_vQ|_{u=v}=(\partial_vQ)_0\hspace{1.5cm}
\end{align*}
Moreover, we can take data at $x=s\in(0,\tfrac{1}{2}(b-a))$ instead, and solve uniquely up to $x=s+\delta$.
\item[(ii)] Given a solution $(r,\Omega^2,Q)$ to the above on $\{0\leq x<s,\hspace{2pt}a\leq u,\hspace{2pt}b\leq v\}$ with $r\in C^{k+2}$, $\Omega^2$, $Q\in C^{k+1}$, suppose moreover that 
$$\sup_{0\leq x<s}r,\hspace{2pt}\sup_{0\leq x<s}r^{-1},\hspace{2pt}\sup_{0\leq x<s}|\hspace{-1pt}\log\Omega^2|,\hspace{2pt}\sup_{0\leq x<s}|\partial_ur|,\hspace{2pt}\sup_{0\leq x<s}|\partial_vr|<\infty $$
Then $r,\Omega^2,Q$ extend continuously, along with their derivatives, to $x=s$.\hfill$\square$
\end{itemize}}
We can of course solve explicitly for $Q$ on the whole characteristic triangle: it is the nonlinearities in the equations for $r,\Omega^2$ that may lead to breakdown. Though we are guaranteed existence over a short interval, this is only the first step. We may obtain a solution on the whole characteristic triangle by looking for perturbations of a suitable region of Reissner-Nordstr\"om, as the following proposition makes precise. In order to represent a perturbation, a smallness assumption on $(\partial_vQ)_0$ must be made. Without this assumption, we do not in general expect existence to hold on the full domain: for large $(\partial_vQ)_0$, anti-trapped surfaces may appear in evolution, as a result of which $r\to0$ may occur in the domain's interior. This appearance of anti-trapped spheres is not a contradiction to Raychaudhuri's equation, because the evolution is `sideways'.\\ \\
We are ready to begin the proof of \hyperlink{thm:1.1}{Theorem 1.1}.\\ \\
\textit{Proof of Theorem 1.1.} Our principal task is to construct region $(I)$, which we view as a perturbation of the Reissner-Nordstr\"om spacetime $\mathcal{M}_{RN[\varpi_0,Q_0]}$ on the characteristic triangle exterior to $\gamma$. As discussed in \hyperlink{sec:2.1}{Section 2.1}, let $(u,v)$ be double null coordinates with respect to which this triangle corresponds to $\triangleright:=\{a\leq u\leq v\leq b\} $ and such that $\Omega^2|_{u=v}=1$. In this coordinate choice, the lapse function, area radius and integrated charge of the unperturbed $\mathcal{M}_{RN[\varpi_0,Q_0]}$ we now denote as $\overline r(u,v)$, $\overline{\Omega^2}(u,v)$ and $Q_0$.
This triple solves the PDE system \hyperlink{eqn:3.1}{(3.1)}-\hyperlink{eqn:3.3}{(3.3)} with the choice of data
\begin{align*}
r|_{u=v}&=r_0; & \log\Omega^2|_{u=v}&=0; \hspace{3.4cm}Q|_{u=v}=Q_0; \\
\partial_vr|_{u=v}&=(\partial_vr)_0; & \partial_v\log\Omega^2|_{u=v}&= (\partial_v\log\Omega^2)_0; \hspace{1.3cm}\partial_vQ|_{u=v}=0 
\tag{\hypertarget{eqn:3.4}{3.4}}
\end{align*}
We set $$\delta=\inf_{a\leq u\leq v\leq b}\overline r>0\vspace{2mm}$$
\ul{Step 1. Setting up the Cauchy stability argument}. We now employ a standard Cauchy stability argument to obtain the solution $(r,\Omega^2,Q)$, with the help of \hyperlink{pro:3.1}{Proposition~3.1}. We do not need precise information about the right-hand sides of \hyperlink{eqn:3.1}{(3.1)}-\hyperlink{eqn:3.2}{(3.2)}, so we put for simplicity
$$\partial_u\partial_vr=f(r,\log\Omega^2,\partial_ur,\partial_vr,Q);\qquad \partial_u\partial_v\log\Omega^2=g(r,\log\Omega^2,\partial_ur,\partial_vr,Q);$$
for smooth functions $f,g$ defined on $\mathbb{R}_{>0}\times\mathbb{R}^4$. We adopt the notation $[\hspace{2pt}\cdot\hspace{2pt}]$ for the difference between variables in $(r,\Omega^2,Q)$ and their counterparts in $(\overline{r},\overline{\Omega^2},Q_0)$, so that for instance
$$[r](u,v)=r(u,v)-\overline{r}(u,v) $$
To avoid $r=0$, and to fix bounds for $f,g$, we consider only those $(r,\Omega^2,Q)$ which remain close to $(\overline r,\overline{\Omega^2},Q_0)$ in the sense that
\begin{equation*}|[r]|,\hspace{2pt}|[\partial_ur]|,\hspace{2pt}|[\partial_vr]|,\hspace{2pt}|[\log\Omega^2]|,\hspace{2pt}|[\partial_u\log\Omega^2]|,\hspace{2pt}|[\partial_v\log\Omega^2]|,\hspace{2pt}|[Q]|\leq \tfrac{1}{2}\delta \tag{\hypertarget{eqn:3.5}{3.5}}\end{equation*}
Under this condition, the functions $f,g$ are only ever evaluated in the compact subset
$$C:=\Big\{(r,\log\Omega^2,\partial_ur,\partial_vr,Q)\hspace{2pt}\Big|\hspace{2pt}\tfrac{1}{2}\delta\leq r\leq \|\overline r\|_{C^0}+\tfrac{1}{2}\delta,\hspace{2pt}|\hspace{-1pt}\log\Omega^2|\leq \|\hspace{-1pt}\log\overline{\Omega^2}\|_{C^0}+\tfrac{1}{2}\delta,\hspace{1.5cm}\vspace{-2mm}$$ $$\hspace{1.5cm}|\partial_ur|\leq\|\partial_u\overline r\|_{C^0}+\tfrac{1}{2}\delta,\hspace{2pt}|\partial_vr|\leq\|\partial_v\overline r\|_{C^0}+\tfrac{1}{2}\delta,\hspace{2pt}|Q|\leq Q_0+\tfrac{1}{2}\delta\Big\}\subset\mathbb{R}_{>0}\times\mathbb{R}^4 $$
We now write
$$\|\partial f\|_{C^0(C)}:=\sum^5_{i=1}\sup_C|\partial_if|,\qquad \|\partial g\|_{C^0(C)}:=\sum^5_{i=1}\sup_C|\partial_ig|$$
\ul{Step 2. Executing the bootstrap argument}. We make the following claim.\\ \\
\hspace*{0.02\textwidth}
\begin{minipage}{0.96\textwidth}\textsc{Claim.} \textit{Let $k\in\mathbb{N}_0$ be a non-negative integer. There exists $\varepsilon>0$, depending on the data $r_0,(\partial_vr)_0,(\partial_v\log\Omega^2)_0$ but \ul{not on $k$}, such that, if $\sigma(t)>0$ satisfies the bound
$$\int^b_a\sigma(t)dt\leq\varepsilon $$
then the unique solution $r\in C^{k+2}$, $\Omega^2,Q\in C^{k+1}$ to equations \hyperlink{eqn:3.1}{(3.1)}-\hyperlink{eqn:3.3}{(3.3)} and taking the data
\begin{align*}
r|_{u=v}&=r_0; & \log\Omega^2|_{u=v}&=0; \hspace{3.4cm}Q|_{u=v}=Q_0; \\
\partial_vr|_{u=v}&=(\partial_vr)_0; & \partial_v\log\Omega^2|_{u=v}&= (\partial_v\log\Omega^2)_0; \hspace{1.3cm}\partial_vQ|_{u=v}=\tfrac{1}{2}\sigma(t) 
\tag{\hypertarget{eqn:3.6}{3.6}}
\end{align*}
extends to the full domain $\triangleright$.} \\
\end{minipage}\\
Using \hyperlink{pro:3.1}{Proposition~3.1}(i), the solution $(r,\Omega^2,Q)$ to the problem \hyperlink{eqn:3.1}{(3.1)}-\hyperlink{eqn:3.3}{(3.3)} exists at least on $0\leq x< \delta$ say, and we aim to establish the \textsc{Claim} by comparing the solution with $(\overline r,\log\overline{\Omega^2},Q_0)$. To this end, for each $\varepsilon>0$ and $M>0$, we define
$$\mathcal{A}_{\varepsilon,M}:=\Big\{x_*\in[0,\tfrac{1}{2}(b-a)]:\hspace{2pt}(r,\Omega^2,Q)\text{ exists on }0\leq x\leq x_*, \text{ and }|[r]|,\hspace{2pt}|[\partial_ur]|,\hspace{2pt}|[\partial_vr]|,\hspace{2pt}$$
$$|[\log\Omega^2]|,\hspace{2pt}|[\partial_u\log\Omega^2]|,\hspace{2pt}|[\partial_v\log\Omega^2]|\leq 2\varepsilon e^{2Mx}\text{ and }|Q-Q_0|\leq\varepsilon\Big\} \subset[0,\tfrac{1}{2}(b-a)]$$
which encodes our bootstrap assumption, and we claim that, for some $\varepsilon,M$, we have $\mathcal{A}_{\varepsilon,M}=[0,\tfrac{1}{2}(b-a)]$. In fact, we choose immediately
$$M=1+2\|\partial f\|_{C^0(C)}+2\|\partial g\|_{C^0(C)},\qquad \varepsilon=\tfrac{1}{4}\delta e^{-M(b-a)}$$
This choice of $\varepsilon$ guarantees that condition \hyperlink{eqn:3.5}{(3.5)} always holds. Here our assumption on $\sigma$ emerges, because existence for $Q$ holds on the full domain already, and the bootstrap assumption on $Q$ is satisfied because
$$|Q(u,v)-Q_0|\leq \int^b_a|(\partial_vQ)_0(t)|dt $$
The above $\mathcal{A}_{\varepsilon,M}$ (in fact, any $\mathcal{A}_{\varepsilon,M}$) is non-empty because it contains 0. Since closed intervals are connected, it suffices to show that $\mathcal{A}_{\varepsilon,M}$ is both open and closed.\\ \\
\ul{$\mathcal{A}_{\varepsilon,M}$ is closed:} Suppose we have $x^{k}_*\in \mathcal{A}_{\varepsilon,M}$ for which $x^k_*\nearrow x_*\in[0,\tfrac{1}{2}(b-a)]$. Then $(r,\Omega^2,Q)$ exist uniquely on $0\leq x< x_*$. Since condition \hyperlink{eqn:3.5}{(3.5)} holds, $r$ is bounded away from zero, and $r,\log\Omega^2$ remain bounded, together with their first derivatives, on $0\leq x< x_*$. We can then apply \hyperlink{pro:3.1}{Proposition~3.1}(ii), which yields that $r$ (resp. $\Omega^2,Q$) extend, as $C^{k+2}$ (resp. $C^{k+1}$) functions, to $x= x_*$. Finally, the other bounds involved in the definition of $\mathcal{A}_{\varepsilon,M}$ continue to hold, by continuity, up to $x=x_*$. So $x_*\in\mathcal{A}_{\varepsilon,M}$.\\ \\
\ul{$\mathcal{A}_{\varepsilon,M}$ is open:} Suppose $x_*\in\mathcal{A}_{\varepsilon,M}$. We now show there is room in the estimate defining  $x_*\in\mathcal{A}_{\varepsilon,M}$ to evolve a little further. By applying the mean value inequality to equations \hyperlink{eqn:3.1}{(3.1)}-\hyperlink{eqn:3.2}{(3.2)}, and the definition of $\|\partial f\|_{C^0(C)}$, $\|\partial g\|_{C^0(C)}$, we have
$$|\partial_u\partial_v[r]|\leq \|\partial f\|_{C^0(C)}\cdot2\varepsilon e^{2Mx},\qquad |\partial_u\partial_v[\log\Omega^2]|\leq \|\partial g\|_{C^0(C)}\cdot2\varepsilon e^{2Mx},$$
Here we have used again that condition \hyperlink{eqn:3.5}{(3.5)} holds, so that we can indeed evaluate $f,g$ on $C$ only. Now since data for $r,\Omega^2$ at $x=0$ coincide for the two solutions, we may integrate from $x=0$ to obtain
$$|[\partial_ur](u,v)|,\hspace{2pt}|[\partial_vr](u,v)|\leq\bigg(\frac{2\|\partial f\|_{C^0(C)}}{M}\bigg)\varepsilon e^{2Mx} $$
$$|[\partial_u\log\Omega^2](u,v)|,\hspace{2pt}|[\partial_v\log\Omega^2](u,v)|\leq\bigg(\frac{2\|\partial g\|_{C^0(C)}}{M}\bigg)\varepsilon e^{2Mx} $$
Integrating again, we have
$$|[r](u,v)|\leq \bigg(\frac{2\|\partial f\|_{C^0(C)}}{M^2}\bigg)\varepsilon e^{2Mx}\qquad |[\hspace{1pt}\log\Omega^2](u,v)|\leq \bigg(\frac{2\|\partial g\|_{C^0(C)}}{M^2}\bigg)\varepsilon e^{2Mx}$$
It is easy to see that our earlier choice of $M$ makes every bracketed term at most 1. We may now apply \hyperlink{pro:3.1}{Proposition~3.1}(i), starting from $x=x_*$, to continue $r,\Omega^2,Q$ to $0\leq x\leq x_*+\delta$. By continuity, and after possibly shrinking $\delta>0$, the bootstrap assumption continues to hold on $0\leq x\leq x_*+\delta$. We conclude that $x_*+\delta\in\mathcal{A}_{\varepsilon,M}$, which completes the proof of the \textsc{Claim}. Applying the \textsc{Claim} for each $k\in\mathbb{N}_0$, we obtain a smooth solution $(r,\Omega^2,Q)$ on $\triangleright$.\\ \\
\ul{Step 3: Recovering the hydrodynamic interpretation}. We now use the positivity of $(\partial_vQ)_0=\tfrac{1}{2}\sigma(t)$ on $t\in(a,b)$, which plays an important role in recovering meaningful hydrodynamic variables in the interacting region, at least in the interior $$\mathring{\triangleright}=\{a<u<v<b\}$$
As discussed in \hyperlink{sec:2.3}{Section~2.3}, this requires of $(r,\Omega^2,Q)$ that 
$$\partial_uQ<0\qquad \partial_vQ>0\qquad \partial_u\bigg(\frac{\partial_ur}{\Omega^2}\bigg)<0\qquad\partial_v\bigg(\frac{\partial_vr}{\Omega^2}\bigg)<0 $$
hold throughout $\mathring{\triangleright}$. The former conditions follow immediately from
$$-(\partial_uQ)_0(t)=(\partial_vQ)_0(t)>0\qquad\text{ on }t\in(a,b) $$
and equation \hyperlink{eqn:2.18}{(2.18)}. (The equality here is due to $Q=Q_0$ on $u=v$.)\\ \\
For the latter conditions, we use equations \hyperlink{eqn:2.27}{(2.27)}-\hyperlink{eqn:2.28}{(2.28)}. Since
$$\partial_u\bigg(\frac{\partial_ur}{\Omega^2}\bigg)\bigg|_{u=v}=0\quad\text{ and }\quad\partial_uQ=0\quad\text{ on }\mathring{\triangleright} $$
integration in $v$ yields that
$$r\Omega^2\partial_u\bigg(\frac{\partial_ur}{\Omega^2}\bigg)<0\quad\text{ on }\mathring{\triangleright} $$
Since $r,\Omega^2$ are positive, the claim holds. The argument is identical for $\partial_v(\partial_vr/\Omega^2)$: note that we have to integrate \textit{backwards} in $u$. If we now \textit{define} $N^u_\text{in}$, $T^{uu}_\text{in}$, $N^v_\text{out}$, $T^{vv}_\text{out}$ through \hyperlink{eqn:2.7}{(2.7)}-\hyperlink{eqn:2.10}{(2.10)}, then each quantity is positive on $\mathring{\triangleright}$. We then recover \textit{bona fide} hydrodynamic variables $\rho_\text{out}$, $\rho_\text{in}$, $k^v_\text{out}$, $k^u_\text{in}$ by the expressions \hyperlink{eqn:2.15}{(2.15)}.\\ \\
\ul{Step 4. Constructing regions $(I\hspace{-2.5pt}I)$-$(V)$}. We have obtained region $(I)$, and we proceed to construct regions $(I\hspace{-2.5pt}I)$-$(V)$. To obtain a neighbourhood of the region defined on $\triangleright$, we can simply extend our data. Since $\gamma$ is a smooth timelike curve segment in $\mathcal{M}_{RN[\varpi,Q]}$, we can smoothly extend it to a longer timelike segment defined on $[a-\delta,b+\delta]$, and obtain as before the transverse derivatives $(\partial_vr)_0,(\partial_v\log\Omega^2)_0$. We then extend $(\partial_vQ)_0$ \textit{by zero} on this larger interval. The resulting function $(\partial_vQ)_0$ has regularity $C^{k-1}[a-\delta,b+\delta]$, where $k$ is the order of both zeroes of $(\partial_vQ)_0$ at $t=a,b$. \\ \\
We now solve \hyperlink{eqn:3.1}{(3.1)}-\hyperlink{eqn:3.3}{(3.3)} on the larger triangle $\{a-\delta\leq u\leq v\leq b+\delta\}$, with these data\footnote[13]{There is no further issue of existence here, since---as we will see---the new regions are Reissner-Nordstr\"om and Vaidya patches which can be extended to null infinity. However, the concerned reader may simply carry out the extension of $\gamma$ \textit{before} Steps 1 and 2, so the issue is subsumed into the previous Cauchy stability argument.}. We have the following picture, divided up according to the values of $\partial_uQ,\partial_vQ$.\\ \\
\begin{minipage}{0.67\textwidth}\hyperlink{lem:2.2}{Lemma~2.2} applies directly to region $(I\hspace{-2.5pt}I)$. Indeed, take 
$$\Sigma:=\{a-\delta< u< a\}\cap\{a-\delta< v< b\}\cap\{u< v\}$$
and $S=\overline{\Sigma}\cap\{u=v\}$. Then the hypotheses of \hyperlink{lem:2.2}{Lemma~2.2} apply and $(I\hspace{-2.5pt}I)$ is isometric to an ingoing Vaidya spacetime. Likewise, region $(I\hspace{-1.5pt}V)$ is isometric to an outgoing Vaidya spacetime. On regions $(I\hspace{-2.5pt}I\hspace{-2.5pt}I)$ and $(V)$, $Q$ is constant, and so we may infer by the charged Birkhoff's theorem that they are locally isometric to suitable Reissner-Nordstr\"om spacetimes.\\ \\
As for the claimed gluing regularity across the various interfaces, we simply use the fact that $Q\in C^{k}$. Since $r,\Omega^2$ have smooth data, and $Q$ appears (undifferentiated) on the right-hand side of \hyperlink{eqn:2.5}{(2.5)}-\hyperlink{eqn:2.6}{(2.6)}, we infer that $r,\Omega^2\in C^{k+1}$ as claimed.\vspace{1mm}
\end{minipage}
\hfill
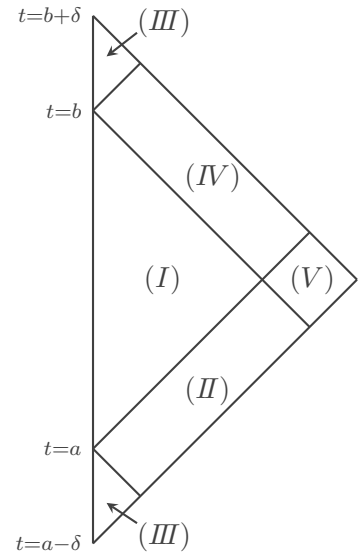
\begin{minipage}{0.33\textwidth}
\vspace{-4mm}
\begin{figure}[H]
\begin{center}
\begin{tikzpicture}[scale=0.9]
\draw [darkgray, thick] (0.0, -1.4) -- (0.0, 6.4);
\draw [darkgray, thick] (0, 0) -- (3.2,3.2);
\draw [darkgray, thick] (0, -1.4) -- (3.9,2.5);
\draw [darkgray, thick] (3.2,1.8) -- (0, 5.0);
\draw [darkgray, thick] (3.9,2.5) -- (0, 6.4);
\draw [darkgray, thick] (0,0) -- (0.7, -0.7);
\draw [darkgray, thick] (0,5) -- (0.7, 5.7);
\draw [darkgray, thick, -stealth] (0.65,6.15) -- (0.2, 5.8);
\draw [darkgray, thick, -stealth] (0.65,-1.1) -- (0.2, -0.8);
\node[darkgray,rotate=0] at (0.5, 6.3) [anchor = west] {$(I\hspace{-2.5pt}I\hspace{-2.5pt}I)$};
\node[darkgray,rotate=0] at (0.5, -1.3) [anchor = west] {$(I\hspace{-2.5pt}I\hspace{-2.5pt}I)$};
\node[darkgray, align=left] at (0,0) [anchor = east] {$_{t=a}$};
\node[darkgray, align=left] at (0,5) [anchor = east] {$_{t=b}$};
\node[darkgray, align=left] at (0,-1.4) [anchor = east] {$_{t=a-\delta}$};
\node[darkgray, align=left] at (0,6.4) [anchor = east] {$_{t=b+\delta}$};
\node[darkgray, align=left] at (1.2,4.0) [anchor = west] {$(I\hspace{-1.5pt}V)$};
\node[darkgray, align=left] at (1.2,0.85) [anchor = west] {$(I\hspace{-2.5pt}I)$};
\node[darkgray, align=left] at (2.75,2.5) [anchor = west] {$(V)$};
\node[darkgray, align=left] at (0.6,2.5) [anchor = west] {$(I)$};
\end{tikzpicture}
\end{center}
\end{figure}
\end{minipage}
\\
\ul{Step 5. Extending across $\mathcal{B}$}. Next, we solve \textit{to the left} for the Reissner-Nordstr\"om region on the other side of $\mathcal{B}$. This is obtained by solving the system \hyperlink{eqn:3.1}{(3.1)}-\hyperlink{eqn:3.3}{(3.3)} to the left ($u>v$) with the data \hyperlink{eqn:3.4}{(3.4)}. Analogously to \hyperlink{pro:3.1}{Proposition 3.1}, this is certainly possible on a small `slab', with the possibility of termination due to $r\to0$ in the Minkowski case ($\varpi_0=Q_0=0$). The data for $r,\Omega^2,Q$ are smooth, so the solution on the resulting slab smoothly glues together the existing regions labelled $(I\hspace{-2.5pt}I\hspace{-2.5pt}I)$ in the above Figure.\\ \\
Examining the regularity across $\mathcal{B}=\{u=v\}$, we see that, on each side of $x=0$, the variables $r,\Omega^2,Q$ are $C^\infty$ with extension to $x=0$, because, when we solve the wave equations \hyperlink{eqn:3.1}{(3.1)}-\hyperlink{eqn:3.2}{(3.2)} in either direction, we pose smooth data. Hence we just need to check to what order the limiting derivatives from each side match. We will use the easily verifiable fact that any continuous function $f:\mathbb{R}^2\to\mathbb{R}$ which belongs to $C^\infty(\{x\geq0\})$ and $C^\infty(\{x\leq0\})$ is locally Lipschitz. First, the derivative $\partial_xQ$ jumps discontinuously, but $\partial_tQ$ is continuous, so $Q,\partial_tQ\in C^{0,1}$ while $Q\notin C^1$. Meanwhile, $r,\Omega^2$ admit (the same) $C^1$ limits on $\mathcal{B}$, so are $C^1$. The right-hand sides of \hyperlink{eqn:3.1}{(3.1)}-\hyperlink{eqn:3.2}{(3.2)} are continuous across $\mathcal{B}$, so $\partial_u\partial_vr$, $\partial_u\partial_v\log\Omega^2$ are continuous across $\mathcal{B}$. Then, continuity of $\partial^2_ur,\partial^2_vr$ follows from our choice of data \hyperlink{eqn:2.24}{(2.24)}-\hyperlink{eqn:2.25}{(2.25)}, which was equivalent to conditions \hyperlink{eqn:2.20}{(2.20)}-\hyperlink{eqn:2.21}{(2.21)}. With $r\in C^2$, we may differentiate \hyperlink{eqn:3.2}{(3.2)} in $\partial_t$. The right-hand side, as a result of the above, is continuous across $\mathcal{B}$. It follows that $\partial_t\log\Omega^2$ is $C^1$ across $\mathcal{B}$. Together with continuity of $\partial_u\partial_v\log\Omega^2$, this is sufficient for $\Omega^2\in C^2$. By earlier comments, we then in fact have $r,\Omega^2\in C^{2,1}$. Note that $g_{\mu\nu}$ is \textit{not} $C^3$-regular, since, if $r\in C^3$, equation \hyperlink{eqn:2.5}{(2.5)} would imply that $Q\in C^1$, which would be a contradiction.\\ \\
Finally, we may extend regions $(I\hspace{-2.5pt}I)$-$(V)$ suitably to past and future null infinity, completing the global construction of $(\mathcal{M},g_{\mu\nu})$.\hfill$\square$
\\ \\ \\
\large\hypertarget{sec:3.2}{\textbf{3.2\hspace{4mm}Examples of termination at a null point}}\label{3.2}\normalsize\\ \\
We begin by recording the analogous result to \hyperlink{pro:3.1}{Proposition~3.1} above, but for the $(r,\kappa,Q)$ system. The boundary conditions \hyperlink{eqn:2.30}{(2.30)} obeyed by this system determine the derivatives $(\partial_vr)_0$ and $(\partial_u\kappa)_0$, which is why these particular transverse derivatives appear in the statements below. We specify $(\partial_uQ)_0$ rather than $(\partial_vQ)_0$, but this is arbitrary as the two quantities sum to zero, so in particular live in the same regularity class.\\ \\
\hypertarget{pro:3.2}{\textbf{Proposition 3.2 }}(Local well-posedness and blowup criterion for the $(r,\kappa,Q)$ system)\textbf{.} \\ \hspace*{4mm}\textit{Let $k\in\mathbb{N}_0$ be a non-negative integer.
\begin{itemize}
\item[(i)] Let data $r_0,\kappa_0,Q_0\in C^{k+1}[a,b]$ and $(\partial_vr)_0,(\partial_u\kappa)_0,(\partial_uQ)_0\in C^k[a,b]$ be given with $r_0,\kappa_0$ positive on $[a,b]$. Then there exists $\delta>0$ and a unique solution $(r,\kappa,Q)$ on $\{0\leq x<\delta,\hspace{2pt}a\leq u,\hspace{2pt}b\leq v\}$ with $r,\kappa,Q\in C^{k+1}$ to the problem
\begin{align*}
\partial_u\partial_vr&=\frac{\kappa\partial_ur}{r}-\frac{\partial_ur\partial_vr}{r}-\frac{\kappa\partial_ur Q^2}{r^3}\tag{\hypertarget{eqn:3.7}{3.7}}\\[0.3em]
\partial_u\partial_v\kappa&= \frac{\partial_u\kappa\partial_v\kappa}{\kappa}+\frac{2\kappa^2Q\partial_uQ}{r^3}-\frac{\kappa\partial_u\kappa}{r}+\frac{\kappa\partial_u\kappa Q^2}{r^3}\tag{\hypertarget{eqn:3.8}{3.8}}\\[0.7em]
\partial_u\partial_vQ&=0\tag{\hypertarget{eqn:3.9}{3.9}}\\[0.4em]
r|_{u=v}=r_0;\hspace{7mm}&\hspace{8.2mm}\kappa|_{u=v}=\kappa_0;\hspace{14.7mm} Q|_{u=v}=Q_0;\\[0.2em]
\partial_vr|_{u=v}=(\partial_vr)_0;&\quad\partial_u\kappa|_{u=v}=(\partial_u\kappa)_0;\quad\partial_uQ|_{u=v}=(\partial_uQ)_0\hspace{1.5cm}
\end{align*}
Moreover, we can take data at $x=s\in(0,\tfrac{1}{2}(b-a))$ instead, and solve uniquely up to $x=s+\delta$.
\item[(ii)] Given a solution $(r,\kappa,Q)$ to the above on $\{0\leq x<s,\hspace{2pt}a\leq u,\hspace{2pt}b\leq v\}$ with $r,\kappa,Q\in C^{k+1}$, suppose moreover that 
$$\sup_{0\leq x<s}r,\hspace{2pt}\sup_{0\leq x<s}r^{-1},\hspace{2pt}\sup_{0\leq x<s}|\partial_ur|,\hspace{2pt}\sup_{0\leq x<s}|\partial_vr|,\hspace{2pt}\sup_{0\leq x<s}\kappa,\hspace{2pt}\sup_{0\leq x<s}\kappa^{-1},\hspace{2pt}\sup_{0\leq x<s}|\partial_u\kappa|,\hspace{2pt}\sup_{0\leq x<s}|\partial_v\kappa|<\infty $$
Then $r,\kappa,Q$ extend continuously, along with their derivatives, to $x=s$.\hfill$\square$
\end{itemize}}
The proof of \hyperlink{thm:1.2}{Theorem~1.2} is similar to what we have just seen for \hyperlink{thm:1.1}{Theorem~1.1}, working now with the $(r,\kappa,Q)$ system. The only difference is that we impose $C^0$ (rather than $L^1$) bounds on $(\partial_uQ)_0$. This is because the derivative $\partial_uQ$ appears in the equation for $\kappa$, so we need $\partial_uQ$ also to be $C^0$-small. We are now ready for the proof of \hyperlink{thm:1.2}{Theorem~1.2}.\\ \\
\textit{Proof of Theorem 1.2.} Following \hyperlink{sec:2.4}{Section~2.4}, we take double null coordinates $(\tilde u,v)$ covering the characteristic triangle exterior to $\gamma$ in $\mathcal{M}_{RN[\varpi,Q]}$. Without loss of generality, the $\tau$ parametrizing $\gamma$ is this advanced time $v$, and so $v$ is valued in $[a,b]$, with $\gamma$ represented by $\{\tilde u=f(v)\}$. Implicitly defining $u$ through $\tilde u=f(u)$, we then have an improper coordinate chart in which the characteristic triangle exterior to $\gamma$ corresponds to $\triangleright=\{a\leq u\leq v\leq b\} $. The values of $\overline{r},\overline{\kappa}$ (that is, from the unperturbed $\mathcal{M}_{RN[\varpi,Q]}$) are unchanged by the reparametrization of $u$, so in particular the values of $r_0,\kappa_0$ (that is, evaluating $\overline{r},\overline\kappa$ at $u=v$) are strictly positive. We view the triple $(\overline{r},\overline{\kappa},Q_0)$ as the unique solution to \hyperlink{eqn:3.7}{(3.7)}-\hyperlink{eqn:3.9}{(3.9)} with the data
\begin{align*}
r|_{u=v}=r_0;\hspace{7mm}&\hspace{12.0mm}\kappa|_{u=v}=\kappa_0;\hspace{13mm} Q|_{u=v}=Q_0;\\[0.2em]
\partial_vr|_{u=v}=(\partial_vr)_0;&\hspace{8mm}\partial_u\kappa|_{u=v}=0;\hspace{1.2cm}\partial_uQ|_{u=v}=0\hspace{1.5cm}\tag{\hypertarget{eqn:3.10}{3.10}}
\end{align*}
where we recall from \hyperlink{eqn:2.30}{(2.30)} $$(\partial_vr)_0=\kappa_0\Big(1-\frac{2\varpi_0}{r_0}+\frac{Q^2_0}{r^2_0}\Big)$$
As in the proof of \hyperlink{thm:1.1}{Theorem~1.1}, we formulate our Cauchy stability argument as follows.\\ \\
\hspace*{0.02\textwidth}
\begin{minipage}{0.96\textwidth}\textsc{Claim.} \textit{Let $k\in\mathbb{N}_0$ be a non-negative integer. There exists $\varepsilon>0$, depending on the data $r_0,\kappa_0$ but \ul{not on $k$}, such that, if $(\partial_uQ)_0<0$ satisfies the bound
$$|(\partial_uQ)_0|\leq\varepsilon $$
then the unique solution $r,\kappa,Q\in C^{k+1}$ to equations \hyperlink{eqn:3.7}{(3.7)}-\hyperlink{eqn:3.9}{(3.9)} and taking the data
\begin{align*}
r|_{u=v}=r_0;\hspace{7mm}&\hspace{12.0mm}\kappa|_{u=v}=\kappa_0;\hspace{13mm} Q|_{u=v}=Q_0;\\[0.2em]
\partial_vr|_{u=v}=(\partial_vr)_0;&\hspace{8mm}\partial_u\kappa|_{u=v}=0;\hspace{1.2cm}\partial_uQ|_{u=v}=(\partial_uQ)_0\hspace{1.5cm}\tag{\hypertarget{eqn:3.11}{3.11}}
\end{align*}
extends to the full domain $\triangleright$.} \\
\end{minipage}\\
We omit the proof of this \textsc{Claim}, which is almost identical to the corresponding proof in \hyperlink{sec:3.1}{Section~3.1}, now using \hyperlink{pro:3.2}{Proposition~3.2}. With the \textsc{Claim} in hand, let $\varepsilon>0$ be the constant obtained. We now choose $(\partial_uQ)_0\in C^\infty[a,b]$ to be identically $-\varepsilon$ in a neighbourhood of $t=b$, vanish to infinite order at $t=a$, and satisfy $(\partial_uQ)_0(t)\in[-\varepsilon,0)$ on $t\in(a,b]$. The \textsc{Claim} yields a solution $(r,\kappa,Q)$ to the system \hyperlink{eqn:3.7}{(3.7)}-\hyperlink{eqn:3.9}{(3.9)} on $\triangleright$. Since we may apply the \textsc{Claim} with an arbitrary $k\in\mathbb{N}_0$, this solution is in fact $C^\infty$. In the region $\triangleleft:=\{a\leq v\leq u\leq b\}$, we retain the original Reissner-Nordstr\"om solution.\\ \\
\begin{minipage}{0.55\textwidth}\vspace{1mm}Transforming back into the original coordinates $(\tilde u,v)$, we have a proper coordinate system covering a characteristic rectangle. (In the opposite Figure, the small rectangles represent equally-sized coordinate squares in $(u,v)$ coordinates.) Following the same argument as in \hyperlink{sec:3.1}{Section~3.1}, we recover hydrodynamic variables for ingoing and outgoing charged dusts on the interior of $\triangleright$. This yields the desired region $(I)$. Extending further to the past can now be undertaken as in the proof of \hyperlink{thm:1.1}{Theorem~1.1}, with $C^\infty$ gluing between regions $(I)$ and $(I\hspace{-2.5pt}I)$, in virtue of $(\partial_uQ)_0$ vanishing to infinite order at $t=a$.\vspace*{0mm}\\
\end{minipage}
\hfill
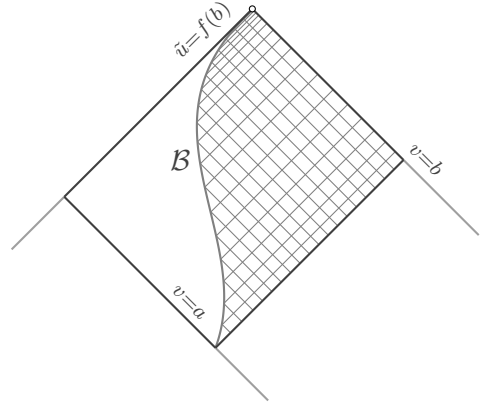
\begin{minipage}{0.45\textwidth}
\vspace{-14mm}
\begin{figure}[H]
\begin{center}
\begin{tikzpicture}
\draw [gray, thick, domain=0:4.5, samples=150] plot ({((\x)/4.5)*(0.5-(4.5-\x)+0.28*(4.5-\x)^2)},{\x});
\draw [gray] (0.2,0.2) -- (0.09,0.31);
\draw [gray] (0.35,0.35) -- (0.11,0.59);
\draw [gray] (0.5,0.5) -- (0.11,0.89);
\draw [gray] (0.625,0.625) -- (0.07,1.18);
\draw [gray] (0.75,0.75) -- (0.01,1.49);
\draw [gray] (0.875,0.875) -- (-0.06,1.81);
\draw [gray] (1,1) -- (-0.14,2.14);
\draw [gray] (1.125,1.125) -- (-0.2,2.45);
\draw [gray] (1.25,1.25) -- (-0.24,2.74);
\draw [gray] (1.375,1.375) -- (-0.25,3);
\draw [gray] (1.5,1.5) -- (-0.23,3.23);
\draw [gray] (1.625,1.625) -- (-0.19,3.44);
\draw [gray] (1.75,1.75) -- (-0.13,3.63);
\draw [gray] (1.875,1.875) -- (-0.05,3.8);
\draw [gray] (2,2) -- (0.04,3.96);
\draw [gray] (2.125,2.125) -- (0.14,4.11);
\draw [gray] (2.25,2.25) -- (0.25,4.25);
\draw [gray] (2.375,2.375) -- (0.37,4.38);
\draw [gray] (0.09,0.31) -- ({0.5*(5+0.09-0.31)},{0.5*(5-0.09+0.31)});
\draw [gray] ({0.5*(5+0.11-0.59)},{0.5*(5-0.11+0.59)}) -- (0.11,0.59);
\draw [gray] ({0.5*(5+0.11-0.89)},{0.5*(5-0.11+0.89)}) -- (0.11,0.89);
\draw [gray] ({0.5*(5+0.07-1.18)},{0.5*(5-0.07+1.18)}) -- (0.07,1.18);
\draw [gray] ({0.5*(5+0.01-1.49)},{0.5*(5-0.01+1.49)}) -- (0.01,1.49);
\draw [gray] ({0.5*(5-0.06-1.81)},{0.5*(5+0.06+1.81)}) -- (-0.06,1.81);
\draw [gray] ({0.5*(5-0.14-2.14)},{0.5*(5+0.14+2.14)}) -- (-0.14,2.14);
\draw [gray] ({0.5*(5-0.2-2.45)},{0.5*(5+0.2+2.45)}) -- (-0.2,2.45);
\draw [gray] ({0.5*(5-0.24-2.74)},{0.5*(5+0.24+2.74)}) -- (-0.24,2.74);
\draw [gray] ({0.5*(5-0.25-3)},{0.5*(5+0.25+3)}) -- (-0.25,3);
\draw [gray] ({0.5*(5-0.23-3.23)},{0.5*(5+0.23+3.23)}) -- (-0.23,3.23);
\draw [gray] ({0.5*(5-0.19-3.44)},{0.5*(5+0.19+3.44)}) -- (-0.19,3.44);
\draw [gray] ({0.5*(5-0.13-3.63)},{0.5*(5+0.13+3.63)}) -- (-0.13,3.63);
\draw [gray] ({0.5*(5-0.05-3.8)},{0.5*(5+0.05+3.8)}) -- (-0.05,3.8);
\draw [gray] ({0.5*(5+0.04-3.96)},{0.5*(5-0.04+3.96)}) -- (0.04,3.96);
\draw [gray] ({0.5*(5+0.14-4.11)},{0.5*(5-0.14+4.11)}) -- (0.14,4.11);
\draw [gray] ({0.5*(5+0.26-4.26)},{0.5*(5-0.26+4.26)}) -- (0.26,4.26);
\draw [darkgray, thick] (0,0) -- (-2.01, 2.01);
\draw [darkgray, thick] (0, 0) -- (2.5,2.5);
\draw [darkgray, thick] (2.5,2.5) -- (0.5, 4.5);
\draw [darkgray, thick, opacity=1] (-2.01,2.01) -- (0.49, 4.51);
\draw [darkgray, thick,opacity=0.5] (-2.01,2.01) -- (-2.71,1.31);
\draw [darkgray, thick,opacity=0.5] (0,0) -- (0.7,-0.7);
\draw [darkgray, thick,opacity=0.5] (2.5,2.5) -- (3.5,1.5);
\node[darkgray, align=left, rotate=-45] at (-0.05,0.3) [anchor = east] {$_{v=a}$};
\node[darkgray, align=left] at (-0.2,2.5) [anchor = east] {$\mathcal{B}$};
\node[darkgray, align=left, rotate=-45] at (2.53,2.8) [anchor = west] {$_{v=b}$};
\node[darkgray, align=left, rotate=45] at (-0.6,3.8) [anchor = west] {$_{\tilde u=f(b)}$};
\filldraw[color=black, fill=white](0.49,4.51) circle (0.04);
\end{tikzpicture}
\end{center}
\end{figure}
\end{minipage}\\
It remains to show that $dQ\neq0$ throughout region $(I\hspace{-2.5pt}I)$, and that the outgoing number current $N^v_\text{out}$ diverges at the null point. For the former, the condition that $(\partial_uQ)_0<0$ on $t\in(a,b]$ means that $(\partial_vQ)_0(t)>0$ on $t\in(a,b]$. By equation \hyperlink{eqn:2.18}{(2.18)}, this propagates backwards into every point of region $(I\hspace{-2.5pt}I)$. Finally, for the latter, we have 
$$(\partial_{\tilde u}Q)_0(v)=\frac{(\partial_uQ)_0(v)}{f'(v)} $$
so that, through \hyperlink{eqn:2.9}{(2.9)}, we have, in a neighbourhood of $v=b$,
$$(N^v_\text{out})_0(v)=\frac{2\varepsilon}{\mathfrak{e}r^2_0\tilde{\Omega}^2_0}\cdot\frac{1}{f'(v)}\to\infty\qquad\text{as }v\to b $$
This completes the proof.\hfill$\square$
\\ \\ \\
\Large\hypertarget{sec:4}{\textbf{4\hspace{4mm}Formation of a timelike bounce hypersurface}}\label{4}\normalsize\\ \\
\large\hypertarget{sec:4.1}{\textbf{4.1\hspace{4mm}Setup of the free boundary problem}}\label{4.1}\normalsize\\ \\
In contrast to \hyperlink{thm:1.1}{Theorems~1.1}-\hyperlink{thm:1.2}{1.2}, we will now prove a \textit{local} result, obtaining a timelike bounce $\mathcal{B}$ that persists for an $\varepsilon$ of advanced time. Since $\varepsilon$ will be determined during the proof, we will use the notation $\triangleright_\varepsilon=\{0\leq u\leq v\leq \varepsilon\}$ to keep in mind this as-yet-undetermined quantity. \\ \\
Recall that the starting point of \hyperlink{thm:1.3}{Theorem~1.3} is to receive hard-edge timelike Vaidya seed data---that is, functions $\varpi(v),Q(v)$ possessing the properties in the theorem statement (and parametrized with the Vaidya $v$ coordinate). It is not immediately clear that the set of such data is non-empty. \\ \\
\hypertarget{ex:4.1}{\textbf{Example 4.1.}} \textit{Choose any $\varpi_0\geq Q_0>0$ and $R>\varpi_0+\sqrt{\varpi^2_0-Q^2_0}$. Then $$\varpi(v)=\begin{cases}\varpi_0+v\\\varpi\end{cases}\qquad  Q(v)=\begin{cases}\sqrt{Q^2_0+2Rv}\\ Q_0\end{cases}\qquad\text{for }v\hspace{4pt}\begin{cases}\geq0\\ <0\end{cases}$$
are hard-edge timelike Vaidya seed data with, in fact, $r_b(v)=R$ on $v\geq0$. By choosing $\varpi_0,Q_0$ suitably, and truncating the beam after a short interval in $v$, the extremality ratio $Q/\varpi$ can be kept as small as desired. Moreover, one obtains, through \hyperlink{eqn:1.14}{(1.14)}, expressions for $k,\rho$ that are non-zero and only become unbounded in the expected manner as the bounce is approached.} \\ \\
Alternatively, any example obtained from \hyperlink{thm:1.1}{Theorem~1.1}, with $\gamma(0)$ belonging to the exterior and $\sigma(\tau)$ not vanishing at the lower endpoint $\tau=a$, gives rise to $\varpi(v),Q(v)$ with the stated properties, at least \textit{locally} near $v=0$ (and after recovering, of course, the Vaidya normalization).\\ \\
Recall from \hyperlink{sec:2.5}{Section~2.5} how, given such seed data $\varpi(v),Q(v)$, we obtain characteristic data $r_+(v)$, $\phi'_+(v),Q_+(v)$ for the $(r,\varpi,\phi,Q)$ system. After some algebra, we can use \hyperlink{eqn:2.32}{(2.32)}-\hyperlink{eqn:2.33}{(2.33)} to directly express
$$\phi'_+(v)=\frac{\varpi'(v)}{r^2_+(v)}\cdot\frac{\big(r_+(v)-r_b(v)\big)}{r'_+(v)} $$
from which we deduce that
$$\phi'_+(0)=0\qquad\text{and}\qquad \phi''_+(0)=\frac{\varpi'(0)}{r^2_+(0)}\bigg(1-\frac{r'_b(0)}{r'_+(0)}\bigg) $$
It is here that we use the hard-edge condition: because the factor $\varpi'(0)$ in the numerator is positive, so is $\phi''_+(0)$\footnote[14]{Note also that the factor in large brackets is positive, and would vanish if $\{r=r_b(v)\}$ began with an (outgoing) null point. As we shall see, the $\varepsilon$ of existence depends on lower bounds on $\phi''_+(0)$, and hence depends quantitatively on how close $\{r=r_b(v)\}$ is to having a null point at $v=0$.}. By continuity (coming from $C^2$ differentiability of the seed data), $\phi''_+(v)$ will be positive near $v>0$. Thus, by truncating the characteristic data, we may assume that both $r'_+$ and $\phi''_+$ are bounded away from zero. The next definition formalizes such characteristic data, taking them to be defined on $[0,1]$. This is only for convenience, and is without loss of generality: admissible data $(r_+,\phi'_+,Q_+)$ as defined below, but only on $v\in[0,a]$ ($a<1$), may be easily extended to $[0,1]$ and retain the defining properties.\\ \\
\hypertarget{def:4.1}{\textbf{Definition 4.1 }}(Admissible data)\textbf{.} \textit{Let $r_+\in C^2[0,1]$, $\phi'_+\in C^1[0,1]$ and $Q_+\in C^2[0,1]$. We call the collection $(r_+,\phi'_+,Q_+)$ \ul{admissible data} if the following properties hold:
\begin{itemize}
\item[(i)] $r_+(0)>0$, and for all $v\in[0,1]$, $r'_+(v)>0$
\item[(ii)] $\phi'_+(0)=0$ and there exists $\delta>0$ such that, for all $v\in[0,1]$, $\phi''_+(v)\geq\delta$.
\item[(iii)] $Q_+(0)>0$, and for all $v\in[0,1]$, $Q'_+(v)>0$.
\end{itemize}}
Given admissible data $(r_+,\phi'_+,Q_+)$, one may recover $\varpi_0$ simply by $$\varpi_0=\frac{r_+(0)}{2}\bigg(1-2r'_+(0)+\frac{Q^2_0}{r^2_+(0)}\bigg)$$
If this comes from the seed data $\varpi(v),Q(v)$ of \hyperlink{thm:1.3}{Theorem~1.3}, then this $\varpi_0$ will be positive, but we will not use positivity of $\varpi_0$ in the sequel. It is, however, essential that $Q_0>0$, to obtain below the positive lower bound \hyperlink{eqn:4.12}{(4.12)}---see again remarks after \hyperlink{thm:1.3}{Theorem~1.3}.\\ \\
\hypertarget{pro:4.1}{\textbf{Proposition 4.1}} (Solution to $(r,\varpi,\phi,Q)$ system)\textbf{.} \textit{Let $(r_+,\phi'_+,Q_+)$ be admissible data, with $\delta=\inf\phi''_+>0$. Then there exists $\varepsilon>0$, depending only on $\|r_+\|_{C^2[0,1]}$, $\|\phi'_+\|_{C^1[0,1]}$, $\|Q_+\|_{C^2[0,1]}$, $\varpi_0,Q_0,r_+(0),r'_+(0)$ and $\delta$, and a unique solution $r,\varpi\in C^1(\triangleright_\varepsilon)$, $\phi\in C^1(\triangleright_\varepsilon)\cap C^2_v(\triangleright_\varepsilon)$ on the domain $\triangleright_\varepsilon=\{0\leq u\leq v\leq \varepsilon\}$ to the PDE problem
\begin{align*}
\partial_u r&=-e^{\phi}\bigg(1-\frac{2\varpi}{r}+\frac{Q^2}{r^2}\bigg)&r\big|_{u=0}=r_+\\
\partial_v \varpi&=\tfrac{1}{2}r\partial_v\phi\bigg(1-\frac{2\varpi}{r}+\frac{Q^2}{r^2}\bigg)+\frac{Q\partial_vQ}{r}&\varpi\big|_{u=v}=\varpi_0\\
\partial_u\partial_v\phi&=e^{\phi}\bigg(-\frac{2Q\partial_vQ}{r^3}+\frac{\partial_v\phi}{r}-\frac{Q^2\partial_v\phi}{r^3}\bigg)&\partial_v\phi\big|_{u=v}=0;\quad\partial_v\phi\big|_{u=0}=\phi'\\[0.4em]
\partial_u\partial_vQ&=0&Q\big|_{u=v}=Q_0;\quad Q\big|_{u=0}=Q_+
\end{align*}}
\hspace{-4.8pt}We defer the proof of this to \hyperlink{sec:4.5}{Section~4.5}, after a number of lemmas have been established.\\ \\ \\
\large\hypertarget{sec:4.2}{\textbf{4.2\hspace{4mm}Description of the iteration scheme}}\label{4.2}\normalsize\\ \\
As in the previous chapters, solving for $Q$ is immediate---see equation \hyperlink{eqn:2.34}{(2.34)}. As such, we regard $Q$ as a fixed function, with $r,\varpi,\phi$ to be determined by advancing an iteration scheme which we now outline. With an iterate $(r,\varpi,\phi)$ in hand, we obtain the next iterate $(\overline{r},\overline{\varpi},\overline{\phi})$ by first solving
\begin{align*}
\partial_u \overline{r}&=-e^{\phi}\bigg(1-\frac{2\varpi}{r}+\frac{Q^2}{r^2}\bigg)\hspace{5cm}\overline{r}\big|_{u=0}=r_+\tag{\hypertarget{eqn:4.1}{4.1}}
\end{align*}
and, with $\overline{r}$ in hand, we then solve
\begin{equation*}
\partial_v \overline{\varpi}=\tfrac{1}{2}r\partial_v\phi\bigg(1-\frac{2\varpi}{r}+\frac{Q^2}{r^2}\bigg)+\frac{Q\partial_vQ}{\overline{r}}\hspace{2.8cm}\overline{\varpi}\big|_{u=v}=\varpi_0\tag{\hypertarget{eqn:4.2}{4.2}}
\end{equation*}
\begin{equation*}\partial_u\partial_v\overline{\phi}=e^{\overline{\phi}}f(\overline{r},\partial_v\phi,Q,\partial_vQ)\hspace{2.0cm}\partial_v\overline{\phi}\big|_{u=v}=0\qquad\partial_v\overline{\phi}\big|_{u=0}=\phi'_+ \hspace{0.2cm}\tag{\hypertarget{eqn:4.3}{4.3}}\end{equation*}
where 
$$f(r,\partial_v\phi,Q,\partial_vQ)=-\frac{2Q\partial_vQ}{r^3}+\frac{\partial_v\phi}{r}-\frac{Q^2\partial_v\phi}{r^3} $$
Obtaining $\overline{r},\overline{\varpi}$ is immediate---indeed, each can be written explicitly as an integral in terms of the previous iterates. On the other hand, solving equation \hyperlink{eqn:4.3}{(4.3)} for $\overline{\phi}$ is a sensitive business, since the values of $\partial_v\overline\phi$ are specified at both boundaries $\{u=0\}$ and $\{u=v\}$, even while the equation itself describes how $\partial_v\overline{\phi}$ is propagated in the $\partial_u$ direction. It is necessary that $\overline{\phi}$ also appears on the right-hand side of \hyperlink{eqn:4.3}{(4.3)}. This equation is studied carefully in \hyperlink{lem:4.1}{Lemmas~4.1} and \hyperlink{lem:4.2}{4.2}. The domain needs to be carefully chosen (i$.$e$.$ $\varepsilon$) to ensure the scheme is well-defined. Note also the presence of $\overline{r}$ terms in \hyperlink{eqn:4.6}{(4.6)}-\hyperlink{eqn:4.7}{(4.7)}, which is why we must first obtain $\overline{r}$ before $\overline{\varpi},\overline{\phi}$. These terms are necessary for the contraction estimates in \hyperlink{sec:4.4}{Section~4.4}.\\ \\
Another property of the above system \hyperlink{eqn:4.1}{(4.1)}-\hyperlink{eqn:4.3}{(4.3)} is that no $\partial_u\phi$ term appears on any right-hand side. Together with the way data are imposed for $\phi$, this means that $\phi$ is \textit{more regular} in the $\partial_v$ direction, in the sense that, at the lowest level of regularity, we have control of $\|\partial^2_v\phi\|_{C^0}$ but not $\|\partial_u\phi\|_{C^0}$. This motivates the definition of a function space controlling derivatives in the $\partial_v$ direction only.\newpage
\textbf{Definition 4.2} ($C^k_v$ spaces)\textbf{.} \textit{Let $\varepsilon>0$ and $k\in\mathbb{N}_0$. We write $C^k_v(\triangleright_\varepsilon)$ for the vector space of continuous functions $f:\triangleright_\varepsilon\to\mathbb{R}$ admitting continuous derivatives $\partial^i_vf$ up to order $i=k$. $C^k_v(\triangleright_\varepsilon)$ is a complete vector space under the norm}
$$\|f\|_{C^k_v(\triangleright_\varepsilon)}:=\sum^k_{i=0}\|\partial^i_vf\|_{C^0} $$
We now define the function space which houses our iterates.\\ \\
\hypertarget{def:4.3}{\textbf{Definition 4.3.}} \textit{Let $\varepsilon,C>0$ be positive constants, and let $(r_+,\phi'_+,Q_+,\varpi_0)$ be admissible data, with $\delta=\inf\phi''_+>0$. Then $\mathcal{A}(C,\varepsilon)$ denotes the set of collections $(r,\varpi,\phi)$ with $r,\varpi\in C^1_v(\triangleright_\varepsilon)$, $\phi\in C^2_v(\triangleright_\varepsilon)$, satisfying 
\begin{itemize}
\item[(i)] (upper and lower bounds)
\begin{equation*}
\|r\|_{C^1_v(\triangleright_\varepsilon)}\leq 2\|r_+\|_{C^2[0,1]}\tag{\hypertarget{eqn:4.4}{4.4}}
\end{equation*}
\begin{equation*}
\hspace{3.2cm}\|\varpi\|_{C^1_v(\triangleright_\varepsilon)}\leq 1+2|\varpi_0|+4r_+(0)^{-1}\|Q_+\|^2_{C^1[0,1]}\tag{\hypertarget{eqn:4.5}{4.5}}
\end{equation*}
\begin{equation*}
\|\partial_v\phi\|_{C^1_v(\triangleright_\varepsilon)}\leq 3\|\phi'_+\|_{C^1[0,1]}\tag{\hypertarget{eqn:4.6}{4.6}}
\end{equation*}
\begin{equation*}
r\geq \tfrac{1}{2}r_+(0)\tag{\hypertarget{eqn:4.7}{4.7}}
\end{equation*}
\begin{equation*}
\|\phi\|_{C^0(\triangleright_\varepsilon)}\leq C \tag{\hypertarget{eqn:4.8}{4.8}}
\end{equation*}
\item[(ii)] (boundary and characteristic data)
$$r\big|_{u=0}=r_+,\qquad \varpi\big|_{u=v}=\varpi_0,\qquad \partial_v\phi\big|_{u=v}=0,\qquad \partial_v\phi\big|_{u=0}=\phi'_+ $$
\end{itemize}}
\large\hypertarget{sec:4.3}{\textbf{4.3\hspace{4mm}Existence of the iteration scheme}}\label{4.3}\normalsize\\ \\
We begin by addressing equation \hyperlink{eqn:4.3}{(4.3)}.\\ \\
\hypertarget{lem:4.1}{\textbf{Lemma 4.1}} (Local existence and uniqueness for $\phi$)\textbf{.} \textit{Let $f\in C^1_v(\triangleright)$ satisfy $f<0$ on} \Large$\triangleright$\normalsize \textit{ with $|f|\geq\delta_1>0$, and let $\xi_+\in C^1[0,1]$ satisfy $\xi_+(0)=0$, and $\xi'_+>0$ on $[0,1]$ with $\xi'_+\geq\delta_2>0$. Then there exists $\varepsilon=\varepsilon(\|\partial_vf\|_{C^0(\triangleright)},\|\xi_+\|_{C^1[0,1]},\delta_1,\delta_2)>0$ and a unique $C^1\cap C^2_v$ solution $\phi:\triangleright_\varepsilon\to\mathbb{R}$ to the PDE problem
\begin{align*}&\partial_u\partial_v\phi=e^\phi f\tag{\hypertarget{eqn:4.9}{4.9}}\\ &\partial_v\phi|_{u=v}=0;\qquad \partial_v\phi|_{u=0}=\xi_+ 
\end{align*}
and for which $\phi$ satisfies the estimates
$$\|\phi\|_{C^0(\triangleright_\varepsilon)}\leq C(\|f\|_{C^0(\triangleright)},\|\xi_+\|_{C^1[0,1]},\delta_1,\delta_2) \qquad \qquad\|\partial_v\phi\|_{C^1_v(\triangleright_\varepsilon)}\leq 3\|\xi_+\|_{C^1[0,1]}$$
where $C>0$ depends only on the indicated quantities.}\\ \\
We will prove this lemma by differentiating \hyperlink{eqn:4.9}{(4.9)} in $\partial_v$ and studying the resulting wave equation satisfied by $\partial_v\phi$. Renaming this quantity $\xi$ (hence the renaming of the characteristic data), we first prove a self-contained lemma about $\xi$ before returning to prove \hyperlink{lem:4.1}{Lemma~4.1} itself. \newpage 
\hypertarget{lem:4.2}{\textbf{Lemma 4.2.}} \textit{Let $g\in C(\triangleright)$ and let $\xi_+\in C^1[0,1]$ satisfy $\xi_+(0)=0$. Then:
\begin{itemize}
\item[(i)] There exists $\varepsilon=\varepsilon(\|g\|_{C^0(\triangleright)},\|\xi_+\|_{C^1[0,1]})>0$ and a unique $C^1$ solution $\xi:\triangleright_\varepsilon\to\mathbb{R}$ to the PDE problem\vspace{-3mm}
\begin{align*}&\partial_u\partial_v\xi=\partial_u\xi(\xi+g)\tag{\hypertarget{eqn:4.10}{4.10}}\\ &\xi|_{u=v}=0;\qquad \xi|_{u=0}=\xi_+ 
\end{align*}
and for which $\xi$ satisfies the estimate
$$\|\xi\|_{C^1(\triangleright_\varepsilon)}\leq 3\|\xi_+\|_{C^1[0,1]} $$
\item[(ii)] If furthermore $\xi'_+>0$ on $[0,1]$ with $\xi'_+\geq \delta>0$, then on some possibly smaller $\varepsilon=\varepsilon(\|g\|_{C^0(\triangleright)},\|\xi_+\|_{C^1[0,1]},\delta)>0$, the solution obtained above has $\partial_u\xi<0$ on $\triangleright_\varepsilon$, with 
$$|\partial_u\xi|\geq\tfrac{1}{2}\delta $$
\end{itemize}}
\textit{Proof of Lemma 4.2.} \ul{(i):} Let $\mathcal{C}\subset C^1(\triangleright_{\varepsilon})$ be the closed ball of radius $3\|\xi_+\|_{C^1[0,1]}$ in $C^1(\triangleright_\varepsilon)$ (with the usual norm $\|\cdot\|_{C^1(\triangleright_\varepsilon)}=\|\cdot\|_{C^0(\triangleright_\varepsilon)}+\|\partial_u(\cdot)\|_{C^0(\triangleright_\varepsilon)}+\|\partial_v(\cdot)\|_{C^0(\triangleright_\varepsilon)}$). Consider the map $\xi\mapsto\overline{\xi}$ defined on $\mathcal{C}$ and which acts as 
\begin{equation*}
\overline{\xi}(u,v):=\xi_+(v)-\xi_+(u)+\int^u_0\int^v_u\partial_u\xi(\xi+g)d\tilde{v}d\tilde{u}\tag{\hypertarget{eqn:4.11}{4.11}}
\end{equation*}
\begin{minipage}{0.67\textwidth}\vspace{0.8mm}(This definition is equivalent to replacing $\xi$ with $\overline{\xi}$ on the left-hand side of \hyperlink{eqn:4.10}{(4.10)}.) The integration region is the diamond depicted in the opposite Figure. Clearly $\overline{\xi}\in C^1(\triangleright_\varepsilon)$, and we claim that $\xi\mapsto\overline{\xi}$ actually defines a contractive self-map $\mathcal{C}\to\mathcal{C}$, for $\varepsilon$ sufficiently small.\vspace{4mm}\\
By a straightforward calculation, one has, for $\xi\in\mathcal{C}$:
\end{minipage}
\hspace{0.05\textwidth}
\begin{minipage}{0.25\textwidth}
\vspace{-6mm}
\begin{figure}[H]
\begin{center}
\begin{tikzpicture}
\draw [darkgray, thick] (0,0) -- (0,3.5);
\draw [darkgray, thick] (0,0) -- (2.5,2.5);
\node[darkgray, align=left] at (1.2,2.8) [anchor = south] {$_{(u,v)}$};
\node[darkgray, align=left] at (1.9,2) [anchor = north west] {$_{(0,v)}$};
\node[darkgray, align=left] at (0.7,0.8) [anchor = north west] {$_{(0,u)}$};
\node[darkgray, align=left] at (0,1.6) [anchor = east] {$_{(u,u)}$};
\node[darkgray, align=left] at (1.5,1.8) [anchor = east] {$_{\int^u_0\hspace{-2pt}\int^v_u}$};
\fill [fill=gray, draw=none,opacity=0.1] (1.2,2.8) -- (2,2) -- (0.8,0.8) -- (0,1.6) -- cycle;
\filldraw[color=black, fill=black](1.2,2.8) circle (0.04);
\filldraw[color=black, fill=black](2,2) circle (0.03);
\filldraw[color=black, fill=black](0,1.6) circle (0.03);
\filldraw[color=black, fill=black](0.8,0.8) circle (0.03);
\end{tikzpicture}
\end{center}
\end{figure}
\end{minipage}
\begin{align*}
\|\overline{\xi}\|_{C^1(\triangleright_\varepsilon)}&\leq\|\xi'\|_{C^0[0,1]}(2+\varepsilon)+\varepsilon(2+\varepsilon)\|\xi\|_{C^1(\triangleright_\varepsilon)}(\|\xi\|_{C^1(\triangleright_\varepsilon)}+\|g\|_{C^0(\triangleright)})\\
&\leq \|\xi_+\|_{C^1[0,1]}(2+\varepsilon)\big(1+3\varepsilon(3\|\xi_+\|_{C^1[0,1]}+\|g\|_{C^0(\triangleright)})\big)
\end{align*}
and, for $\xi_1,\xi_2\in\mathcal{C}$:
$$\|\overline{\xi}_1-\overline{\xi}_2\|_{C^1(\triangleright_\varepsilon)}\leq 3\varepsilon(3\|\xi_+\|_{C^1[0,1]}+\|g\|_{C^0(\triangleright)})\|\xi_1-\xi_2\|_{C^1(\triangleright_\varepsilon)}$$
so one checks that, for instance, 
$$\varepsilon\leq \big(4+9(3\|\xi_+\|_{C^1[0,1]}+\|g\|_{C^0(\triangleright)})\big)^{-1}$$
works. By the contraction mapping principle, there exists a unique fixed point $\xi\in\mathcal{C}$. If $\xi$ satisfies the integral equation \hyperlink{eqn:4.11}{(4.11)} with $\overline{\xi}=\xi$, then $\xi$ admits the derivative $\partial_u\partial_v\xi$ and satisfies the desired equation \hyperlink{eqn:4.10}{(4.10)}, so we have obtained the claimed solution $\xi$.\\ \\
\ul{(ii):} The boundary condition $\xi|_{u=v}=0$ means that
$$\partial_u\xi|_{u=v}=-\partial_v\xi|_{u=v} $$
and so we have the identity
\begin{align*}
\partial_u\xi(u,v)&=\partial_u\xi(u,u)+\int^v_u\partial_u\partial_v\xi(u,\tilde{v})d\tilde{v}\\
&=-\xi'_+(u)-\int^u_0\partial_u\partial_v\xi(\tilde{u},u)d\tilde{u}+\int^v_u\partial_u\partial_v\xi(u,\tilde{v})d\tilde{v}
\end{align*}
We now estimate
$$-\partial_u\xi(u,v)\geq\delta-\varepsilon\|\partial_u\partial_v\xi\|_{C^0(\triangleright_\varepsilon)}$$
and so it suffices to choose $\varepsilon$ small enough so that the latter term is at most $\tfrac{1}{2}\delta$. To this end, by the bound established in part (i), we have
$$\|\partial_u\partial_v\xi\|_{C^0(\triangleright_\varepsilon)}\leq 3\|\xi_+\|_{C^1[0,1]}\big(3\|\xi_+\|_{C^1[0,1]}+\|g\|_{C^0(\triangleright)}\big) $$
and so reducing $\varepsilon$, if needed, to satisfy 
$$\varepsilon\leq\tfrac{1}{6}\delta\|\xi_+\|^{-1}_{C^1[0,1]}\big(3\|\xi_+\|_{C^1[0,1]}+\|g\|_{C^0(\triangleright)}\big)^{-1} $$
suffices, completing the proof.\hfill$\square$
\\ \\
\textit{Proof of Lemma 4.1.} Differentiating \hyperlink{eqn:4.9}{(4.9)} with $\partial_v$ reveals that $\xi:=\partial_v\phi$ obeys precisely equation \hyperlink{eqn:4.10}{(4.10)}, with the given conditions and with the choice $g:=\partial_vf/f$. Our strategy is to use \hyperlink{lem:4.2}{Lemma~4.2} to obtain $\xi$ (uniquely). We then recover $\phi$ from equation \hyperlink{eqn:4.9}{(4.9)} and check everything works.\\ \\
More specifically, we insert $g=\partial_vf/f$ into \hyperlink{lem:4.2}{Lemma~4.2}(ii), obtaining $\varepsilon=\varepsilon(\|\partial_vf\|_{C^0(\triangleright)},\|\xi_+\|_{C^1[0,1]},\delta_1,\delta_2)$ $>0$ and a solution $\xi$ to \vspace{-4mm}
\begin{align*}&\partial_u\partial_v\xi=\partial_u\xi\Big(\xi+\frac{\partial_vf}{f}\Big)\\ &\xi|_{u=v}=0;\qquad \xi|_{u=0}=\xi_+ 
\end{align*}
on $\triangleright_\varepsilon$ and satisfying\vspace{-2mm}
$$\|\xi\|_{C^1(\triangleright_\varepsilon)}\leq 3\|\xi_+\|_{C^1[0,1]}\qquad\text{ and }\qquad-\partial_u\xi\geq\tfrac{1}{2}\delta_2 $$
We now set $\phi:=\log|\partial_u\xi|-\log|f|$, which is well-defined and belongs to $C^1(\triangleright_\varepsilon)$. It is easy to recover $\partial_v\phi=\xi$: we compute
$$\partial_v\phi =\frac{\partial_u\partial_v\xi}{\partial_u\xi}-\frac{\partial_vf}{f}=\xi$$
and in particular $\partial_v\phi\in C^1(\triangleright_\varepsilon)$, so $\phi\in C^2_v(\triangleright_\varepsilon)$. Equation \hyperlink{eqn:4.3}{(4.3)}, together with its boundary conditions, is immediately verified, which also implies the existence and continuity of $\partial_u\phi$, so that $\phi\in C^1(\triangleright_\varepsilon)$ as well. The bound on $\|\partial_v\phi\|_{C^1(\triangleright_\varepsilon)}$ comes from the conclusion of \hyperlink{lem:4.1}{Lemma~4.1}, so it remains only to estimate $\|\phi\|_{C^0(\triangleright_\varepsilon)}$. The logarithms require us to use both lower and upper bounds for $|\partial_u\xi|$ and $|f|$, so we write
$$\|\phi\|_{C^0(\triangleright_\varepsilon)}\leq |\hspace{-1pt}\log(3\|\xi_+\|_{C^1[0,1]})|+|\hspace{-1pt}\log(\tfrac{1}{2}\delta_2)|+|\hspace{-1pt}\log\|f\|_{C^0(\triangleright)}|+|\hspace{-1pt}\log\delta_1| $$
and this defines our $C(\|f\|_{C^0(\triangleright)},\|\xi_+\|_{C^1[0,1]},\delta_1,\delta_2)$.\hfill$\square$
\newpage
We are now ready to show that the iteration scheme is well-defined as a self-map on $\mathcal{A}(C,\varepsilon)$ for suitably chosen $C,\varepsilon$. In the following lemma, we do not prove the \textit{non-emptiness} of such a $\mathcal{A}(C,\varepsilon)$.\\ \\
\hypertarget{lem:4.3}{\textbf{Lemma 4.3.}} \textit{There exist positive constants $C,\varepsilon>0$, depending only on $\|r_+\|_{C^2}$, $\|\phi'_+\|_{C^1}$, $\|Q_+\|_{C^2}$, $\varpi_0,Q_0,\delta,r_+(0)$ and $r'_+(0)$, such that the iteration scheme described above is a well-defined self-map $$\Psi:\mathcal{A}(C,\varepsilon)\to \mathcal{A}(C,\varepsilon)$$
and for the intermediate collection $(\overline{r},\overline{\varpi},\phi)$, we have the lower bound
\begin{equation*}-f(\overline{r},\partial_v\phi,Q,\partial_vQ)\geq I:=\frac{Q_0Q'_+(0)}{r_+(0)^3}>0 \tag{\hypertarget{eqn:4.12}{4.12}}
\end{equation*}
The same statement holds for any smaller $\varepsilon>0$, with $C>0$ fixed. Moreover, the statement also holds for any larger $C$, albeit for a smaller $\varepsilon>0$ now also depending on $C>0$.}\\ \\
\textit{Proof.} We first fix a value of $C$---recall that this is to be our $C^0$ bound \hyperlink{eqn:4.8}{(4.8)} on $\phi$. With
$$f(r,\partial_v\phi,Q,\partial_vQ)=-\frac{2Q\partial_vQ}{r^3}+\frac{\partial_v\phi}{r}-\frac{Q^2\partial_v\phi}{r^3}$$
we see that the positive finite quantity
$$S:=\sup\Big\{\big|f(r,\partial_v\phi,Q,\partial_vQ)\big|\hspace{2pt}\Big|\hspace{2pt}\tfrac{1}{2}r_+(0)\leq r\leq 2\|r_+\|_{C^2},\hspace{2pt}|\partial_v\phi|\leq 2\|\phi'_+\|_{C^1},\hspace{2pt}|Q|,|\partial_vQ|\leq\|Q_+\|_{C^2}\Big\} $$
depends only on $\|r_+\|_{C^2}$, $\|\phi'_+\|_{C^1}$, $\|Q_+\|_{C^2}$ and $r_+(0)$. Our $C^0$ estimate for $\phi$ will be given by \hyperlink{lem:4.1}{Lemma~4.1}, and so we now set
$$C=C(S,\|\phi'_+\|_{C^1[0,1]},I,\delta)>0 $$
where this $C$ is from the statement of \hyperlink{lem:4.1}{Lemma~4.1}. As mentioned in the statement, one may also choose their own (larger) $C>0$, and the proof below is unaffected, except that the final $\varepsilon>0$ will also depend on this choice of $C$.\\ \\
With $C$ fixed, we now study the iteration scheme itself, and derive conditions on $\varepsilon$. Given an element $(r,\varpi,\phi)\in\mathcal{A}(C,\varepsilon)$, we immediately obtain solutions to the defining equations \hyperlink{eqn:4.1}{(4.1)}-\hyperlink{eqn:4.2}{(4.2)} for $\overline{r},\overline{\varpi}$ by integration. One easily verifies that $\overline{r},\overline{\varpi}\in C^1_v(\triangleright_\varepsilon)$. Indeed, for $\overline{\varpi}$, equation \hyperlink{eqn:4.2}{(4.2)} itself is for the derivative $\partial_v\varpi$, which belongs to $C^0(\triangleright_\varepsilon)$. Meanwhile for $\overline{r}$, the right-hand side of \hyperlink{eqn:4.1}{(4.1)} belongs to $C^1_v(\triangleright_\varepsilon)$, and the characteristic data belongs to $C^1[0,1]$, because $r_+\in C^2[0,1]$. Hence it follows that $\overline{r}$ attains the claimed $C^1_v$ regularity. \\ \\
\ul{Lower and upper bounds (4.4)-(4.5) and (4.7) on $\overline{r},\overline{\varpi}$:} Starting with $\overline{r}$, we study equation \hyperlink{eqn:4.1}{(4.1)}, with which we may estimate $\partial_u\overline{r}$---and also $\partial_u\partial_v\overline{r}$ by differentiating in $\partial_v$---in terms of the collection $(r,\varpi,\phi)$. Since $(r,\varpi,\phi)\in\mathcal{A}(C,\varepsilon)$, this collection satisfies all the bounds \hyperlink{eqn:4.4}{(4.4)}-\hyperlink{eqn:4.8}{(4.8)}. It follows that each of $|\partial_u\overline{r}|,|\partial_u\partial_v\overline{r}|$ is bounded by a constant depending only on $\|r_+\|_{C^2[0,1]}$, $\|\phi'_+\|_{C^0[0,1]}$, $\|Q_+\|_{C^1[0,1]}$, $\varpi_0$ and $r_+(0)$. Hence we may choose $\varepsilon>0$ small enough to recover \hyperlink{eqn:4.4}{(4.4)} for $\overline{r}$. For instance, we have
$$|\overline{r}(u,v)|\leq |r_+(v)|+u\|\partial_u\overline{r}\|_{C^0}\leq \|r_+\|_{C^0}+\varepsilon\|\partial_u\overline{r}\|_{C^0}$$ and so a small enough $\varepsilon$, depending only on $\|r_+\|_{C^2}$, $\|\phi'_+\|_{C^0}$, $\|Q_+\|_{C^1}$, $\varpi_0$ and $r_+(0)$, suffices.\\ \\
Next, for $\overline{\varpi}$, we estimate the derivative $\partial_v\overline{\varpi}$ directly (not by integration). Here we use the fact that $\partial_v\phi$ vanishes on $u=v$, so we have
$$|\partial_v\phi(u,v)| \leq(v-u)\|\partial^2_v\phi\|_{C^0}\leq3\varepsilon\|\phi'_+\|_{C^2[0,1]}$$
$$|\partial_vQ(u,v)|\leq |\partial_vQ(0,v)|\leq \varepsilon\|Q_+\|_{C^2[0,1]} $$
Applying these bounds in \hyperlink{eqn:4.2}{(4.2)}, we have
$$|\partial_v\overline{\varpi}|\leq 3\varepsilon\|r_+\|_{C^2[0,1]}\|\phi'_+\|_{C^1[0,1]}\Big(1+4r_+(0)^{-1}(1+2|\varpi_0|)+20r_+(0)^{-2}\|Q_+\|^2_{C^1[0,1]}\Big)\vspace{-2mm}$$
$$\qquad+2r_+(0)^{-1}\|Q_+\|^2_{C^1[0,1]} $$
so that because $\|\overline\varpi\|_{C^1_v(\triangleright_\varepsilon)}\leq |\varpi_0|+(1+\varepsilon)\|\partial_v\overline{\varpi}\|_{C^0}$, we likewise recover \hyperlink{eqn:4.5}{(4.5)} for $\overline{\varpi}$, with a small enough $\varepsilon>0$ depending only on $\|r_+\|_{C^2[0,1]}$, $\|\phi'_+\|_{C^1[0,1]}$, $\|Q_+\|_{C^2[0,1]}$, $\varpi_0$ and $r_+(0)$.\\ \\
With \hyperlink{eqn:4.4}{(4.4)} established, we can also recover the lower bound for $\overline{r}$:
\begin{align*}
|\overline{r}(u,v)-r_+(0)|&\leq |\overline{r}(u,v)-\overline{r}(0,v)|+|\overline{r}(0,v)-r_+(0)|\\
&\leq\varepsilon(\|\partial_u\overline{r}\|_{C^0}+\|r'_+\|_{C^0})\\
&\leq\varepsilon\Big(e^C\big(1+4r_+(0)^{-1}(1+2|\varpi_0|)+20r_+(0)^{-2}\|Q_+\|^2_{C^2[0,1]}\big)+\|r_+\|_{C^2[0,1]}\Big)
\end{align*}
which, for small enough $\varepsilon$, is at most $\tfrac{1}{2}r_+(0)$. We obtain \hyperlink{eqn:4.7}{(4.7)}.
\\ \\
\ul{Lower bound (4.12):} 
To establish \hyperlink{eqn:4.12}{(4.12)}, note that the lower bound $I$ is precisely half the value of $|f|$ evaluated at $(u,v)=(0,0)$. We have just shown above that
$$|\overline{r}(u,v)-r_+(0)|\lesssim_{\|r_+\|_{C^2},\|Q_+\|_{C^2},C,r_+(0),\varpi_0} \hspace{4pt}\varepsilon$$
and it follows similarly that
$$|Q(u,v)-Q_0|,\hspace{4pt}|\partial_vQ(u,v)-Q'_+(0)|\lesssim_{\|Q_+\|_{C^2}} \hspace{4pt}\varepsilon$$
We now estimate 
\begin{align*}\big|-f(\overline{r},\partial_v\phi,Q,\partial_vQ)-2I\big|&\leq \bigg|\frac{2Q\partial_vQ}{\overline{r}^3}-2I\bigg|+\bigg|\frac{\partial_v\phi}{\overline{r}}-\frac{Q^2\partial_v\phi}{\overline{r}^3}\bigg| 
\end{align*}
We shall show that both collections of terms on the right-hand side are at most $\tfrac{1}{2}I$, for small enough $\varepsilon$. The latter collection of terms each has a factor of $\partial_v\phi$, which, as in the previous step, is bounded above by a multiple of $\varepsilon$. Meanwhile, the first collection of terms is
$$\bigg|\frac{2Q\partial_vQ}{\overline{r}^3}-2I\bigg|=\bigg|\frac{2Q\partial_vQ}{\overline{r}^3}-\frac{2Q_0Q'_+(0)}{r_+(0)^3}\bigg| $$
By the bounds mentioned above for the differences between $\overline{r},Q,\partial_vQ
$ and their values at (0,0), this too is at most $\tfrac{1}{2}I$ for small enough $\varepsilon$. Hence we conclude that the lower bound \hyperlink{eqn:4.12}{(4.12)} holds for the intermediate collection $(\overline{r},\overline{\varpi},\phi)$.\\ \\
\ul{Obtaining $\overline{\phi}$ and conditions (4.6),(4.8):} It is now immediate from \hyperlink{lem:4.1}{Lemma~4.1} that equation \hyperlink{eqn:4.3}{(4.3)} for $\overline{\phi}$ admits a unique solution on $\triangleright_\varepsilon$, taking $\xi_+=\phi'_+$ and $f=f(\overline{r},\partial_v\phi,Q,\partial_vQ)$. We also obtain estimate \hyperlink{eqn:4.6}{(4.6)} and, by our earlier choice of $C$, estimate \hyperlink{eqn:4.8}{(4.8)}. The $\varepsilon$ provided by \hyperlink{lem:4.1}{Lemma~4.1} depends only on $\|r_+\|_{C^2}$, $\|\phi'_+\|_{C^1}$, $\|Q_+\|_{C^2}$, $\varpi_0,Q_0,\delta,r_+(0)$ and $r'_+(0)$ (one checks that $\|\partial_vf\|_{C^0}$ is bounded in terms of all these quantities). We conclude that, by choosing $\varepsilon$ small enough to satisfy all the above conditions, the iteration scheme $\Psi$ is indeed well-defined.\hfill$\square$\\ \\ \\
\large\hypertarget{sec:4.4}{\textbf{4.4\hspace{4mm}Contraction estimates}}\label{4.4}\normalsize\\ \\
We now turn to establishing contraction estimates for the iteration scheme $\Psi$. We will consider two collections $(r_i,\varpi_i,\phi_i)_{i=1,2}$ belonging to $\mathcal{A}(C,\varepsilon)$ and study the difference between the new iterates $(\overline r_i,\overline\varpi_i,\overline\phi_i)_{i=1,2}$. For this we adopt the notation $[\hspace{2pt}\cdot\hspace{2pt}]$, so that for instance
$$[r](u,v):=r_1(u,v)-r_2(u,v) $$
and use this consistently in the sequel without further comment.
\\ \\
\hypertarget{lem:4.4}{\textbf{Lemma 4.4.}} \textit{For $i=1,2$, let $\overline{r}_i\in C^1_v(\triangleright_\varepsilon)$ and $\phi_i\in C^2_v(\triangleright_\varepsilon)$ satisfy \hyperlink{eqn:4.4}{(4.4)} and \hyperlink{eqn:4.6}{(4.6)}-\hyperlink{eqn:4.8}{(4.8)}, with $\overline{r}_i$ arising from the iteration scheme $\Psi$ (in particular, the lower bound \hyperlink{eqn:4.12}{(4.12)} holds). Let $\overline{\phi}_i\in C^2_v(\triangleright_\varepsilon)$ be the unique solution to 
$$\partial_u\partial_v\overline{\phi}_i=e^{\overline{\phi}_i}f(\overline{r}_i,\partial_v\phi_i,Q,\partial_vQ)\qquad \partial_v\overline{\phi}_i\big|_{u=v}=0\qquad\partial_v\overline{\phi}_i\big|_{u=0}=\phi'_+ $$ on $\triangleright_\varepsilon$. Then we have the estimate
$$\|[\overline{\phi}]\|_{C^2_v(\triangleright_\varepsilon)} \lesssim \|[\overline{r}]\|_{C^1_v(\triangleright_\varepsilon)}+\varepsilon\|[\partial_v\phi]\|_{C^1_v(\triangleright_\varepsilon)}$$
where the constant in the inequality depends only on $\|r_+\|_{C^2[0,1]}$, $\|\phi'_+\|_{C^1[0,1]}$, $\|Q_+\|_{C^2[0,1]}$, $Q_0$, $\delta$, $r_+(0)$ and $r'_+(0)$.}\\ \\
\textit{Proof.} We use only the fact that $f$ is a smooth function of its arguments on $\mathbb{R}_{>0}\times\mathbb{R}^3$ and is only evaluated on a compact subset of this domain, and that for $i=1,2$ the lower bound $-f(\overline{r}_i,\partial_v\phi_i,Q,\partial_vQ)\geq I$ holds on $\triangleright_\varepsilon$. In the sequel, the symbol $\lesssim$ shall always be used as in the lemma statement, i$.$e$.$ with the implicit constant depending (only) on the quantities listed in the statement.\\ \\
To begin with, we observe that 
\begin{equation*}
\|[f(\overline{r},\partial_v\phi,Q,\partial_vQ)]\|_{C^0}\lesssim\|[\overline{r}]\|_{C^0}+\|[\partial_v\phi]\|_{C^0}\tag{\hypertarget{eqn:4.13}{4.13}}
\end{equation*}
\begin{equation*}
\|\partial_v[f(\overline{r},\partial_v\phi,Q,\partial_vQ)]\|_{C^0}\lesssim \|[\overline{r}]\|_{C^1_v}+\|[\partial_v\phi]\|_{C^1_v}\tag{\hypertarget{eqn:4.14}{4.14}}
\vspace{2mm}\end{equation*}
In the second estimate, when we apply the derivative $\partial_v$ (which commutes with $[\hspace{2pt}\cdot\hspace{2pt}]$), the chain rule yields an expression in first derivatives of $f$, and $\partial_v$ derivatives of $\overline{r}$, etc. In computing the difference between $i=1,2$, we use the mean value inequality to estimate differences in the $\partial f$ terms in terms of second derivatives of $f$. As just mentioned, $f$ is evaluated on a compact domain determined by $\|r_+\|_{C^2}$, $\|\phi'_+\|_{C^1}$, $\|Q_+\|_{C^2}$, $r_+(0)$, so its derivatives are bounded, yielding \hyperlink{eqn:4.14}{(4.14)} with the claimed (implicit) constant.\\ \\
We now outline the strategy before giving the detailed argument. In estimating $[\overline{\phi}]$, the key step is using the boundary conditions $[\partial_v\overline{\phi}]|_{u=v}=[\partial_v\overline{\phi}]|_{u=0}=0$ to obtain an identity for $\overline{\phi}$. Indeed, these boundary conditions entail that, for each $v\in[0,\varepsilon]$,
$$[\partial_v\phi](v,v)-[\partial_v\phi](0,v)\equiv\int^v_0[\partial_u\partial_v\overline{\phi}](\tilde{u},v)d\tilde{u}\equiv \int^v_0[e^{\overline{\phi}}f(\overline{r},\partial_v\phi,Q,\partial_vQ)](\tilde{u},v)d\tilde{u} =0$$
Differentiating in $v$ yields a boundary term:
$$[e^{\overline{\phi}}f(\overline{r},\partial_v\phi,Q,\partial_vQ)](v,v)+\int^v_0\partial_v[e^{\overline{\phi}}f(\overline{r},\partial_v\phi,Q,\partial_vQ)](\tilde{u},v)d\tilde{u}=0 $$
which allows us to directly estimate $[\overline{\phi}]$.\\ \\
In detail now, we follow Christodoulou's argument/notation in \hyperlink{Chr96a}{[Chr96a]}, and estimate $[\overline{\phi}]$ and $[\partial_v\overline{\phi}]$ simultaneously, writing
$$\triangle(u,v):=|[\overline{\phi}](u,v)|+|[\partial_v\overline{\phi}](u,v)| $$
Since $[\partial_v\overline{\phi}]|_{u=0}=0$, we have
$$[\partial_v\overline{\phi}](u,v)=\int^u_0[e^{\overline{\phi}}f(\overline{r},\partial_v\phi,Q,\partial_vQ)](\tilde{u},v)d\tilde{u} $$
so
\begin{align*}
|[\partial_v\overline{\phi}](u,v)|&\leq \int^u_0|[e^{\overline{\phi}}]f(\overline{r}_1,\partial_v\phi_1,Q,\partial_vQ)|(\tilde{u},v)d\tilde{u}+\varepsilon \|e^{\overline{\phi}_2}[f(\overline{r},\partial_v\phi,Q,\partial_vQ)]\|_{C^0}\\
&\leq e^CS\int^u_0\triangle(\tilde{u},v)d\tilde{u}+\varepsilon e^C \|[f(\overline{r},\partial_v\phi,Q,\partial_vQ)]\|_{C^0}\\
&\lesssim \int^u_0\triangle(\tilde{u},v)d\tilde{u}+\varepsilon\Big( \|[\overline{r}]\|_{C^0}+\|[\partial_v\phi]\|_{C^0}\Big)
\end{align*}
In the second line, we used the mean value estimate
$|[e^{\overline{\phi}}]|\leq e^C|[\overline{\phi}]| $ and in the third, we used \hyperlink{eqn:4.13}{(4.13)}.\\ \\
Moving onto $[\overline{\phi}]$, we integrate from $u=v$, to obtain
$$[\overline\phi](u,v)=\int^v_u[\partial_v\overline\phi](u,\tilde{v})d\tilde{v}+[\overline\phi](u,u) $$ 
We now use the identity mentioned above, writing
$$[e^{\overline\phi}](u,u)f(\overline{r}_1,\partial_v\phi_1,Q,\partial_vQ)(u,u)+e^{\overline\phi_2(u,u)}[f(\overline{r},\partial_v\phi,Q,\partial_vQ)](u,u) $$
$$\qquad\qquad+\int^u_0\partial_v[e^{\overline\phi}f(\overline{r},\partial_v\phi,Q,\partial_vQ)](\tilde{u},u)d\tilde{u}=0 $$
(It is the second term that will prevent us from obtaining a factor of $\varepsilon$ on the right-hand side of our final estimate.) We can make $[e^{\overline\phi}](u,u)$ the subject of this, and use this to estimate $[\overline{\phi}](u,u)$, using our lower bound $|f|\geq I$, and noting that, since $\overline\phi_i\in[-C,C]$, we have
$$|[\overline\phi]|=|[\log(e^{\overline\phi})]|\leq e^C|[e^{\overline\phi}]| $$
Now we estimate
\begin{align*}
|[\overline\phi](u,v)|&\leq \int^v_0|[\partial_v\overline\phi(u,\tilde{v})]|d\tilde{v}+|[\overline\phi](u,u)|\\
&\leq \int^v_0\triangle(u,\tilde{v})d\tilde{v}+e^C|[e^{\overline\phi}](u,u)|
\end{align*}
\begin{align*}
&\leq \int^v_0\triangle(u,\tilde{v})d\tilde{v}+e^CI^{-1}\bigg(e^C|[f(\overline{r},\partial_v\phi,Q,\partial_vQ)](u,u)|\\
&\hspace{6mm}+\int^u_0\Big|\partial_v[e^{\overline\phi}]f(\overline{r}_1,\partial_v\phi_1,Q,\partial_vQ)+[e^{\overline\phi}]\partial_v\big(f(\overline{r}_1,\partial_v\phi_1,Q,\partial_vQ)\big)\\
&\hspace{12mm}+(\partial_ve^{\overline\phi_2})[f(\overline{r},\partial_v\phi,Q,\partial_vQ)]+e^{\overline\phi_2}\partial_v[f(\overline{r},\partial_v\phi,Q,\partial_vQ)]\Big|(\tilde{u},u)d\tilde{u}\bigg)\\
&\lesssim \int^v_0\triangle(u,\tilde{v})d\tilde{v}+\int^u_0\triangle(\tilde{u},u)d\tilde{u}+\|[\overline{r}]\|_{C^1_v}+ \varepsilon\|[\partial_v\phi]\|_{C^1_v}
\end{align*}
In arriving at the final line, the difference terms in $\overline\phi$ contribute the integral $\int^u_0\triangle(\tilde{u},u)d\tilde{u}$, while the difference terms in $f$ are estimated by \hyperlink{eqn:4.13}{(4.13)} and \hyperlink{eqn:4.14}{(4.14)}. We draw special attention to the only term not integrated, namely $[f(\overline{r},\partial_v\phi,Q,\partial_vQ)](u,u)|$. We do \textit{not} use \hyperlink{eqn:4.13}{(4.13)} to estimate this, or we would be left with a term in $\|[\partial_v\phi]\|_{C^1_v}$ without a factor of $\varepsilon$. Instead, we note carefully that, since $[\partial_v\phi]|_{u=v}=0$, this term is controlled only by the differences in $\overline{r}$, which is why only $\|[\partial_v\overline\phi]\|_{C^1_v}$ appears with a factor of $\varepsilon$ in the final line.\\ \\
Combining our estimates for $[\overline\phi]$ and $[\partial_v\overline\phi]$, we have
$$\triangle(u,v)\lesssim \int^u_0\triangle(\tilde{u},v)d\tilde{u}+\int^v_0\triangle(u,\tilde{v})d\tilde{v}+\int^u_0\triangle(\tilde{u},u)d\tilde{u}+\|[\overline{r}]\|_{C^1_v}+  \varepsilon\|[\partial_v\phi]\|_{C^1_v} $$
We can now use this to apply Gronwall's inequality to $\sup_{u+v\leq2t}\triangle$, which yields
$$|[\overline{\phi}](u,v)|+|[\partial_v\overline{\phi}](u,v)| \lesssim \|[\overline{r}]\|_{C^1_v}+ \varepsilon\|[\partial_v\phi]\|_{C^1_v}$$
Finally, estimating $[\partial^2_v\overline{\phi}]$ is almost immediate from \hyperlink{eqn:4.14}{(4.14)} and the fact that $[\partial^2_v\overline{\phi}]|_{u=0}=0$. Indeed, integrating from $u=0$ gives
$$[\partial^2_v\overline{\phi}](u,v)=\int^u_0\partial_v[e^{\overline{\phi}}f(\overline{r},\partial_v\phi,Q,\partial_vQ)](\tilde{u},v)d\tilde{u} $$
and so 
$$|[\partial^2_v\overline{\phi}](u,v)|\lesssim \varepsilon\Big(\|[\overline{r}]\|_{C^1_v}+\|[\partial_v\phi]\|_{C^1_v}\Big)$$
Combining this with the above estimates for $|[\overline{\phi}]|$, $|[\partial_v\overline{\phi}]|$, we arrive at the claimed estimate for $\|[\overline{\phi}]\|_{C^2_v}$.\hfill$\square$
\\ \\ 
\hypertarget{lem:4.5}{\textbf{Lemma 4.5.}} \textit{For a possibly smaller $\varepsilon>0$ than in \hyperlink{lem:4.3}{Lemma~4.3}, still depending only on $\|r_+\|_{C^2}$, $\|\phi'_+\|_{C^1}$, $\|Q_+\|_{C^2}$, $\varpi_0,Q_0,\delta,r_+(0)$ and $r'_+(0)$, the iteration scheme $\Psi$ obtained in \hyperlink{lem:4.3}{Lemma~4.3} is contractive with respect to the norm on $C^1_v(\triangleright_\varepsilon)^3\times C^2_v(\triangleright_\varepsilon)$: for two iterates $(r_1,\varpi_1,\phi_1),(r_2,\varpi_2,\phi_2)$, one has
$$\|[\overline{r}]\|_{C^1_v}+\|[\overline{\varpi}]\|_{C^1_v}+\|[\overline{\phi}]\|_{C^2_v} \leq\tfrac{1}{2}\big(\|[r]\|_{C^1_v}+\|[\varpi]\|_{C^1_v}+\|[\phi]\|_{C^2_v}\big)$$}
\hspace{-4pt}\textit{Proof.} As in the previous proof, whenever we use the symbol $\lesssim$ in the sequel, the implied constant depends on all the quantities listed in the lemma statement.\\ \\
With \hyperlink{lem:4.4}{Lemma~4.4} in hand, we only need to prove that
$$\|[\overline{r}]\|_{C^1_v},\hspace{2pt}\|[\overline{\varpi}]\|_{C^1_v}\lesssim \varepsilon\Big(\|[r]\|_{C^1_v}+\|[\varpi]\|_{C^1_v}+\|[\phi]\|_{C^2_v}\Big) $$
and then the existence of an $\varepsilon>0$, depending on the correct constants, and small enough to guarantee a contraction, follows.\\ \\
For $[\overline{r}]$, we simply examine the right-hand side of \hyperlink{eqn:4.1}{(4.1)} to see that
$$|\partial_u[\overline{r}]|\lesssim \|[r]\|_{C^0}+\|[\varpi]\|_{C^0}+\|[\phi]\|_{C^0} $$
and, by differentiating \hyperlink{eqn:4.1}{(4.1)},
$$|\partial_u[\partial_v\overline{r}]|\lesssim \|[r]\|_{C^1_v}+\|[\varpi]\|_{C^1_v}+\|[\phi]\|_{C^1_v} $$
Since $[\overline{r}]|_{u=0}=[\partial_v\overline{r}]|_{u=0}=0$, integrating in $u$ gives us the desired factor of $\varepsilon$.\\ \\
It remains to give the estimate for $[\overline{\varpi}]$ and, as in \hyperlink{lem:4.3}{Lemma~4.3}, we need to use the smallness of $\partial_v\phi$, namely
$$\|[\partial_v\phi]\|_{C^0}\leq\varepsilon\|[\partial^2_v\phi]\|_{C^0},\qquad \|\partial_v\phi\|_{C^0}\leq\varepsilon\|\partial^2_v\phi\|_{C^0} $$
We also make use of the $\overline{r}$ on the right-hand side of \hyperlink{eqn:4.2}{(4.2)}, which is used to gain a factor of $\varepsilon$. From equation \hyperlink{eqn:4.2}{(4.2)}, we estimate
\begin{align*}
\|[\partial_v\overline\varpi]\|_{C^0}&\lesssim\|[\partial_v\phi]\|_{C^0}+\|\partial_v\phi\|_{C^0}\big(\|[r]\|_{C^0}+\|[\varpi]\|_{C^0}\big)+\|[\overline r]\|_{C^0}\\
&\leq \varepsilon\|[\partial^2_v\phi]\|_{C^0}+\varepsilon\|\partial^2_v\phi\|_{C^0}\big(\|[r]\|_{C^0}+\|[\varpi]\|_{C^0}\big)+\varepsilon\big(\|[r]\|_{C^0}+\|[\varpi]\|_{C^0}+\|[\phi]\|_{C^0}\big)
\end{align*}
Since also $[\overline\varpi]|_{u=v}=0$, we have\vspace{-3mm}
$$\|[\overline{\varpi}]\|_{C^1_v}\lesssim \varepsilon\Big(\|[r]\|_{C^1_v}+\|[\varpi]\|_{C^1_v}+\|[\phi]\|_{C^2_v}\Big) $$ as required.\hfill$\square$\\ \\ \\
\large\hypertarget{sec:4.5}{\textbf{4.5\hspace{4mm}Proof of Theorem 1.3}}\label{4.5}\normalsize\\ \\
With properties of the iteration scheme now established, we now combine all the foregoing lemmas to provide a proof of \hyperlink{pro:4.1}{Proposition~4.1}, on the existence and uniqueness of solutions to the $(r,\varpi,\phi,Q)$ system.\\ \\
\textit{Proof of Proposition 4.1.} We first examine \hyperlink{def:4.3}{Definition~4.3}, seeing that, for any $\varepsilon\in(0,1]$ and $C>0$, the function space $\mathcal{A}(C,\varepsilon)$ is a closed subset of the Banach space $C^1_v(\triangleright_\varepsilon)^3\times C^2_v(\triangleright_\varepsilon)$. Next, from \hyperlink{lem:4.3}{Lemmas~4.3} and \hyperlink{lem:4.5}{4.5}, we obtain $\varepsilon,C>0$, depending only on $\|r_+\|_{C^2}$, $\|\phi'_+\|_{C^1}$, $\|Q_+\|_{C^2}$, $\varpi_0,Q_0,\delta,r_+(0)$ and $r'_+(0)$, and a contractive self-map $\Psi$ defined on $\mathcal{A}(C,\varepsilon)$. If we can argue that $\mathcal{A}(C,\varepsilon)$ is non-empty, then the Banach contraction mapping principle yields a unique fixed point $(r,\varpi,\phi)$.\\ \\
To this end, define the collection $(r,\varpi,\phi)$ on $\triangleright$ by 
$$r(u,v)=r_+(v),\qquad \varpi(u,v)=\varpi_0 $$
$$\phi(u,v)=\int^v_u\phi'_+(\tilde{v})d\tilde{v}-(v-u)\phi'_+(u) $$
The choice of $\phi$ here is equivalent to the choices
$$\partial_v\phi(u,v)=\phi'_+(v)-\phi'_+(u),\qquad \phi|_{u=v}=0$$
which are a straightforward way to ensure that $\partial_v\phi$ attains its boundary values. Except for condition \hyperlink{eqn:4.8}{(4.8)}, it is immediate to verify that this collection defines an element of $\mathcal{A}(C,\varepsilon)$. However, condition \hyperlink{eqn:4.8}{(4.8)} may be arranged by simply enlarging $C>0$, at the cost of possibly reducing $\varepsilon>0$ (see comments at the end of \hyperlink{lem:4.3}{Lemma~4.3}). Since this adjustment to $C>0$ is determined by the size of $\|\phi'_+\|_{C^0}$ only, the final $\varepsilon>0$ still depends on the advertised quantities only.\\ \\
We now have a contractive self-map defined on a non-empty function space $\mathcal{A}(C,\varepsilon)$, and so existence and uniqueness is established.\hfill$\square$\\ \\
\textit{Proof of Theorem 1.3.} Given hard-edge seed data $\varpi(v),Q(v)$ as in the theorem statement, we obtain admissible characteristic data $(r_+,\phi'_+,Q_+)$ in the sense of \hyperlink{def:4.1}{Definition~4.1} (though possibly on a smaller domain), as detailed in \hyperlink{sec:2.5}{Sections~2.5} and \hyperlink{sec:4.1}{4.1}.\\ \\
By \hyperlink{pro:4.1}{Proposition~4.1}, we obtain $\varepsilon>0$ and a unique solution $r,\varpi\in C^1(\triangleright_\varepsilon)$, $\phi\in C^1(\triangleright_\varepsilon)\cap C^2_v(\triangleright_\varepsilon)$ to the $(r,\varpi,\phi,Q)$ system with these characteristic data. Since we chose seed data forming a bounce in the exterior region, we have $\partial_vr(0,0)>0$, and so, after possibly reducing $\varepsilon$ further, $\partial_vr>0$ holds throughout $\triangleright_\varepsilon$. We may therefore recover a solution to the $(r,\Omega^2,Q)$ system on $\triangleright_\varepsilon$, as detailed again in \hyperlink{sec:2.5}{Section~2.5}. This completes the construction of region $(I)$.\\ \\
Finally, attaching regions $(I\hspace{-2.5pt}I)$-$(I\hspace{-2.5pt}I\hspace{-2.5pt}I)$ is identical to the same process in the proof of \hyperlink{thm:1.1}{Theorem~1.1}, in the lowest regularity case ($C^1$ gluing, and no better, between regions $(I)$ and $(I\hspace{-2.5pt}I)$). We have therefore obtained the desired $(\mathcal{M},g_{\mu\nu})$.\hfill$\square$
\newpage
\Large\textbf{References}\normalsize \\ \\
\setlength{\tabcolsep}{0em}
\begin{tabularx}{\textwidth} { 
   >{\raggedright\arraybackslash \hsize=.1\textwidth}X 
   >{\raggedright\arraybackslash}X  }
$[$Bic25$]$ & D. Bick. \hypertarget{Bic25}{Caustics in the spherically symmetric Einstein-dust system.} arXiv:2512.07812, 2025.\vspace*{2mm} \\
$[$BI91$]$ & C. Barrab\`es and W. Israel. \hypertarget{BI91}{Thin shells in general relativity and cosmology: the lightlike limit.} \textit{Physical Review D}, 43:1129, 1991.\vspace*{2mm} \\
$[$Bur89$]$ & G. A. Burnett. \hypertarget{Bur89}{The high-frequency limit in general relativity.} \textit{J. Math. Phys.}, 30(1):90–96, 1989.\vspace*{2mm} \\
$[$BV70$]$ & W. B. Bonnor and P. C. Vaidya. \hypertarget{BV70}{Spherically symmetric radiation of charge in Einstein-Maxwell theory.} \textit{General Relativity and Gravitation}, 1(2):127-130, 1970.\vspace*{2mm} \\
$[$CB17$]$ & B. Creelman and I. Booth. \hypertarget{CB17}{Collapse and bounce of null fluids.} \textit{Physical Review D}, 95:124033, 2017.\vspace*{2mm} \\
$[$CD87$]$ & C. J. S. Clarke and T. Dray. \hypertarget{CD87}{Junction conditions for null hypersurfaces.} \textit{Classical and Quantum Gravity}, 4:265-275, 1987.\vspace*{2mm} \\
$[$Chr95$]$ & D. Christodoulou. \hypertarget{Chr95}{Self-gravitating relativistic fluids: A two-phase model.} \textit{Arch. Ration. Mech. Anal.}, 130(4):343–400, 1995.\vspace*{2mm} \\
$[$Chr96a$]$ &  D. Christodoulou. \hypertarget{Chr96a}{Self-gravitating relativistic fluids: The continuation and termination of a free phase boundary.} \textit{Arch. Ration. Mech. Anal.}, 133(4):333–398, 1996.\vspace*{2mm} \\
$[$Chr96b$]$ & D. Christodoulou. \hypertarget{Chr96b}{Self-gravitating relativistic fluids: The formation of a free phase boundary in the phase transition from soft to hard.} \textit{Arch. Ration. Mech. Anal.}, 134(2):97–154, 1996.\vspace*{2mm} \\
$[$CGV16$]$ & S. Chatterjee, S. Ganguli and A. Virmani. \hypertarget{CGV16}{Charged Vaidya solution satisfies weak energy condition} \textit{General Relativity and Gravitation}, 48:91, 2016.\vspace*{2mm} \\
$[$CL14$]$ & D. Christodoulou and A. Lisibach. \hypertarget{CL14}{Self-Gravitating Relativistic Fluids: The Formation of a Free Phase Boundary in the Phase Transition from Hard to Soft.} arXiv:1411.4888, 2014.\vspace*{2mm} \\
$[$Daf14$]$ & M. Dafermos. \hypertarget{Daf14}{Black holes without spacelike singularities.} \textit{Commun. Math. Phys.}, 332:729–757, 2014.\vspace*{2mm} \\
$[$DI67$]$ &  V. De la Cruz and W. Israel. \hypertarget{DI67}{Gravitational bounce.} \textit{Il Nuovo Cimento A (1965-1970)}, 51.3:744–760, 1967.\vspace*{2mm} \\
$[$Dra90$]$ & T. Dray. \hypertarget{Dra90}{Bouncing shells.} \textit{Classical and Quantum Gravity}, 7:131, 1990.\vspace*{2mm} \\
$[$Eva10$]$ & L. C. Evans. \hypertarget{Eva10}{\textit{Partial Differential Equations (2nd Ed)}}. American Mathematical Society, 2010.\vspace*{2mm} \\
$[$FH79$]$ & C. J. Farrugia and P. Hajicek. \hypertarget{FH79}{The third law of black hole mechanics: a counterexample.} \textit{Comm. Math. Phys}, 68.3:291-299, 1979.\vspace*{2mm} \\
$[$HE73$]$ & S. W. Hawking and G. F. R. Ellis. \hypertarget{HE73}{\textit{The large scale structure of space-time.}} Cambridge University Press, 1973. \vspace*{2mm} \\
$[$His81$]$ & W. Hiscock. \hypertarget{His81}{Evolution of the interior of a charge black hole.} \textit{Physics Letters A}, 83A(3):110-112, 1981.\vspace*{2mm} \\
\end{tabularx}
\begin{tabularx}{\textwidth} { 
   >{\raggedright\arraybackslash \hsize=.1\textwidth}X 
   >{\raggedright\arraybackslash}X  }
$[$Isr65$]$ & W. Israel. \hypertarget{Isr65}{Singular hypersurfaces and thin shells in general relativity.} \textit{Nuovo Cimento B (1965-1970)}, 44:1-14, 1965.\vspace*{2mm} \\
$[$Kha11$]$ &  S. Khakshournia, \hypertarget{Kha11}{Collapsing spherical null shells in general relativity.} \textit{Iranian Journal of Physics Research}, 10:4, 2011.\vspace*{2mm} \\
$[$KU22$]$ & C. Kehle and R. Unger. \hypertarget{KU22}{Gravitational collapse to extremal black holes and the third law of black hole thermodynamics.} \textit{J. Eur. Math. Soc.}, 2022.\vspace*{2mm} \\
$[$KU24$]$ &  C. Kehle and R. Unger, \hypertarget{KU24}{Extremal black hole formation as a
critical phenomenon.} arXiv:2402.10190.\vspace*{2mm} \\
$[$LR20$]$ & J. Luk and I. Rodnianski. \hypertarget{LR20}{High-frequency limits and null dust shell solutions in general relativity.} arXiv:2009.08968.\vspace*{2mm} \\
$[$LZ91$]$ & K. Lake and T. Zannias. \hypertarget{LZ91}{Structure of singularities in the spherical gravitational collapse of a charged null fluid.} \textit{Physical Review D,} 43.6:1798, 1991.\vspace*{2mm}\\
$[$Mos17$]$ & G. Moschidis. \hypertarget{Mos17}{The Einstein–null dust system in spherical symmetry with an inner mirror: structure of the maximal development and Cauchy stability.} arXiv: 1704.08685, 2017.\vspace*{2mm} \\
$[$Mos20$]$ & G. Moschidis. \hypertarget{Mos20}{A proof of the instability of AdS for the Einstein-null dust system with an inner mirror.} \textit{Anal. Part. Diff. Eq.}, 13.6:1671–1754, 2020.\vspace*{2mm} \\
$[$Ori91$]$ & A. Ori. \hypertarget{Ori91}{Charged null fluid and the weak energy condition.} \textit{Classical and Quantum Gravity}, 8:1559–1575, 1991.\vspace*{2mm} \\
$[$PI90$]$ & E. Poisson and W. Israel. \hypertarget{PI90}{Internal structure of black holes.} \textit{Physical Review D}, 41(6):1796-1809, 1990.\vspace*{2mm} \\
$[$Pr\'o83$]$ & M. Pr\'oszy\'nski. \hypertarget{Pro83}{Thin charged shells and the violation of the third law of black hole mechanics.} \textit{General Relativity and Gravitation}, 15(5):403–415, 1983.\vspace*{2mm} \\
$[$PS68$]$ & J. Plebanski and J. Stachel. \hypertarget{PS68}{Einstein tensor and spherical symmetry.} \textit{Journal of Mathematical Physics}, 9(2):269–283, 1968.\vspace*{2mm} \\
$[$SI80$]$ & B. T. Sullivan and W. Israel. \hypertarget{SI80}{The third law of black hole mechanics: What is it?} \textit{Physical Letters A}, 79.(5-6):371–372, 1980.\vspace*{2mm} \\
$[$Tou25$]$ & A. Touati. \hypertarget{Tou25}{The reverse Burnett conjecture for null dusts.} \textit{Annals of PDE}, 11(22), 2025.\vspace*{2mm} \\
$[$Vai51$]$ & P.C. Vaidya. \hypertarget{Vai51}{The gravitational field of a radiating star.} \textit{Proc. Indian Acad. Sci.}, A33:264, 1951.
\end{tabularx}

\end{document}